\documentclass[binding=0.6cm,LaM,noexaminfo]{sapthesis}
\usepackage{microtype}
\usepackage[T1]{fontenc}
\usepackage{amsmath}
\usepackage{amssymb}
\usepackage{amsfonts}
\usepackage{mathtools}
\usepackage{caption}
\usepackage{slashed}
\usepackage{xcolor}
\usepackage{cite}
\usepackage{tikz} \usetikzlibrary{shapes,arrows,positioning,automata,backgrounds,calc,er,patterns} \usepackage{tikz-feynman} \tikzfeynmanset{compat=1.1.0}
\usepackage{hyperref}
\usepackage{autobreak}
\usepackage{scalerel}
\usepackage{marginnote}
\usepackage[indention=0ex,tableposition=top]{caption}

\DeclareMathOperator{\tr}{Tr}
\let\oldeqref\eqref
\renewcommand{\eqref}[1]{Eq.~\oldeqref{#1}}
\newcommand{\figref}[1]{Fig.~\ref{#1}}
\newcommand{\ord}[1]{\mathcal{O} (\alpha_s^{#1})}
\newcommand{\MSbar}[0]{\overline{\mbox{MS}}}

\hypersetup{pdftitle={Thesis-Laurenti},pdfauthor={Niccolò Laurenti}}
\title{Construction of a next-to-next-to-next-to-leading order approximation for heavy flavour production in deep inelastic scattering with quark masses}
\author{Niccolò Laurenti}
\IDnumber{1750131}
\course{Fisica}
\courseorganizer{Faculty of Mathematics, Physics, and Natural Sciences}
\submitdate{2020/2021}
\AcademicYear{2020/2021}
\copyyear{2021}
\advisor{Dr. Marco Bonvini}
\authoremail{laurenti.1750131@studenti.uniroma1.it}
\versiondate{October 18, 2021}
\examdate{15/10/2021}
\begin{document}
%/tikzfeynman/warn luatex=false
\frontmatter
\maketitle
%\dedication{Dedicated to }
\begin{abstract}
The subject of this thesis is the construction of an approximation for the next-to-next-to-next-to leading order (N$^3$LO) deep inelastic scattering (DIS) massive coefficient function of the gluon for $F_2$ in heavy quark pair production. Indeed, this object is one of the ingredients needed for the construction of any variable flavour number (factorization) scheme at $\ord{3}$. The construction of such scheme is crucial for the improvement of the accuracy of the extraction of the parton distribution functions from the experimental data, that in turn will provide an improvement of the accuracy of all the theoretical predictions in high energy physics.

Despite the function we are interested in is not known exactly, its expansion in some kinematic limits is available. In particular the high-scale limit ($Q^2 \gg m^2$), high-energy limit ($z\rightarrow 0$, where $z$ is the argument of the coefficient function) and threshold limit ($z\rightarrow z_{\rm max}=1/(1+4m^2/Q^2)$) of the exact coefficient function are all known, with the exception of some terms that we will provide in approximate form. Therefore, combining these limits in a proper way, we will construct an approximation for the unknown term of the N$^3$LO gluon coefficient function, that describes the exact curve in the whole range of $z$.

Our approach consists first of all in the construction of an asymptotic limit: this limit approximates the exact coefficient function in the small-$z$ region for all the values of $Q^2$. It will be constructed using the high-scale limit in which we will reinsert the neglected power terms in the small-$z$ limit. In this way we will make sure that the asymptotic limit will approach the exact curve.   

Once that we have the asymptotic limit we will combine it with the threshold limit using two damping functions so that the final result will approach the exact coefficient function both for $z\rightarrow 0$ and for $z\rightarrow z_{\rm max}$. For intermediate values of $z$ the agreement between the approximate and the exact curves will depend on the choice of the damping functions. In order to choose such functions, we will apply our approximation procedure to the NLO and NNLO coefficient functions, that are exactly known, and we will choose the functional form that provides the best agreement between the exact and the approximate curve. 

Since other approximations for the N$^3$LO gluon coefficient functions are present in the literature, we will conclude by comparing our final approximate coefficient functions with such approximations. We will show a comparison both for the NNLO, whose exact function is known, and for the N$^3$LO.
With our approach, we expect our results to be more accurate than previous approximations, thus providing a sufficient precision for a complete description of DIS at N$^3$LO and the consequent determination of N$^3$LO PDFs.

\end{abstract}
\tableofcontents
\mainmatter
\chapter{Introduction} \label{intro}

In the search of physics beyond the Standard Model (BSM) a high accuracy for the study of the Standard Model (SM) processes, both in the experimental measurements and in the theoretical predictions, is required. In fact, if these ones did not have enough precision, in presence of data that slightly deviate from the SM predictions, we wouldn't be able to understand whether this deviation is due to BSM phenomena or not, since it would be within the uncertainty band. 
For this reason the high-energy experimental physics is going towards a high precision phase. Indeed, in the High Luminosity LHC (HL-LHC) era the objective will be to increase the LHC luminosity by a factor of 10 beyond its actual value. This will bring a great improvement in the precision of the experimental measurement, such that the goal will be to reach an accuracy of the order of percent or better. For this reason the theoretical predictions will have to be at least of the same precision, in order to compare theory
with experiments.

In high-energy physics the hadronic cross sections are computed within the factorization theorem as a convolution between the partonic cross sections, in which the external particles are the elementary constituents of the hadron (i.e.\ quarks and gluons) and not the hadron itself, and the parton distribution functions (PDFs), that contain information on the internal structure of the hadron, as it will be explained in the following chapters.
Since at present time the PDFs are one of the main sources of uncertainty in theoretical predictions, PDF determination will have to be more accurate than the one that have been performed so far.

For energies smaller than the typical hadronization scale, i.e.\ $Q \lesssim 1 \,\,\mbox{GeV}$, Quantum Chromodynamics becomes non perturbative. Therefore the PDFs cannot be computed but must be extracted from data. Such procedure is usually called PDF fit. 
It means that in order to increase the precision of the prediction of the cross sections, on the one hand we have to increase the precision in the computation of partonic cross sections, computing them at the order required by the accuracy we want to reach and on the other hand we have to increase the precision of the PDF fit.
A crucial observation is that the PDFs are process-independent objects, which means that they depend only on the energy scale Q characteristic of the interaction and on the fraction of longitudinal proton’s momentum carried by the parton, but not on the particular process the proton is involved into. Thus the strategy is to extract the PDFs from a very clean process, i.e.\ a process that is easy to describe theoretically and to reconstruct experimentally, and then to use them to obtain predictions in more complicated ones, for example like the proton-proton collision at LHC.
Usually the PDFs are fitted from deep inelastic scattering (DIS) data, in which an electron and a proton collide at energy high enough to break the internal structure of the proton and form new hadrons in the final state. The data used in the fits come mostly from DIS experiments at HERA collider. For a generic structure function $F$ for DIS, the factorization theorem takes the form
\begin{equation}
F(x,Q^2)=x\sum_{i=q,\bar{q},g}\int_x^1\frac{dz}{z}\, f_i\Bigl(\frac{x}{z}, Q^2\Bigr)C_i(z,Q^2), \qquad q=\text{light quarks},\label{PDFfit}
\end{equation}
where $f_i$ is the PDF of a parton of type $i$ (quark or gluon), $x$ is the Bjorken's scaling variable and $C_i$ is the coefficient function (the partonic cross section for DIS) that is computed in perturbation theory. 
DIS and \eqref{PDFfit} with its various factors will be further explained in Sec.~\ref{part}. 
All we have to observe now is that, using \eqref{PDFfit}, we can extract the PDFs: measuring the left hand side and computing the $C_i$ up to a given fixed order, we can fit $f_i$ at the scale $Q$.
The uncertainty of our fit will depend on different sources: first of all, on the precision of the computation of the coefficient functions, i.e.\ on the order in perturbation theory at which we have computed them; then, it will depend on the experimental errors associated to the measurement of the left hand side of \eqref{PDFfit}; last, the fit procedure itself, i.e.\ the way the PDFs are extracted from experimental data, is a source of uncertainty. In fact, the functional form that is used to extract the PDFs from \eqref{PDFfit} can in some cases be too constrained and bias the final result.

Another important ingredient for the PDF determination is the Dokshitzer-Gribov-Lipatov-Altarelli-Parisi (DGLAP) evolution equation, that governs the scale dependence of the PDFs. It will be described in Sec.~\ref{DGLAP:sec}. DGLAP equation depends on the splitting functions which are computed in perturbation theory. Solving it with a certain initial condition at the scale $\mu_0$, we can obtain the PDFs at every scale $\mu$. %, i.e.\ its solution gives
%\begin{equation}
%f_i(\mu)=\sum_{j=q,\bar{q},g}U_{ij}(\mu,\mu_0)f_j(\mu_0), \qquad i=q,\bar{q},g,
%\end{equation}
%where $U_{ij}(\mu,\mu_0)$ are the DGLAP evolution factors from the scale $\mu_0$ to the scale $\mu$.
Hence, after fitting the PDFs from \eqref{PDFfit} at a given initial scale $Q_0$, we can obtain them at every scale using DGLAP evolution. Obviously, in order to obtain consistent results, we have to compute DGLAP evolution with the same level of accuracy of the coefficient functions, i.e. at the same order in
perturbation theory.

Nowadays the PDFs are extracted using next-to-next-to-leading order (NNLO) theory: it means that both the coefficient functions and DGLAP evolution are computed at NNLO. However, this level of precision is not enough if we want the accuracy of theoretical predictions to be of the order of percent. Therefore, if we want to reach this accuracy we have to perform a next-to-next-to-next-to-leading order (N$^3$LO) fit. In order to do it we need the coefficient functions and DGLAP evolution at $\ord{3}$. However, not all the ingredients required for such a fit are available in the literature yet.
So far the quark and gluon DIS coefficient functions in which the massive effects are neglected are known exactly up to $\ord{3}$ \cite{Vermaseren_2005}.
Instead, the coefficient functions with the exact quark mass dependence are known exactly at $\ord{2}$ from Ref.~\cite{LAENEN1993162}, but at $\ord{3}$ they are still unknown, even if partial information is available in the literature \cite{Laenen_1999, van_Neerven_2000}. Therefore, if we want to perform a N$^3$LO fit we need either to compute exactly the unknown parts of the quark and gluon exact massive coefficient functions at $\ord{3}$ or to find good approximations for them.
Regarding N$^3$LO DGLAP evolution, in order to compute it at $\ord{3}$ we need the N$^3$LO splitting functions. However, only partial information is available in the literature so far \cite{Moch_2017}.

In this thesis we will construct an approximation for the unknown part of the N$^3$LO massive DIS coefficient function of the gluon for $F_2$ in heavy quark pair production.
Such approximation will be constructed combining the threshold limit ($s \simeq 4m^2$, where $\sqrt s$ is the partonic center-of-mass energy) of the exact coefficient function, with an asymptotic limit ($z\rightarrow 0$) that we will construct in Chapter~\ref{approx}. 
Although we will treat only the gluon coefficient function, because the gluon PDF is dominant at small-$x$, i.e.\ at high-energy, with respect to the quark PDFs, all our methods can be applied to the quark coefficient function to find an analogous approximation.

This thesis is organized as follows: in Chapter \ref{strong:int} we will introduce the basics of Quantum Chromodynamics like renormalization, renormalization group equation, variable flavour number (renormalization) schemes and the parton model. 
In Chapter \ref{fact} we will describe the factorization of collinear divergences, DGLAP equations, factorization of mass logarithms and variable flavour number (factorization) schemes.
In Chapter \ref{approx} we will construct three kinematics limits of the DIS exact massive gluon coefficient function for $F_2$ at $\ord{3}$ in heavy flavour pair production, i.e.\ the high-scale limit ($Q^2 \gg m^2$), the high-energy expansion ($s\rightarrow \infty$) and the threshold limit. Such limits are known with the exception of some terms, for which we will provide approximate results.
Then these limits will be combined using two damping functions to find the final approximation of the unknown term of the $\ord{3}$ exact gluon coefficient function.
In conclusion, in Chapter \ref{results} such approximation will be tested and tuned on the NLO and on the NNLO coefficient functions, that are known, and we will give the results at N$^3$LO. Moreover, we will compare our approximation with the ones already present in the literature.

%%%%%%%%%%%%%%%%%%%%%%%%%%%%%%%%%%%%%%%%%%%%%%%%%%%%%%%%%%%%%%%%%%%%%%%%%%%%

\chapter{Strong Interactions}\label{strong:int}

The Standard Model is the theory that describes the fundamental interactions of the elementary particles. It is a quantum field theory and it is based on the gauge symmetry group
\begin{equation}
    SU(3)_C \otimes SU(2)_W \otimes U(1)_Y,
\end{equation}
where $SU(3)_C$ is the subgroup of the strong interactions and $SU(2)_W \otimes U(1)_Y$ is the one of electroweak interactions. Since in this discussion we are interested in the strong interactions, we will focus only on the strong sector $SU(3)_C$. 

The part of the Standard Model that describes the strong interactions is called Quantum Chromodynamics (QCD). As mentioned before, it is a field theory based on the non abelian gauge symmetry group $SU(3)$ and its charge is called \textit{color}. The fundamental fields of QCD are the ones of the \textit{quarks} and of the \textit{gluons}. The quarks are the matter fields: they are spin $1/2$ particles which belong to the fundamental representation of $SU(3)$, i.e.\ to a triplet. In other words we have three different colors for the fermions (sometimes called blue, red and green). The different kinds of quarks are called \textit{flavours}: all of them have the same quantum numbers under the Lorentz and color groups but different masses. It means that the strong interaction is flavour independent and the differences that arise from the different masses of the quarks are only of kinematical nature.
In addition to the quarks there are their antiparticles that are called \textit{antiquarks}. They belong to the antitriplet representation of $SU(3)$ and their charge is sometimes called \textit{anticolor}.
Then there are the gluons: they are the carriers of the strong interactions, i.e.\ they are the gauge bosons of the $SU(3)$ symmetry. It means that they are spin 1 massless particles that belong to the adjoint representation of $SU(3)$, i.e.\ to an octect. Therefore, there are eight possible colors  for the gluons. We already notice a big difference between QCD and Quantum Electrodynamics (QED): in QED the carrier of the interaction (i.e.\ the photon) does not have a charge under the $U(1)_Q$ symmetry group. It means that we cannot have photon-photon interactions at tree-level (but we can have such interaction at loop level with a fermion loop). Instead, in QCD the gluons are charged under color. Therefore, we can have vertices with three or four gluons, as we will see in Sec.~\ref{QCD}. This difference comes from the fact that for QED the gauge symmetry is abelian, while for QCD it is non abelian.

An important property of strong interactions is \textit{color confinement}: it is the phenomenon that colored particles cannot be observed experimentally. It means that the asymptotic states of the theory can only be singlets of $SU(3)$, i.e.\ colorless particles.
Such particles are called \textit{hadrons} and are composite by two or three quarks.
This makes the computation of the cross sections impossible in the context of perturbative QCD, because of the failure of perturbation theory for energy scales typical of the hadronization scale. A possible way to avoid this problem is the introduction of the so-called \textit{parton model} that will be described in Sec.~\ref{part}. Another important consequence of color confinement is that we can only have indirect experimental evidences of the existence of color: for example in the decay $\pi^0\rightarrow 2\gamma$ and in the $e^+e^-$ annihilation into hadrons, color appears as a multiplicative factor for the reaction rates but it cannot be observed directly. See for example Ref.~\cite{Muta} for a more detailed description.

Another fundamental property of QCD is that the strong coupling goes to zero at high energies. This property is called \textit{asymptotic freedom}.
In fact, in contrast to QED, in which the electromagnetic coupling is almost constant in a wide range of energies, the strong coupling has a strong dependence on the energy scale of the process we are considering. In particular if $\Lambda$ is the typical energy of the hadronic physics, i.e.\ $\Lambda\sim1\,\mbox{GeV}$, then for energies much bigger than $\Lambda$ the strong coupling goes to zero, while for energies comparable or smaller than $\Lambda$, it becomes of $\mathcal{O}(1)$. This has very important consequences: at high energies, being the coupling small, QCD is perturbative and therefore we can compute our observables up to a certain fixed order in perturbation theory, neglecting all the higher orders since they are a small correction.
Instead, at energy scales typical of the hadronization scale QCD becomes non perturbative because in the power series expansion every term is of the same order (or bigger) of the previous one, and therefore we should include infinite terms in order to have reliable predictions. It means that at small energies our theory cannot give predictions.
%In Sec.~\ref{ren} asymptotic freedom will be derived by means of the renormalization group equation.

In this chapter we will describe the basics of QCD, presenting its lagrangian and its main properties.
In Sec.~\ref{QCD} we will write down the various terms composing the lagrangian of QCD, explaining where they come from and their implications.
In Sec.~\ref{ren} we will describe the basics of renormalization. In Sec.~\ref{re:group:eq} we will derive the renormalization group equation and we will explain how it implies asymptotic freedom. In Sec.~\ref{VFNS:ren} we will describe why a variable flavour number (renormalization) scheme is needed and how it is constructed.
In conclusion, in Sec.~\ref{part} we will describe the parton model that is fundamental in order to obtain predictions in hadronic physics.

\section{Quantum Chromodynamics}
\label{QCD}
The QCD interactions are governed by the QCD lagrangian that is
\begin{equation}\label{LQCD}
    \mathcal{L}_{\text{QCD}}=\mathcal{L}_{\text{gauge}}+\mathcal{L}_{\text{CP}}+\mathcal{L}_{\substack{\text{gauge} \\\text{fixing}}}+\mathcal{L}_{\text{ghost}}.
\end{equation}
The first term contains the kinetic terms of both quarks and gluons, the mass terms of the quarks and both the quark-gluon and the gluon-gluon interaction vertices. Its expression is
\begin{equation}\label{Lkin}
    \mathcal{L}_{\text{gauge}}=\sum_{f=\text{flavours}}\overline{\psi}_i^{(f)}\Bigl(i\slashed{D}_{ij} - m^{(f)} \delta_{ij}\Bigr)\psi^{(f)}_j-\frac{1}{4}F_{\mu \nu}^a F^{a,\mu \nu},
\end{equation}
where
\begin{align}
   D_{\mu,ij}&=\delta_{ij} \partial_\mu  - ig_s T^a_{ij}A_\mu^a, \label{der:cov} \\
    F^a_{\mu \nu}&=\partial_\mu A_\nu^a - \partial_\nu A_\mu^a + g_s f^{abc}A_\mu^b A_\nu^c. \label{Fmunu}
\end{align}
Plugging the definitions of Eqs.~(\ref{der:cov}) and (\ref{Fmunu}) into \eqref{Lkin} we find that
\begin{align}
    \mathcal{L}_{\text{gauge}}={}&\sum_{f=\text{flavours}}\Bigl(i\overline{\psi}_i^{(f)}\slashed{\partial}\psi_i^{(f)} - m^{(f)}\overline{\psi}_i^{(f)}\psi_i^{(f)} +g_s\overline{\psi}_i^{(f)}\slashed{A}_{ij}\psi_j^{(f)}\Bigr) \notag \\
    &+\frac{1}{2}A^{a,\mu}(g_{\mu \nu}\partial^2 - \partial_\mu\partial_\nu)A^{a,\nu} - g_sf^{abc}A^{a,\mu}A^{b,\nu}\partial_\mu A^{c}_{\nu} \notag \\
    &-\frac{1}{4}g^{2}_sf^{abe}f^{cde}A^{a,\mu}A^{b,\nu}A^{c}_{\mu} A^d_{\nu},\label{Lkin2}
\end{align}
with $\slashed{A}_{ij}=\gamma^\mu A_{\mu,ij}=\gamma^\mu A_\mu^aT^a_{ij}$, where $\gamma^\mu$ are the Dirac matrices \cite{Peskin:1995ev}.
The objects $\psi^{(f)}_i$ and $A_\mu^a$ appearing in Eqs.~(\ref{Lkin}-\ref{Lkin2}) are respectively the quark and gluon fields. Notice that we have omitted the spinor indices of $\psi^{(f)}_i$ and the dependence on the space-time point $x\equiv x^\mu\equiv (x^0,x^1,x^2,x^3)$ of both $\psi^{(f)}_i$ and $A_\mu^a$.
The sum over the repeated indices is understood. In particular $i,j=1,2,3$ and $a,b,c=1,\dots,8$ are the color indices respectively of the triplet and of the octect of $SU(3)_C$.
%Notice that in passing from \eqref{Lkin} to \eqref{Lkin2} we omitted the color indices but they are still present.
$T^a_{ij}$ are the generators of the fundamental representation of $SU(3)$ and $f^{abc}$ are the structure constants, i.e.\ they satisfy the Lie algebra of the group
\begin{equation}\label{Lie}
    [T^a,T^b]=if^{abc}\,T^c, \qquad a,b,c=1,\dots,8.
\end{equation} Fixing the normalization of the $T^a$ such that 
\begin{equation}\label{TF}
\tr\Bigl(T^a T^b\Bigr)=T_F \delta_{ab},\qquad T_F=\frac{1}{2}.
\end{equation}
we can identify the matrices $T^a$ with the Gell-Mann matrices~\cite{Georgi}. An important property that will be useful in the following is that
\begin{equation}\label{CF}
    T^a_{ik}T^a_{kj}=C_F\delta_{ij}, \qquad C_F=\frac{4}{3},
\end{equation}
where $C_F$ is called \textit{Casimir} of the fundamental representation. Obviously we can consider others representations of $SU(3)$ in addition to the fundamental. A very important one is the \textit{adjoint representation}: if we define
\begin{equation}\label{Adj}
    (T^a_{\text{Adj}})_{bc}=-if^{abc}, \qquad a,b,c=1,\dots,8,
\end{equation}
it is easy to show that they satisfy the algebra of $SU(3)$, i.e.\ \eqref{Lie}. They are the generators of the adjoint representation of $SU(3)$. With this definition we have that
\begin{equation}\label{CA}
    T^a_{bc}T^a_{cd}=C_A\delta_{bd}, \qquad C_A=3,
\end{equation}
and $C_A$ is called Casimir of the adjoint representation.
In \eqref{Lkin2} we can clearly see that the two final terms are the ones that give the three and four gluon vertices that we mentioned at the beginning of this chapter. In QED instead, these terms are absent because, being it an abelian gauge theory, we have that $f^{abc}=0$. This is why in QED we don't have photon-photon interaction at tree-level.

The transformation properties of the fundamental fields are the following:
\begin{align}
   & \psi'_i=U_{ij}\psi_j, \label{psi:trans}\\
    & T^a A'^a_\mu=U\Bigl(T^a A^a_\mu-\frac{i}{g_s} U^{-1}\partial_\mu U\Bigr)U^{\dagger}, \label{A:trans}
\end{align}
where
\begin{equation}
   U\equiv U(x)_{ij}=\Bigl(e^{-i\theta^a(x)T^a}\Bigr)_{ij}.
\end{equation}
Using \eqref{A:trans} we can show that
\begin{equation}
    F'^a_{\mu \nu}T^a=UF^a_{\mu \nu}T^aU^{\dagger},
\end{equation}
that is the transformation rule of the adjoint representation.
With these transformation properties we can easily show that \eqref{Lkin} is invariant under the $SU(3)_C$ gauge symmetry group. It is very convenient to consider infinitesimal transformations, i.e.\ transformations in which the group parameters $\theta^a(x)$ are small so that
\begin{equation}
    U_{ij}\simeq \delta_{ij} - i\theta^aT^a_{ij},
\end{equation}
where we have omitted the dependence of $U$ and $\theta$ on the space-time point $x$.
With this expansion the transformation properties of the fields become
\begin{align}
\psi'_i&=\psi_i - i\theta^aT^a_{ij}\psi_j, \\
A'^a_\mu &=A_\mu^a + f^{abc}\theta^b A_\mu^c -\frac{1}{g_s}\partial_\mu\theta^a. \label{Ainf}
\end{align}
From Eqs.~(\ref{Ainf}) and (\ref{Adj}) we can derive that
\begin{equation}
    F'^a_{\mu \nu}=(\delta_{ab}+f^{abc}\theta^b)F^c_{\mu \nu},
\end{equation}
that is the infinitesimal transformation rule of the adjoint representation.

Another term that is allowed by the symmetries is
\begin{equation}
    \mathcal{L}_{\text{CP}}=\theta \frac{\alpha_s}{4 \pi}\widetilde{F}^a_{\mu \nu}F^{a,\mu \nu},
\end{equation}
where 
\begin{equation}
    \widetilde{F}^{a}_{\mu \nu}=\frac{1}{2}\epsilon_{\mu \nu \rho \sigma}F^{a,\rho \sigma},
\end{equation}
and $\theta$ is a dimensionless parameter. This term violates CP because of the presence of the completely antisymmetric tensor $\epsilon_{\mu \nu \rho \sigma}$. Since for strong interactions CP violation has never been observed, either this term is absent or $\theta$ is extremely small. Experimental estimates give the bound $\theta \lesssim 10^{-10}$. This is known as strong CP problem.

Then we have the gauge fixing term, that is
\begin{equation}
\mathcal{L}_{\substack{\text{gauge} \\\text{fixing}}} = -\frac{1}{2\xi}\sum_{a=1}^8\bigl(\partial_\mu A^{a,\mu}\bigr)^2.
\end{equation}
This term is needed in order to correctly quantize the theory, avoiding the infinities that arise in the functional integral formalism from the invariance of the lagrangian over all possible infinite gauge transformations. Therefore choosing the parameter $\xi$ we can choose a particular gauge. It follows that this term is not gauge invariant, as it can be easily verified. For example setting $\xi=1$ we recover the Feynman gauge, while setting $\xi=0$ we find the so-called Landau gauge.

The last term of \eqref{LQCD} is 
\begin{equation}
    \mathcal{L}_{\text{ghost}}=(\partial^\mu \chi^{a})^*D_\mu^{ab}\chi^b,
\end{equation}
where the fields $\chi$ are complex scalars obeying Fermi statistics and are called \textit{Faddev-Popov ghosts}.
$D_\mu^{ab}$ is the covariant derivative in the adjoint representation, therefore
\begin{equation}
 D_\mu^{ab}=\delta^{ab}\partial_\mu - g_s f^{abc}A_\mu^c .
\end{equation}
This term, as the previous one, is required in order to correctly quantize the theory. It comes from the fact that QCD is a non abelian gauge theory. In fact we don't have a similar term in QED.
Since the ghosts are not ``real'' particles they cannot appear in the final states but only in virtual corrections and can be eliminated with a convenient gauge choice. Moreover, ghosts are needed because when we sum over the gluon polarizations, the unphysical longitudinal polarizations do not cancel as it happens in QED. But when we include the ghost contributions this cancellations is complete and we are left just with the physical transverse polarizations.

\section{Renormalization}\label{ren}
%renormalization, dimensional regularization $\overline{\mbox{MS}}$ scheme, renorm. group. eq., asymptotic freedom.

Whenever we compute a cross section in the SM we have to deal with the appearance of divergences. Being the cross section an observable, it must be finite and therefore these divergences are unphysical. It means that we have to find a way to remove them. In this section we will focus on the divergences that come from the virtual corrections to the Feynman diagrams, that are the corrections coming from internal loops, i.e.\ that don't change the number of particles in the final state. Such divergences can be of two types: \textit{ultraviolet} (UV) or \textit{infrared} (IR). The first ones appear when the momentum flowing inside a loop diagram goes to infinity. In these diagrams we have loop integrals of the form
\begin{equation}\label{div}
    \int_\kappa^\infty dk\, k^{D-1}.
\end{equation}
The lower limit of integration $\kappa$ has been put in order to underline that we are focused on the behavior for $k\rightarrow \infty$. For $D<0$ the integral is finite, but for $D \geq0$ it is divergent.
The way in which UV divergences are removed is called renormalization. 
The IR divergences arise when the momentum flowing in the loop goes to zero. This kind divergences are present whenever we have to deal with massless particles like photons or gluons, but also when we neglect the mass of particle like a light quark. They are cured including all diagrams in which we have the emission of one or more photon (or gluon) from one of the external legs. In this chapter we will focus on the UV divergences, while IR divergences will be described in a more complete way in Chapter~\ref{fact}. 

In order to deal with divergent quantities we have to regularize them: it means that we have to modify our theory introducing a regulator that prevents every quantity from being divergent.
Once we have regularized all the divergences we perform all the computations in this ``modified'' theory assuring that the ``divergent'' quantities (divergent in the limit in which the regulator is removed) are canceled. Then we can remove the regulator recovering the ``real'' theory that now is free from divergences. For example in \eqref{div} we can integrate up to a certain cutoff $\Lambda$. In this way every integral is perfectly finite and the divergences are recovered in the limit $\Lambda\rightarrow \infty$. 
Now we can perform the renormalization to cancel the ``divergent'' quantities. After that, taking the limit $\Lambda\rightarrow\infty$ gives no problem and the result is perfectly finite. 
When we regularize the theory we have to choose a regularization that keeps all the important properties of our theory. This is not the case for the cutoff regulator since it spoils the gauge invariance.
A much better regularization is the so called \textit{dimensional regularization}: in this case we regularize the theory performing all the momentum integrations in $d=4-2\epsilon$ dimensions, instead of the physical $d=4$. In this way both the UV and IR divergences are regularized and appear as poles in $\epsilon$. The advantage of dimensional regularization is that the gauge invariance is preserved in the regularized theory.

%So far we said that renormalization is needed in order to remove the divergences of the theory. Actually it will be necessary even if all the diagrams would be finite. In fact the need of renormalization comes from the fact that, due to the presence of interaction, the parameters that appear in the lagrangian like $m$ or $e$ are not the physical ones but they have a correction, and from the fact that the fields appearing in the lagrangian are not well normalized, i.e.\ the propagator has not residue $1$ in $p^2=m^2$.

In order to renormalize the theory first of all we call every field and every coupling appearing in \eqref{LQCD} \textit{bare fields} and \textit{bare couplings} and address them with a subscript $0$, e.g.\ $\psi_0$, $A_{0\mu}^a$, $m_0$ etc. They are the fields and the couplings of the divergent theory. Then the \textit{bare lagrangian} is the lagrangian written in terms of the bare quantities.
Therefore, considering for example just the first term of \eqref{LQCD}, we have that the bare lagrangian is
\begin{align}
    \mathcal{L}_{\text{gauge}}={}&\sum_{f=\text{flavours}}\Bigl(i\overline{\psi}^{(f)}_{0i}\slashed{\partial}\psi^{(f)}_{0i} - m^{(f)}_0\overline{\psi}^{(f)}_{0i}\psi^{(f)}_{0i} +g_{s0}\overline{\psi}^{(f)}_{0i}\slashed{A}_{0ij}\psi^{(f)}_{0j}\Bigr) \notag \\
   & +\frac{1}{2}A^{a,\mu}_0(g_{\mu \nu}\partial^2 - \partial_\mu\partial_\nu)A^{a,\nu}_0 - g_{s0}f^{abc}A^{a,\mu}_0A^{b,\nu}_0\partial_\mu A^{c}_{0\nu} \notag \\
   & -\frac{1}{4}g^{2}_{s0}f^{abe}f^{cde}A^{a,\mu}_0A^{b,\nu}_0A^{c}_{0\mu} A^d_{0\nu}.\label{Lkinbare}
\end{align}
Then we rescale all these quantities defining
\begin{equation}\label{renormalized}
    \psi= \frac{\psi_0}{\sqrt{Z_2}},
    \quad A^a_\mu=\frac{A^a_{0\mu}}{\sqrt{Z_3}},
    \quad m=\frac{Z_2}{Z_m}m_0,
    \quad g_s=\frac{Z_2\sqrt{Z_3}}{Z_1}\mu^{-\epsilon}g_{s0},
\end{equation}
that are called renormalized fields and constants. The objects $Z_i$ are dimensionless and are called \textit{renormalization constants}.
$\mu$ is a scale with dimension of energy and has been introduced because if we compute the dimensions of the various terms of the lagrangian in dimensional regularization, i.e.\ with $d=4-2\epsilon$, we find that, in unity of energy
\begin{align}
    [\psi_0]&=\frac{3}{2}-\epsilon ,\\
    [A_0]&=1-\epsilon, \\
    [g_{s0}]&=\epsilon.
\end{align}
Now the bare coupling $g_{s0}$ is no more dimensionless. Therefore, since we want to perform an expansion in terms of a dimensionless parameter, the scale $\mu$ has been introduced so that the renormalized strong coupling $g_s$ is dimensionless. With this definition we have that
\begin{equation}\label{alphaMS}
    \alpha_{s0}=\frac{Z_1^2}{Z_2^2Z_3}\mu^{2\epsilon} \alpha_{s},
\end{equation}
where we have used that
\begin{equation}
    \alpha_s=\frac{g_s^2}{4\pi},
\end{equation}
and the same holds for the bare coupling. $\mu$ is called \textit{renormalization scale} (sometimes addressed with $\mu_R$) and it is a completely arbitrary scale: it means that any observable cannot depend on the choice of $\mu$.
With these definitions \eqref{Lkinbare} becomes
\begin{align}
    \mathcal{L}_{\text{gauge}}={}&\sum_{f=\text{flavours}}\Bigl(iZ_2\overline{\psi}_i^{(f)}\slashed{\partial}\psi_i^{(f)} - Z_m m^{(f)}\overline{\psi}_i^{(f)}\psi_i^{(f)} +Z_1\mu^{\epsilon}g_s\overline{\psi}_i^{(f)}\slashed{A}_{ij}\psi^{(f)}_j\Bigr) \notag \\
    &+\frac{1}{2}Z_3A^{a,\mu}\bigl(g_{\mu \nu}\partial^2 - \partial_\mu\partial_\nu\bigr)A^{a,\nu} - Z_{3g}\mu^{\epsilon}g_sf^{abc}A^{a,\mu}A^{b,\nu}\partial_\mu A^{c}_\nu \notag \\
   & -\frac{1}{4}Z_{4g}\mu^{2\epsilon}g_s^2f^{abe}f^{cde}A^{a,\mu}A^{b,\nu}A^{c}_\mu A^d_\nu,
\end{align}
where $Z_{3g}$ and $Z_{4g}$ are combinations of $Z_1$, $Z_2$ and $Z_3$.
Now we can compute order by order in perturbation theory the renormalization constants $Z_i$ in order to cancel the poles in $\epsilon$ from the matrix elements. Once we have done it, we can remove the regulator taking the limit $\epsilon\rightarrow 0$ and all the diagrams will be perfectly finite since we have removed all the ``divergent'' terms. In other words we are reabsorbing all the divergences into a redefinition of the bare couplings and the bare fields, that therefore become infinite. This does not represent a problem since, being the bare couplings the ones of the divergent theory, they are unphysical and therefore unmeasurable.

In addition to removing the divergences, we still can choose how much of the finite part of the amplitudes we can subtract in the renormalization.
This gives rise to different \textit{renormalization schemes}. Obviously the choice of the scheme does not modify the physical quantities like cross sections or decay rates.
The one that we have described so far is the so-called minimal subtraction (MS) scheme. In this scheme we are removing only the poles $1/\epsilon$, leaving all the finite parts untouched.
Instead of using the definitions in Eqs.~(\ref{renormalized}) and (\ref{alphaMS}) we can use slightly modified ones: since in the amplitudes $\mu^2$ is always multiplied by a term $e^\gamma/4\pi$ we can define
\begin{align}
     g_s={}&\frac{Z_2\sqrt{Z_3}}{Z_1}\tilde{\mu}^{-\epsilon} g_{s0}, \\
     \alpha_{s0}={}&\frac{Z_1^2}{Z_2^2Z_3}\tilde{\mu}^{2\epsilon}\alpha_{s}, \label{alphaMSbar}
\end{align}
with
\begin{equation}\label{muMSbar}
    \tilde{\mu}^2=\frac{\mu^2e^\gamma}{4\pi}.
\end{equation}
This scheme is called $\overline{\mbox{MS}}$ scheme. After we have done the redefinition in \eqref{muMSbar}, the $\rm \overline{MS}$ scheme is realized subtracting only the poles in $\epsilon$. This is equivalent in using the variable $\mu$ and subtracting the poles in $\epsilon$, plus the finite term $\log(4\pi)- \gamma$.

In conclusion the values of the renormalization constants at the first order in the expansion of $\alpha_s$ in the $\MSbar$ scheme are
\begin{align}
    Z_1&=1-(C_F+C_A)\frac{\alpha_s}{4\pi}\frac{1}{\epsilon} + \ord{2}, \label{Z1}\\
    Z_2&=1-C_F\frac{\alpha_s}{4\pi}\frac{1}{\epsilon}+\ord{2} ,\label{Z2}\\
    Z_3&=1+\Bigl(\frac{5}{3}C_A-\frac{4}{3}n_fT_F\Bigr)\frac{\alpha_s}{4\pi}\frac{1}{\epsilon}+\ord{2},\label{Z3}
\end{align}
where $T_F$, $C_F$ and $C_A$ are the color factors defined in Eqs.~(\ref{TF}),(\ref{CF}) and (\ref{CA}), while $n_f$ is the number of flavors of our theory, i.e.\ $n_f=6$. Obviously, in order to have finite observables, renormalization must be applied to the other terms of \eqref{LQCD} too.

\section{Renormalization group equation}
\label{re:group:eq}
In order to cancel the poles in $\epsilon$ from the amplitudes we had to introduce a completely arbitrary scale $\mu$ with the dimension of energy. Obviously, physical observables do not have to depend on the choice of $\mu$. Imposing the independence of the observables from $\mu$ we find that the only possibility is that the renormalized strong coupling $g_s$, or equivalently $\alpha_s$, is $\mu$-dependent.
Instead of imposing the independence from $\mu$ on the physical observables, like cross sections, we can observe that, being the bare lagrangian independent from $\mu$, the bare coupling must be independent from $\mu$ as well. Therefore we can impose that
\begin{equation}
    \mu^2\frac{d}{d\mu^2}g_{s0}=0,
\end{equation}
or equivalently 
\begin{equation}\label{RGE:bare}
    \mu^2\frac{d}{d\mu^2}\alpha_{s0}=0.
\end{equation}
Plugging \eqref{alphaMSbar} into \eqref{RGE:bare} we find
\begin{equation}
    \mu^2\frac{d}{d\mu^2}\biggl(\frac{Z_1^2}{Z_2^2 Z_3}\tilde{\mu}^{2\epsilon}\alpha_s\biggr)=0.
\end{equation}
Now using Eqs.~(\ref{Z1}), (\ref{Z2}), (\ref{Z3}) we get that
\begin{equation}\label{RGE}
    \mu^2\frac{d}{d\mu^2}\alpha_s=\beta(\alpha_s),
\end{equation}
that is called \textit{renormalization group equation} (RGE) and gives the running of $\alpha_s$ with the renormalization scale, i.e.\ its solution gives $\alpha_s(\mu^2)$. $\beta(\alpha_s)$ is called \textit{beta function} of QCD and is computed in perturbation theory. In particular one can show that
\begin{equation}
    \beta(\alpha_s)=-\epsilon\alpha_s - \bigl(\beta_0\alpha_s^2 + \beta_1\alpha_s^3 + \beta_2\alpha_s^4 + \dots \bigr).
\end{equation}
Solving \eqref{RGE} at $\ord{2}$, with a certain initial condition $\alpha_s(\mu_0^2)$, one finds that at leading order (and with $\epsilon=0$)
\begin{equation}\label{RGE:sol}
    \alpha_s(\mu^2)=\frac{\alpha_s(\mu^2_0)}{1+\alpha_s(\mu_0^2)\beta_0\log\bigl(\frac{\mu^2}{\mu_0^2}\bigr)}=\frac{1}{\beta_0\log(\frac{\mu^2}{\Lambda^2})},
\end{equation}
where we have defined
\begin{equation}\label{Lambda}
    \Lambda^2=\mu_0^2e^{-\frac{1}{\beta_0\alpha_s(\mu_0^2)}}.
\end{equation}
$\Lambda$ is called Landau pole. It is the value of $\mu$ for which the coupling $\alpha_s$ diverges. Depending on the sign of $\beta_0$ we can have two different situations:
\begin{itemize}
    \item For $\beta_0>0$ the coupling $\alpha_s(\mu^2)$ goes to zero for $\mu^2 \gg \Lambda^2$. In this case the theory is said UV free.
    \item For $\beta_0<0$ the coupling $\alpha_s(\mu^2)$ goes to zero for $\mu^2 \ll \Lambda^2$. In this case the theory is said IR free.
\end{itemize}
Computing $\beta_0$ in QCD yields the result of
\begin{equation}
    \beta_0=\frac{11C_A-4n_fT_F}{12\pi}.
\end{equation}
Since as far as we know the number of quark flavours in the SM is $n_f=6$ we are in the case $\beta_0>0$ and QCD is UV free. This is where asymptotic freedom comes from. In contrast, if we compute $\beta_0$ in QED we find that $\beta_0^{QED}=-\frac{4n_f}{12\pi}$. Therefore in QED we are in opposite case. The value of $\Lambda$ can be computed measuring $\alpha_s$ at a certain scale $\mu_0$ and inserting it in \eqref{Lambda}. Using that $\alpha_s(M_Z^2)=0.118$ \cite{PDG} we find that $\Lambda \sim 200\,\mbox{MeV}$.
%For energies of the order of the mass of the proton, QCD becomes heavily coupled, so that the quarks tend to form bound state, i.e.\ hadrons.
It means that as $\mu$ becomes of order $\Lambda$, $\alpha_s$ becomes of $\mathcal{O}(1)$. This invalidates perturbation theory since every term in the expansion becomes of $\mathcal{O}(1)$ and therefore we should resum the whole infinite series in order to get reliable results.
If we perform the computation of $\Lambda$ in QED we find, using $\alpha_{em}\bigl(\mu^2=m_e^2=(0.511\,\mbox{MeV})^2\bigr)=1/137$, that  $\Lambda_{QED}\sim 10^{286}\,\mbox{eV}$. This is why in QED the dependence of $\alpha_{em}$ on the renormalization scale, at the energies we usually deal with, is much weaker than the one of $\alpha_s$. In fact, we have, for example, that $\alpha_{em}(M_Z^2)=1/128$.

Now the question is: how do we choose $\mu$? Since it is completely arbitrary we can in principle choose any value. However, also in this case some problems arise.
For example, in the computation of the virtual corrections to the couplings (e.g.\ $g_s$ or $e$) we always have the insurgence of logarithms of the form $\log \mu^2/\mu_0^2$, where $\mu_0$ is a certain energy scale at which the coupling is known (for example from an experimental measurement). It means that choosing $\mu \sim \mu_0$ the logarithms are small and we can safely use perturbation theory since every order in the perturbative expansion is much smaller than the previous ones. If instead $\mu^2 \gg \mu_0^2$ the logarithms are big and perturbation theory breaks down. In this second case, in order to obtain reliable results, it is mandatory to resum all the perturbative series. This is completely analogous to using the RGE to compute the coupling at the scale $\mu$ with the difference that we don't have to compute all the infinite diagrams that give a correction to this coupling, that in many cases can be impossible. 
%In general in the computation of all the virtual diagrams we have the insurgence of logarithmically enhanced terms of the form $\log^k \mu^2/E^2$, where $E^2$ is the typical energy of the process. These logarithms are big whenever we want to compute these diagrams at a scale $\mu$ different from $E$. The RGE allows us to resum such big logarithms instead of computing all the diagrams in perturbation theory.

%To conclude this section we can observe that in writing the solution of the RGE we swapped a dimensionless constant $g_s$ with a dimensional scale $\Lambda$, \eqref{RGE:sol}. This is called \textit{dimensional transmutation}. In other words, QCD is not fundamentally defined by the coupling $g_s$ but by the constant $\Lambda$. Analogous considerations hold for QED. \textcolor{red}{Magari leva questa parte}

To conclude this section we can observe that the same arguments that have led to the conclusion that the strong coupling runs with the renormalization scale, can be applied to the mass. In fact, requiring that the bare mass is independent from $\mu$, we can obtain a RGE for the renormalized mass. Hence, we get that
\begin{equation}\label{RGE:mass}
    \mu^2\frac{d}{d\mu^2}m=\gamma_m m,
\end{equation}
where $\gamma_m$ is called \textit{anomalous dimension} and is computed in perturbation theory. The solution of \eqref{RGE:mass} gives the running of $m$ with the renormalization scale $\mu$. For this reason $m$ is also called \textit{running mass}.
However, we usually work with the physical mass that is a constant of nature and is the value that we measure experimentally. It is defined as the pole in the propagator of the quark (or in general of any particle). For this reason it is also called \textit{pole mass}. Since the renormalized propagator of a fermion is
\begin{equation}
    iG_R(\slashed{p})=\frac{i}{\slashed{p}-m+\Sigma(\slashed{p})},
\end{equation}
where $\Sigma(\slashed{p})$ is the fermion self-energy and is computed in perturbation theory, the relation between the physical mass and the renormalized mass is given by solving $\bigl(iG_R(\slashed{p}=m_P)\bigr)^{-1}=0$, where $m_P$ is the physical mass. It means that we have to solve
\begin{equation}\label{polemass:vs:renormmass}
    m_P -m + \Sigma(m_P)=0.
\end{equation}
In the remaining of this thesis we will always work in terms of physical masses, that from now on will be addressed with $m$.

\section{Running of \texorpdfstring{$\alpha_s$}{alphas} with different flavour numbers}
\label{VFNS:ren}

In the last section we saw the RGE, \eqref{RGE}, that depends on the $\beta$-function. Since the computation of this object involves quark loops of all possible flavours, it depends on the total number of flavours, i.e.\ as far as we know $n_f=6$.
Moreover, if there are heavy quarks that we don't know yet, we should include them in the computation of the beta function as well.
This is not an ideal situation since we expect that physically, if a particle is much heavier than the scales of the considered process, then its contribution to physical observables must be negligible. It is the content of the Applequist-Carazzone decoupling theorem \cite{Appelquist:1974tg}.
The decoupling of heavy flavours can be achieved for example with the decoupling renormalization scheme (DS): if we perform a subtraction at \textit{zero momentum} then the propagators of the flavours with $m^2 \gg \mu^2$ go to zero. 
Therefore, the heavy flavours vanish from the amplitudes and from physical observables like cross sections and decay rates.
However, the $\beta$ function is not a physical observable. It means that Applequist-Carazzone decoupling theorem does not apply and also in the DS the RGE must be written in terms of $n_f=6$.
This problem can be avoided using effective field theories to make decoupling explicit. For example let us consider QCD with $n_l=n_f-1$ light flavours and one heavy flavour with mass $m_h^2\gg\mu^2$ ($n_f$ flavour scheme). Then we consider a modified theory with $n_l$ light flavours and no heavy flavours ($n_f-1$ flavour scheme).
It means that we have integrated out the heavy flavour so that we are left with a smaller number of degrees of freedom that take part to the interactions.
Therefore, in this second scheme we will write the RGE in terms of a smaller number of flavours.
The equivalence between the two theories is obtained requiring that physical observables must be equal in both of them.
This leads to matching conditions between the couplings in the two theories, namely $\alpha_s^{[n_f-1]}$ and $\alpha_s^{[n_f]}$.
We find that, up to NNLO \cite{Chetyrkin_1997},
\begin{align}
    \frac{\alpha_s^{[n_f-1]}(\mu^2)}{\alpha_s^{[n_f]}(\mu^2)}={}&1-\frac{\alpha_s^{[n_f]}(\mu^2)}{6\pi}\log \frac{\mu^2}{m_h^2} + \frac{\bigl(\alpha_s^{[n_f]}(\mu^2)\bigr)^2}{6\pi^2}\Biggl(\frac{1}{6}\log^2\frac{\mu^2}{m_h^2} - \frac{19}{4}\log \frac{\mu^2}{m_h^2} -\frac{7}{4} \Biggr) \notag \\
    &+\mathcal{O}\Bigl(\bigl(\alpha_s^{[n_f]}\bigr)^3\Bigr). \label{alpha:matching}
\end{align}
From \eqref{alpha:matching} we can see that at LO and NLO $\alpha_s^{[n_f]}(m_h^2)=\alpha_s^{[n_f-1]}(m_h^2)$, but at NNLO it is no longer true.
Moreover, we introduce a matching scale $\mu_h$ that is the threshold such that if $\mu>\mu_h$ we use the $n_f$ flavour scheme and and if $\mu<\mu_h$ we use the $n_f-1$ flavour scheme.
Obviously we must have that $\mu_h \sim \mathcal{O}(m_h)$, otherwise we would have unresummed large logarithms.

In conclusion, we have constructed a scheme where the number of active flavours varies with the renormalization scale $\mu$.
In fact, whenever $\mu$ crosses the threshold of a given heavy quark, we have to switch from the $n_f$ flavour scheme to the $n_f-1$ flavour scheme (or vice-versa if $\mu$ is increasing).
This is called variable flavour number (renormalization) scheme (VFNS).
%\marginnote{\textcolor{red}{LANDAU POLE???}}
%In addition, we must pay attention to the fact that also the Landau pole $\Lambda$ depends on the number of active flavours. We find that in the $\MSbar$ scheme, using the expansion of $\beta$ up to $\beta_4$~\cite{PDG}
%\begin{equation*}
 % \Lambda^{[3]}=332\,\mbox{MeV},  \quad  \Lambda^{[4]}=292\,\mbox{MeV} , \quad  \Lambda^{[5]}=210\,\mbox{MeV},  \quad  \Lambda^{[6]}=89\,\mbox{MeV}.  
%\end{equation*}

\section{Parton model}\label{part}
%DIS, scaling, violations from scaling, factorization theorem

So far in this chapter we talked about the fundamental fields of QCD that are quarks and gluons. However, as consequence of color confinement, what is really observed experimentally are the hadrons.
Due to the failure of perturbation theory in studying the confinement of quarks and gluons into hadrons, computing cross sections with hadrons in the initial and final states is impossible in the context of perturbatitive QCD. Therefore, we need the introduction of the so-called parton model. In this section it will be described starting from the deep inelastic scattering (DIS) and then it will be extended to proton-proton collision. 

\begin{figure}[!t]
\centering
\begin{tikzpicture}
\begin{feynman}[large]
\vertex (i1) {\(e^-\)};
\vertex[right=6cm of i1] (o1) {\(e^-\)}; 
\vertex[below=4cm of i1] (i2) ;
\vertex[below=0.1cm of i2] (i3);
\vertex[below=0.1cm of i3] (i4);
\vertex[left=0.1cm of i4] (i5) {\(p\)};
\vertex[below=4cm of o1] (o2) ;
\vertex[below=0.1cm of o2] (o3);
\vertex[below=0.1cm of o3] (o4) ;
\vertex[right=0.1cm of o4] (o5) {\(p\)};
\vertex[right=3cm of i1] (tmp1);
\vertex[below=1cm of tmp1] (v1);
\vertex[below=2.5cm of tmp1,blob, fill=black] (v2) {};
\vertex[below=0.1cm of v2] (v3);
\vertex[below=0.1cm of v3] (v4);
\diagram* {
(i1) -- [fermion, edge label=\(k\)] (v1) -- [fermion,edge label=\(k'\)] (o1),
(v1) -- [boson,edge label=\(q\)] (v2),
(i3) --  (v3) -- (o3),
(i4) --  (v4) -- (o4),
(i2) -- [edge label=\(P\)] (v2) -- [edge label=\(P'\)](o2),
};
\end{feynman}
\end{tikzpicture}
\caption{Elastic scattering of an electron with a proton.}
\label{DIS1}
\end{figure}
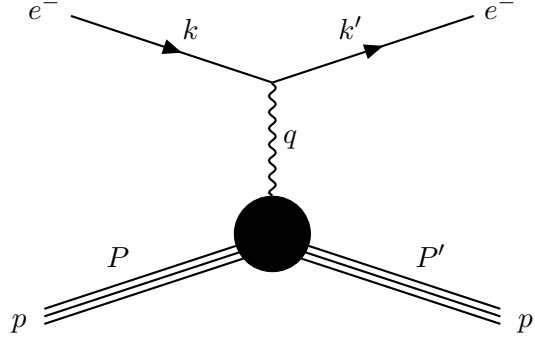

Let us consider the scattering of an electron off a proton via the exchange of a virtual photon.
We can define the following quantities
\begin{align}
    Q^2&=-q^2, \\
    \nu&=P \cdot q, \\
    x&=\frac{Q^2}{2P \cdot q}=\frac{Q^2}{2\nu}, \\
    y&=\frac{q \cdot P}{k \cdot P},     
\end{align}
where $k$ and $k'$ are respectively the momenta of the initial and final electrons, $P$ the one of the initial proton and $P'$ is the one of the final hadronic state that can be still composed by the proton (elastic scattering) or can be a multi-hadron state (inelastic scattering).
The variable $x$ is called Bjorken's scaling variable and $q=k-k'=P'-P$ is the momentum carried by the photon. The differential cross section can be written as 
\begin{equation}\label{DIS:crosssect}
    \frac{d\sigma}{dxdy}=\frac{4\pi\alpha_{em}^2(S-M^2)}{Q^4}\biggl(y^2xF_1+\Bigl(1-y-\frac{xyM^2}{S-M^2}\Bigr)F_2\biggr),
\end{equation}
where $F_a\equiv F_a(x,Q^2)$ with $a=1,2$ are called \textit{structure functions} and parametrize the internal structure of the proton target as ``seen'' by the virtual photon. $M$ is the mass of the proton and $\sqrt{S}$ is the hadronic center-of-mass energy, i.e.\ $S=(P+k)^2$. If the scattering is elastic the structure functions will be those of an object with an internal structure. It means that they will depend on a scale $Q_0$ characteristic of the extension of the proton, e.g.\ $Q_0 \sim \sqrt{\langle r^2 \rangle}$. Therefore the structure functions will depend on $Q$ through the ratio $Q/Q_0$ because they must be dimensionless. In this case we have that
\begin{equation}
    M^2=P'^2=(P+q)^2=M^2+q^2+2P \cdot q \quad\implies\quad x=1.
\end{equation}
It is a consequence of the fact that we still have the proton in the final state. \figref{DIS1} shows the elastic process.
If the electron has enough energy it can break the internal structure of the proton, forming new hadrons in the final state. 
Then the process will be $e^-+P\rightarrow e^-+X$, where $X$ is a generic hadronic state.
Neglecting the proton mass we can write \eqref{DIS:crosssect} as
\begin{equation}\label{diff:cross:sec:DIS}
    \frac{d\sigma}{dxdQ^2}=\frac{2\pi\alpha_{em}^2}{xQ^4}\biggl(\bigl(1+(1-y)^2\bigr)F_2-y^2F_L\biggr),
\end{equation}
where we have defined
\begin{equation}
    F_L(x,Q^2)=F_2(x,Q^2)-2xF_1(x,Q^2).
\end{equation}
In this case an approximate scaling law, called \textit{Bjorken scaling}, is observed: in the limit $Q^2,\nu \rightarrow \infty$ with $x$ fixed we have that
\begin{equation}
    F_a(x,Q^2)\rightarrow F_a(x), \qquad a=2,L.
\end{equation}
Bjorken scaling implies that the virtual photon scatters against a point-like free particle inside the proton. In fact, if it scattered against a composite particle, then the structure functions would depend on the ratio between a scale characteristic of the extension of the particle and $Q$, like it was for the elastic scattering. Notice that in this case
\begin{equation}
    M^2<P'^2=(P+q)^2=M^2+q^2+2P \cdot q \quad\implies\quad x<1.
\end{equation}
The first relation of this equation comes from the fact that we no longer have the proton in the final state but instead we have a multi-hadron state. Therefore we conclude that the proton is composed by point-like constituents that we call \textit{partons}. For the moment let us neglect the existence of the gluons. It means that we identify partons with quarks. This is the so-called \textit{``naive'' parton model}. It is the model in which we don't consider gluons and QCD corrections. 

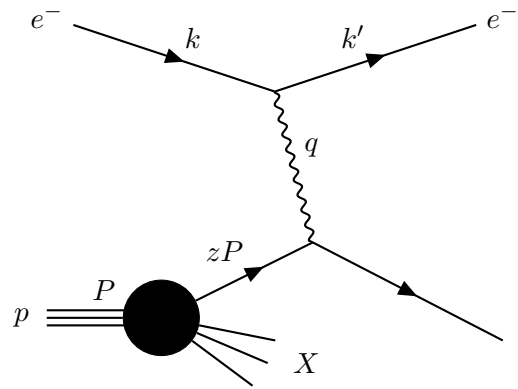
\begin{figure}[!t]
\centering
\begin{tikzpicture}
\begin{feynman}[large]
\vertex (i1) {\(e^-\)};
\vertex[right=6cm of i1] (o1) {\(e^-\)}; 
\vertex[below=4cm of i1] (i2) ;
\vertex[below=0.1cm of i2] (i3);
\vertex[above=0.1cm of i2] (i4);
\vertex[left=0.1cm of i2] (i5) {\(p\)};
\vertex[below=4.3cm of o1] (o2) ;
\vertex[right=3cm of i1] (tmp1);
\vertex[below=1cm of tmp1] (v1);
\vertex[right=1cm of i2,blob, fill=black] (v2) {};
\vertex[below=0.1cm of v2] (v3) ;
\vertex[above=0.1cm of v2] (v4);
\vertex[below=3cm of tmp1] (tmp2);
\vertex[right=0.5cm of tmp2] (v5) ;
\vertex[right=1.5cm of v2] (tmp3) ;
\vertex[below=0.3cm of tmp3] (v6) ;
\vertex[below=0.3cm of v6] (tmp4) ;
\vertex[left=0.1cm of tmp4] (v7) ;
\vertex[below=0.3cm of v7] (tmp5) ;
\vertex[left=0.2cm of tmp5] (v8) ;
\vertex[right=0.2cm of v7] (v9) {\(X\)};
\diagram* {
(i1) -- [fermion, edge label=\(k\)] (v1) -- [fermion,edge label=\(k'\)] (o1),
(v1) -- [boson,edge label=\(q\)] (v5),
(i3) --  (v3),
(i4) -- [edge label=\(P\)] (v4) ,
(i2) -- (v2) ,
(v2) -- [fermion, edge label=\(zP\)] (v5) -- [fermion] (o2),
(v2) -- (v6),
(v2) -- (v7),
(v2) -- (v8),
};
\end{feynman}
\end{tikzpicture}
\caption{DIS of a quark with the virtual photon emitted by the electron.}
\label{DIS3}
\end{figure}

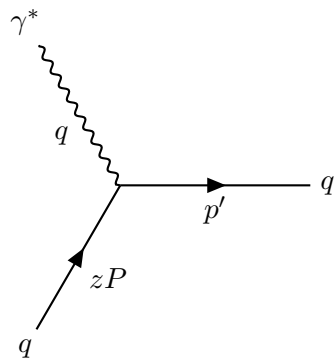
\begin{figure}[!t]
\centering
\begin{tikzpicture}
\begin{feynman}[large]
\vertex (i1) {\(\gamma^*\)};
\vertex[below=2.165cm of i1] (tmp1); 
\vertex[right=1.25cm of tmp1] (v1);
\vertex[below=4.33cm of i1] (i2) {\(q\)};
\vertex[right=2.5cm of v1] (o1) {\(q\)};
\diagram* {
(i1) -- [boson, edge label'=\(q\)] (v1),
(i2) -- [fermion, edge label'=\(zP\)] (v1) -- [fermion, edge label'=\(p'\)] (o1),
};
\end{feynman}
\end{tikzpicture} 
\caption{DIS partonic process $q(zP)+\gamma^*(q)\rightarrow q(p')$ that gives the partonic structure functions in Eqs.~(\ref{scaling}) and (\ref{scalingFL}).}
\label{DIS2}
\end{figure}

We have said that the cross section is the one of the scattering of a virtual photon against a point-like free particle, as shown in \figref{DIS3}. 
Let us suppose that this particle carries a fraction $z$ of the longitudinal momentum of the proton, i.e.\ if $p$ is the momentum of such particle then $p=zP$. We are neglecting the possibility that the partons have a transverse momentum with respect to the one of the proton. 
Then we define the partonic structure functions as the structure functions of the process in which we have the partons as external states and not the proton itself. They will addressed with a hat, e.g.\ $\hat{F}_1$, $\hat{F}_2$, etc.
For example in this case we have that the partonic process is $q(zP)+\gamma^*(q)\rightarrow q(p')$. Its tree-level diagram is shown in \figref{DIS2}. 
However, the partonic processes cannot be observed experimentally because of color confinement.
What we observe are the hadronic processes, in which the proton and the hadrons are the external particles.
In order to compute the hadronic cross sections we have to introduce the \textit{parton distribution functions} (PDFs): $f_i(z)dz$ represents the probability that the virtual photon scatters off a parton of type $i$ that carries a momentum fraction between $z$ and $z+dz$ with $0\leq z \leq 1$. 
Moreover, we will make the assumption that the photon scatters incoherently against the different types of partons.
Therefore, a generic structure function $F_a(x)$, will be of the form
\begin{equation}\label{hadronic:F2}
    F_a(x)=\sum_{i=q,\bar{q}}\int_0^1dz\, f_i(z)\hat{F}_{a,i}\Bigl(\frac{x}{z}\Bigr), \qquad a=2,L,\quad q=u,d,s.
\end{equation}
$\hat{F}_{a,i}$ is the partonic structure function $\hat{F}_a$ of the parton of type $i$, where $i$ runs over all possible light quarks but also on their antiquarks. 
Since the PDFs contain the information on the internal structure of the proton, they are non-perturbative objects.
Indeed, they are governed by an energy scale that is of the order of the proton mass and therefore the strong coupling becomes of $\mathcal{O}(1)$ and QCD becomes non-perturbative. It means that we cannot predict the PDFs from the theory but we have to extract them from data, as we explained in Chapter~\ref{intro}.
We didn't consider heavy quarks because their mass is higher than the one of the proton and therefore they are are not present as its constituents but they are generated perturbatively. 
Actually, the charm quark has a mass that is of the order of the one of the proton and therefore it is not light enough to be neglected but it is not heavy enough so that we can fully rely on perturbation theory.
Therefore, we cannot rule out a priori the presence of a charm PDF.
In this discussion, in order to avoid complications, we will assume that we don't have intrinsic charm in the proton and therefore we don't have a charm PDF. 

In Sec.~\ref{massless:comp} we will show how the structure functions are extracted from the matrix element of the Feynman diagram of a certain process.
Computing the partonic structure functions of DIS of a quark with a virtual photon, i.e.\ the process shown in \figref{DIS2}, we find that
\begin{align}
    \hat{F}_2\Bigl(\frac{x}{z}\Bigr)&=e_q^2\delta \Bigl(1-\frac{x}{z}\Bigr), \label{scaling}\\
    \hat{F}_L\Bigl(\frac{x}{z}\Bigr)&=0.\label{scalingFL}
\end{align}
where $e_q$ is the charge of the quark involved in the scattering.
%It means that the photon interacts with a quark of momentum fraction $z=x$.
Plugging these relations into \eqref{hadronic:F2} we find
\begin{align}
    F_2(x)&=\sum_{i=q,\bar{q}}\int_0^1dz\, f_i(z)e^2_ix\delta(z-x)= \sum_{i=q,\bar{q}}e_i^2xf_i(x), \label{naive:F2}\\
    F_L(x)&=0.\label{naive:FL}
\end{align}
\eqref{naive:FL} implies that $F_2(x)=2xF_1(x)$, that is called \textit{Callan-Gross relation}.

So far we didn't consider the existence of gluons and QCD corrections. Adding these contributions we find the full parton model (sometimes called \textit{improved parton model}). In particular we have to include a gluon PDF that interacts with the virtual photon via splitting in quark-antiquark pairs, and we have to compute the partonic processes, like $\gamma^*+q\rightarrow q$, to higher orders in perturbation theory. If we consider fully inclusive cross sections we will have emissions of other particles in the final state. In this case one observes that the scaling is broken by the QCD corrections. Moreover, in the full parton model the PDFs acquire a dependence on the energy scale $Q^2$ of the process we are considering. Such dependence will be further discussed in Sec.~\ref{coll:div}. A consequence is that also the structure functions acquire a dependence from $Q^2$.
When we consider the full parton model we expect that the PDFs satisfy certain sum rules: the first one is related to the proton's momentum conservation
\begin{equation}
    \int_0^1dz\sum_{i=q,\bar{q},g}z f_i(z,Q^2)=1.
\end{equation} 
Then, imposing proton flavour conservation, we have that
\begin{align}
    \int_0^1dz\Bigl(f_u(z,Q^2)-f_{\bar{u}}(z,Q^2)\Bigr)&=2 ,\\
    \int_0^1dz\Bigl(f_d(z,Q^2)-f_{\bar{d}}(z,Q^2)\Bigr)&=1.
\end{align}

If we want to compute a hadronic cross section, then it will be of the form
\begin{equation}\label{fact:theo}
\sigma(x,Q^2)=\sum_{i=q,\bar{q},g}\int_0^1dz\,f_i(z,Q^2)\hat{\sigma}_i\Bigl(\frac{x}{z},\alpha_s(Q^2)\Bigr),
\end{equation}
%where $q$ runs over all the light quarks, i.e.\ $q=u,d,s$, and obviously $\bar{q}$ over their antiquarks. 
%This is the so called \textit{factorization theorem}.
where $\hat{\sigma}_i$ is the partonic cross section of a parton of type $i$ and it is computed in perturbation theory, so $\hat{\sigma}=\sum_k\hat{\sigma}^{(k)}\alpha_s^k$. Taking the leading order expansion we recover the ``naive'' parton model.
\eqref{fact:theo} involves various simplifications: first of all the partons in the initial state that constitute the proton are considered free. This is true only if the virtuality $Q^2$ of the virtual photon is much larger than the hadronization scale. It means that \eqref{fact:theo} neglects terms of $\mathcal{O}(\Lambda^2/Q^2)$.
A second simplification is that in the computation of the partonic cross sections the initial and final partons are always considered \textit{on-shell}. Obviously it is not true since, due to color confinement, they never appear as final particles but give rise to parton showers that end with hadrons in the final state. 
However, being the calculation of the confinement of the partons into hadrons impossible due to the failure of perturbation theory, such simplification is used to obtain useful results. Also in this case we are throwing away terms of $\mathcal{O}(\Lambda^2/Q^2)$.
Third, so far we always neglected the possibility that the interaction between the electron and the parton is mediated by a weak vector boson (i.e.\ $W^{\pm}$ or $Z^0$). If the energy of the electron is high enough we have to consider such contributions and this gives rise to different structure functions.

If we consider structure functions instead of cross sections, using $\mu=Q$, \eqref{fact:theo} takes the form
\begin{equation}\label{coeff:func}
    F_a(x,Q^2)=x\sum_{i=q,\bar{q},g}\int_x^1\frac{dz}{z}f_i(z,Q^2)C_{a,i}\Bigl(\frac{x}{z},\alpha_s(Q^2)\Bigr), \qquad a=2,L
\end{equation}
where $C_{a,i}$ are called \textit{coefficient functions} and are defined as
\begin{equation}\label{coeff:funct}
    \frac{x}{z}C_{a,i}\Bigl(\frac{x}{z},\alpha_s(Q^2)\Bigr)=\hat{F}_{a,i}\Bigl(\frac{x}{z},\alpha_s(Q^2)\Bigr).
\end{equation}
They are computed in perturbation theory and so we have that
\begin{equation}\label{Coeff:funct:exp}
    C_{a,i}\bigl(z,\alpha_s\bigr)=C_{a,i}^{(0)}(z)+\alpha_s C_{a,i}^{(1)}(z)+\alpha_s^2 C_{a,i}^{(2)}(z)+\mathcal{O}(\alpha_s^3).
\end{equation}
The $k$-th order expansion of \eqref{Coeff:funct:exp}, i.e.\ the function $C^{(k)}_{a,i}$, is usually called N$^k$LO coefficient function. 
The lower limit of integration of \eqref{coeff:func} is $x$ and not 0 because the partonic variable $\frac{Q^2}{2p \cdot q}=\frac{Q^2}{2zP \cdot q}=\frac{x}{z}$ must be smaller than $1$, for the same reason that $x$ must be smaller than 1 in the inelastic scattering.
Obviously, at zeroth order in perturbation theory we get the results in Eqs.~(\ref{naive:F2}) and (\ref{naive:FL}).
However, it is not true beyond LO. Comparing these definitions with what we have said in this section, it's easy to show that
\begingroup
\allowdisplaybreaks
\begin{align}
    C_{2,q}^{(0)}(z)&=e_q^2\delta(1-z) , \\
    C_{L,q}^{(0)}(z)&=0 , \\
     C_{2,g}^{(0)}(z)&=0,  \\
      C_{L,g}^{(0)}(z)&=0 .
\end{align}
\endgroup
The expressions of the coefficient functions at NLO will be given in the next chapter.
A notation that will be very useful in the following chapters is
\begin{equation}\label{conv}
    \bigl(f \otimes g\bigr)(x)=\int_x^1\frac{dz}{z}f(z)g\Bigl(\frac{x}{z}\Bigr)=\int_x^1\frac{dz}{z}f\Bigl(\frac{x}{z}\Bigr)g(z),
\end{equation}
so that \eqref{coeff:func}, omitting the $x$ and $Q^2$ dependence, can be written in the compact form
\begin{equation}
    F_a=x\sum_{i=q,\bar{q},g}f_i \otimes C_{a,i}.
\end{equation}

\begin{figure}[!b]
\centering
\begin{tikzpicture}
\begin{feynman}[large]
\vertex (i1);
\vertex [left=0.1cm of i1] (tmp1) {\(H_1\)};
\vertex[below=0.1cm of i1]  (i2);
\vertex[above=0.1cm of i1]  (i3);
\vertex[right=1cm of i1, blob, fill=black] (v1) {};
\vertex[right=1cm of i2] (v2);
\vertex[right=1cm of i3] (v3);
\vertex [below=4cm of i1] (i4);
\vertex [left=0.1cm of i4] (tmp2) {\(H_2\)};
\vertex[below=0.1cm of i4]  (i5);
\vertex[above=0.1cm of i4]  (i6);
\vertex[right=1cm of i4, blob, fill=black] (v4) {};
\vertex[right=1cm of i5] (v5);
\vertex[right=1cm of i6] (v6);
\vertex[below=2cm of v1] (tmp3);
\vertex[right=2.5cm of tmp3,blob,fill=black] (v7) {\textcolor{white}{$\hat{\sigma}_{ij}$}};
%\vertex[right=2.5cm of tmp3,blob,label={above:$\hat{\sigma}_{ij}$},fill=black] (v7) {};
%%%
\vertex[right=2.4cm of v7] (tmp4);
\vertex[above=0.6cm of tmp4] (tmp5);
\vertex[below=0.6cm of tmp4] (tmp6);
\vertex[left=0.2cm of tmp5] (tmp7);
\vertex[left=0.2cm of tmp6] (tmp8);
%%%
\vertex[right=1.5cm of v4] (tmp9) ;
\vertex[below=0.3cm of tmp9] (v8) ;
\vertex[below=0.3cm of v8] (tmp10) ;
\vertex[left=0.1cm of tmp10] (v9) ;
\vertex[below=0.3cm of v9] (tmp11) ;
\vertex[left=0.2cm of tmp11] (v10) ;
%\vertex[right=0.2cm of v10] (v11) {\(X\)};
%%%
\vertex[right=1.5cm of v1] (tmp12) ;
\vertex[above=0.3cm of tmp12] (v12) ;
\vertex[above=0.3cm of v12] (tmp13) ;
\vertex[left=0.1cm of tmp13] (v13) ;
\vertex[above=0.3cm of v13] (tmp13) ;
\vertex[left=0.2cm of tmp13] (v14) ;
%\vertex[right=0.2cm of v12] (v15) {\(X\)};
\diagram* {
(i1) -- (v1),
(i2) -- (v2),
(i3) --  [edge label=\(P_1\)](v3),
(i4) -- (v4),
(i5) -- (v5),
(i6) --  [edge label=\(P_2\)](v6),
(v1) -- [edge label'=\(z_1 P_1\), edge label=$i$] (v7),
(v4) -- [edge label=\(z_2P_2\), edge label'=$j$] (v7),
(v7) -- (tmp4),
(v7) -- (tmp7),
(v7) -- (tmp8),
(v4) -- (v8),
(v4) -- (v9),
(v4) -- (v10),
(v1) -- (v12),
(v1) -- (v13),
(v1) -- (v14),
};
\end{feynman}
\end{tikzpicture}
\caption{Diagram of the interaction between two initial state hadrons $H_1$ and $H_2$, via their point-like constituents that carry a fraction $z_1$ and $z_2$ of their momentum. $i$ and $j$ can be quarks, antiquarks or gluons.}
\label{LHC}
\end{figure}
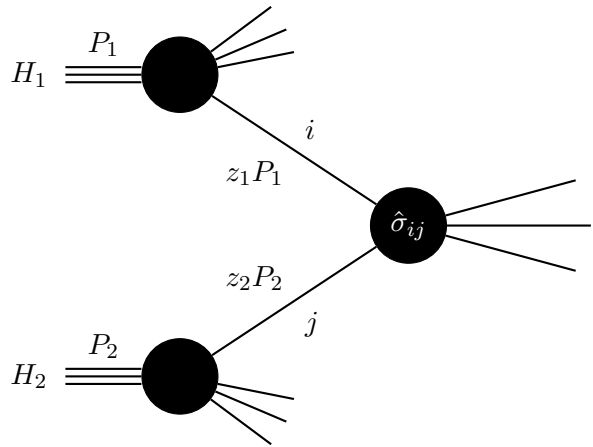
So far we focused our discussion on DIS, where we have one proton in the initial state. For this reason \eqref{fact:theo} contains only one PDF. However we can consider also processes with two protons in the initial state, for example proton-proton collision at LHC.
These processes are much more complex than DIS due to the presence of two hadrons $H_1$ and $H_2$ in the initial state, that interact through their point-like constituents, which carry a fraction $z_1$ and $z_2$ of their momenta. In this case the hadronic cross section is given by
\begin{equation}
    \sigma(Q^2)=\sum_{i,j=q,\bar{q},g}\int_0^1dz_1\int_0^1dz_2\,f_{H_1,i}(z_1,Q^2)f_{H_2,j}(z_2,Q^2)\hat{\sigma}_{ij}\bigl(z_1,z_2,\alpha_s(Q^2)\bigr),
\end{equation}
where $Q^2$ is the typical scale of the process we are considering, $f_{H_1,i}(z_1,Q^2)$ and $f_{H_2,j}(z_2,Q^2)$ are respectively the probabilities densities of finding the parton $i$ or $j$ in the hadron $H_1$ or $H_2$ with longitudinal momentum fraction $z_1$ and $z_2$ at the scale $Q^2$.
$\hat{\sigma}_{ij}$ is the partonic cross section of a certain process with the partons $i$ and $j$ as initial states and it is computed in perturbation theory.
\figref{LHC} shows graphically the process we are considering.

%From what we have said in this section it is clear that the PDFs are a crucial ingredient for the theoretical predictions in the hadronic processes. As we said, they must be extracted from data because the failure of perturbation theory invalidates any theoretical prediction. Moreover the PDFs are process-independent objects: it means that they depend on the physical scale of the process (as it will be shown in the next chapter), on the momentum fraction carried by the parton, but not on the particular process the proton is involved into. Thus the strategy is to measure the PDFs in a very clean process like DIS, i.e.\ a process that is easy to describe theoretically and to reconstruct experimentally, and then use them to obtain predictions in more complicated ones, for example like the proton-proton collision at LHC. \textcolor{red}{lo ho messo uguale in intro} Usually the PDFs are extracted from DIS data since we know the momentum of the initial electron and we can measure the one of the final electron. In this way we can compute the momentum $q$ flowed in the DIS process. This cannot be done for proton-proton collision because don't know the momenta of the two partons that interact.

%%%%%%%%%%%%%%%%%%%%%%%%%%%%%%%%%%%%%%%%%%%%%%%%%%%%%%%%%%%%%%%%%%%%%%%%%%%%%%%%%%%%%%%%%%%%%%%%

\chapter{Factorization}\label{fact}
%fattorizz. div. coll., DGLAP, fatt. grandi log., ?VFNS? 
%\section{Introduction}
%Whenever we compute a cross sections in the Standard Model (SM) we always have to deal with divergences that can be of different types. First of all we have the ultraviolet divergences (UV divergences): they come from virtual corrections to the diagrams and appear when all the momenta of the internal lines go to infinity. They are cured with renormalization, i.e. the divergences are absorbed into a redefinition of the unmeasurable bare coupling constants (for example like $\alpha$ or $\alpha_s$) so that the amplitudes become UV finite. This procedure requires the introduction of an arbitrary scale $\mu_R$ that has the dimension of an energy. Since this scale is arbitrary we can impose that the cross section (or also the bare coupling constants in the lagrangian) is independent from $\mu_R$ and this implies that $\alpha$ and $\alpha_s$ acquire a dependence from $\mu_R$ that is controlled by the renormalization group equation (RGE) (for example for in the case of $\alpha_s$)
%where $\beta(\alpha_s)$ is known as QCD beta function (the same holds for QED). \small{\textcolor{red}{cenni su asymptotic freedom?}}
In Sec.~\ref{ren} we saw that whenever we compute a cross section in the SM we always have to deal with divergences. In that discussion we focused our attention on the UV divergences, that are cured with renormalization. 
In this chapter we will focus on the IR divergences: they appear when we have massless particles in our theory, for example photons and gluons.
Moreover, if we neglect the mass of the quarks (or of any fermion) we have the appearance of other IR divergences. In this case the divergences are an artifact that comes from sending the quark mass to zero and therefore they can be eliminated reinserting the mass dependence. However, being light quarks much lighter than the energies we usually deal with in perturbative QCD, treating them as massless is a very good approximation. Therefore, we have to find a way to remove the IR divergences that come from neglecting the mass of the quarks.
IR divergences can appear both in the virtual corrections and in the real ones. 
For the virtual corrections the divergences appear when the momentum flowing in the propagator goes to zero. 
The real corrections consist in emission of real particles and when such particles have zero mass they gives rise to an IR divergence that can be of two types: \textit{soft} or \textit{collinear}.
The first ones appear when the energy of the emitted particle goes to zero (since we are treating a massless particle the energy can be arbitrarily small) while the second ones appear when the angle between the emitted particle and the emitting one tends to zero.

Now that we have introduced all the different IR divergences we have to find a way to remove them because the final cross sections must be finite (they are physical quantities). A crucial observation is that when the emitted particle becomes very soft or very collinear (or both), the real emission process becomes indistinguishable from the no-emission one, both practically and theoretically. Then Bloch-Nordsieck theorem states that IR singularities cancel between real and virtual diagrams when summing up all resolution-indistinguishable final states. This means that summing virtual and real corrections ensures the cancellation of both soft and collinear singularities from the emissions coming from final state particles.
Moreover, the Kinoshita-Lee-Nauenberg theorem says that mass singularities ($m\rightarrow 0$) of external particles are canceled if all mass-degenerate states are summed up. From this theorem one can derive that the soft singularities cancel also from the emissions coming from initial state state particles. However, collinear divergences coming from the initial state do not cancel, so we have to find a way to deal with this kind of divergences. This is called \textit{factorization} and it is the subject of this chapter.
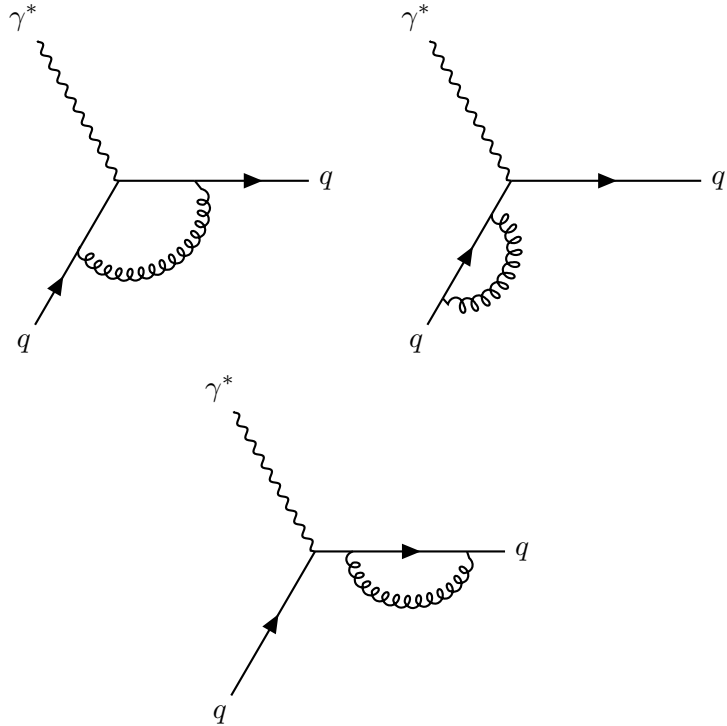
\begin{figure}[t!]
\centering
\begin{tikzpicture}
\begin{feynman}[large]
\vertex (i1) {\(\gamma^*\)};
\vertex[below=2.165cm of i1] (tmp1); 
%\vertex[below=1.0825cm of tmp1] (tmp2);
\vertex[below=0.866cm of tmp1] (tmp2);
%\vertex[right=0.625cm of tmp2] (v2);
\vertex[right=0.74998cm of tmp2] (v2);
\vertex[right=1.25cm of tmp1] (v1);
\vertex[below=4.33cm of i1] (i2) {\(q\)};
\vertex[right=2.5cm of v1] (o1) {\(q\)};
\vertex[right=1cm of v1] (v3) ;
\diagram* {
(i1) -- [boson] (v1),
(i2) -- [fermion] (v2) -- (v1) -- (v3) -- [fermion] (o1),
(v2) -- [gluon, half right] (v3),
};
\end{feynman}
\end{tikzpicture}
\hspace{1em}
\begin{tikzpicture}
\begin{feynman}[large]
\vertex (i1) {\(\gamma^*\)};
\vertex[below=2.165cm of i1] (tmp1); 
\vertex[below=0.433cm of tmp1] (tmp2);
%\vertex[below=1.299cm of tmp2] (tmp3);
\vertex[below=1.1258cm of tmp2] (tmp3);
\vertex[right=1cm of tmp2] (v2);
\vertex[right=0.35cm of tmp3] (v3);
\vertex[right=1.25cm of tmp1] (v1);
\vertex[below=4.33cm of i1] (i2) {\(q\)};
\vertex[right=2.5cm of v1] (o1) {\(q\)};
\diagram* {
(i1) -- [boson] (v1),
(i2) --  (v3) --[fermion] (v2) -- (v1) -- [fermion] (o1),
(v2) -- [gluon, half left] (v3),
};
\end{feynman}
\end{tikzpicture}
\hspace{1em}
\begin{tikzpicture}
\begin{feynman}[large]
\vertex (i1) {\(\gamma^*\)};
\vertex[below=2.165cm of i1] (tmp1); 
\vertex[right=1.25cm of tmp1] (v1);
\vertex[below=4.33cm of i1] (i2) {\(q\)};
\vertex[right=0.5cm of v1] (tmp2);
\vertex[right=1.5cm of tmp2] (tmp3);
\vertex[right=2.5cm of v1] (o1) {\(q\)};
\diagram* {
(i1) -- [boson] (v1),
(i2) -- [fermion] (v1) -- (tmp2) -- [fermion](tmp3) -- (o1),
(tmp2) -- [gluon,half right] (tmp3),
};
\end{feynman}
\end{tikzpicture} 
    \caption{Virtual correction to the diagram in \figref{DIS2}: they are all IR divergent due to the zero mass of the gluon.}
    \label{virtualDIS}
\end{figure}
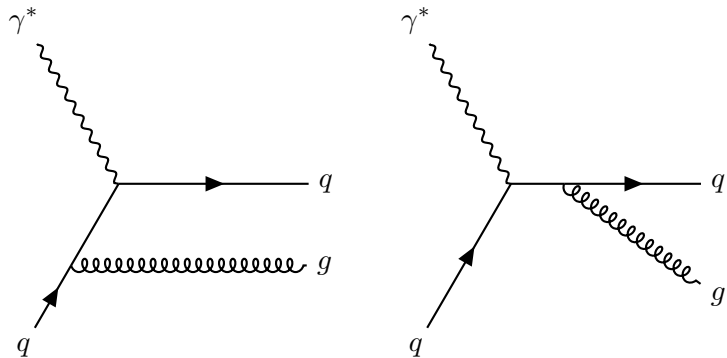
\begin{figure}[t!]
\centering
\begin{tikzpicture}
\begin{feynman}[large]
\vertex (i1) {\(\gamma^*\)};
\vertex[below=2.165cm of i1] (tmp1); 
\vertex[below=1.0825cm of tmp1] (tmp2);
\vertex[right=0.625cm of tmp2] (v2);
\vertex[right=1.25cm of tmp1] (v1);
\vertex[below=4.33cm of i1] (i2) {\(q\)};
\vertex[right=2.5cm of v1] (o1) {\(q\)};
\vertex[right=3.1cm of v2] (o2) {\(g\)};
\diagram* {
(i1) -- [boson] (v1),
(i2) -- [fermion] (v2) -- (v1) -- [fermion] (o1),
(v2) -- [gluon] (o2),
};
\end{feynman}
\end{tikzpicture}
\hspace{1em}
\begin{tikzpicture}
\begin{feynman}[large]
\vertex (i1) {\(\gamma^*\)};
\vertex[below=2.165cm of i1] (tmp1); 
\vertex[right=1.25cm of tmp1] (v1);
\vertex[below=4.33cm of i1] (i2) {\(q\)};
\vertex[right=0.7cm of v1] (tmp2);
\vertex[right=2.5cm of v1] (o1) {\(q\)};
\vertex[below=1.5cm of o1] (o2) {\(g\)};
\diagram* {
(i1) -- [boson] (v1),
(i2) -- [fermion] (v1) -- (tmp2) -- [fermion] (o1),
(tmp2) -- [gluon] (o2),
};
\end{feynman}
\end{tikzpicture} 
\caption{Real correction to the diagram in \figref{DIS2}: the diagram on the left is the initial state emission, while the one on the right is the final state emission.}
\label{realDIS}
\end{figure}

For example if the process we are considering is the scattering of a quark on a virtual photon shown in \figref{DIS2}, then the LO virtual corrections are those in \figref{virtualDIS}, while the real ones are those in \figref{realDIS}. As we were saying, summing up all these divergent diagrams provides the cancellation of all the singularities with the exception of the collinear divergences coming from the first diagram of \figref{realDIS}. 
%Therefore, if we want to isolate the divergent part of this process, the only radiative correction that we will consider is the one on the left of \figref{realDIS}, while the sum of all the other diagrams will give a finite result.

In this chapter we will describe factorization: in Sec.~\ref{coll:div} we will describe how to remove collinear divergences coming from the initial state emission and in Sec.~\ref{DGLAP:sec} we will describe how this implies DGLAP equations. In Sec.~\ref{big:logs} we will use the exact quark mass dependence and we will explain why factorization is still necessary and how to perform it. Then in Sec.~\ref{VFNS} we will describe what a variable flavour number (factorization) scheme (VFNS) is and how to construct it.

\section{Factorization of collinear divergences}\label{coll:div}
\subsection{Discussion in perturbation theory}
Before dealing with the collinear divergences of the initial state we have to regularize them: one possible way is using a cutoff that prevents the transverse momentum of the emitted particle (transverse with respect to the momentum of the emitting particle) from going to zero. Another possibility is using dimensional regularization since it regularizes UV and IR divergences at the same time. 

In the case of the cutoff regularization, the computation of the sum of the diagrams in \figref{DIS2}, \ref{virtualDIS} and \ref{realDIS}, neglecting the quark mass, gives \cite{ellis_stirling_webber_1996}
\begin{equation}\label{F2q}
    \hat{F}_{2,q}(z, Q^2)=e_q^2x\biggl[\delta(1-z) + \alpha_s\Bigr(P_{qq}^{(0)}(z)\log\frac{Q^2}{\kappa^2} + c_{2,q}(z)\Bigr)\biggr],%=e_q^2zC_q(z,Q^2)
\end{equation}
where 
\begin{equation}\label{Pqq}
    P_{qq}^{(0)}(z)=\frac{C_F}{2\pi}\biggl(\frac{1+z^2}{(1-z)_+} +\frac{3}{2}\delta(1-z)\biggr) =\frac{C_F}{2\pi}\biggl(\frac{1+z^2}{1-z}\biggr)_+,
\end{equation}
and $c_{2,q}(x)$ is a calculable function. $\kappa$ is a cutoff that prevents the transverse momentum of the emitted gluon from going to zero, i.e. $|k_T|^2 > \kappa^2$. In \eqref{Pqq} we introduced the \textit{plus distribution} that will be described in Appendix~\ref{plus:appendix}.
Thanks to the \textit{factorization theorem}, the DIS hadronic structure function is
\begin{align}
    F_{2,q}(x, Q^2)&=\sum_{q,\bar{q}}\int_x^1\frac{dz}{z}\,q_0(z)\hat{F}_{2,q}\Bigl(\frac{x}{z},Q^2\Bigr) \notag \\
   & =\sum_{q,\bar{q}}e_q^2x\int_x^1\frac{dz}{z}\,q_0(z)\Biggl[\delta\Bigr(1-\frac{x}{z}\Bigr)+\alpha_s\biggl(P_{qq}^{(0)}\Bigl(\frac{x}{z}\Bigr)\log\frac{Q^2}{\kappa^2} + c_{2,q}\Bigl(\frac{x}{z}\Bigr)\biggr)\Biggr],\label{hadr:F2}
\end{align}
where $q_0$ is the PDF of the quark. Using the same procedure that is adopted for the renormalization, we can regard $q_0$ as a \textit{bare} PDF, which is unmeasurable since it is the PDF of the divergent theory, and redefine it in order to absorb the collinear singularity. In order to do it we must introduce a scale $\mu_F$, called \textit{factorization scale}, that plays a similar role that the renormalization scale $\mu_R$ plays for renormalization. Hence we define a new PDF so that the divergences are canceled from the hadronic structure functions:
\begin{equation}\label{PDF:redef}
    q_{_{DIS}}(x,\mu_F^2) = \int_x^1\frac{dz}{z}\,q_0(z)\Biggl[\delta\Bigr(1-\frac{x}{z}\Bigr)+\alpha_s\biggl(P_{qq}^{(0)}\Bigl(\frac{x}{z}\Bigr)\log\frac{\mu_F^2}{\kappa^2} + c_{2,q}\Bigl(\frac{x}{z}\Bigr)\biggr)\Biggr].
\end{equation}
Therefore, \eqref{hadr:F2} becomes
\begin{equation}\label{F2fact}
    F_{2,q}(x,Q^2)=\sum_{q,\bar{q}}e_q^2x\int_x^1\frac{dz}{z}\,q_{_{DIS}}(z,\mu_F^2)\biggl[\delta\Bigr(1-\frac{x}{z}\Bigr)+\alpha_sP_{qq}^{(0)}\Bigl(\frac{x}{z}\Bigr)\log\frac{Q^2}{\mu_F^2} \biggr].
\end{equation}
In this way we have factorized the singular part of \eqref{F2q} into a redefinition of the PDF. In addition to it we have also factorized in the PDF the whole regular part $c_{2,q}$. However, we could have factorized only a term $c'_{2,q}$ that would have left in \eqref{F2fact} a finite contribution $c_{2,q}-c'_{2,q}$.
In fact, while factorization provides a prescription for dealing with the collinear divergences, we still have an arbitrariness in how the finite contribution is treated. How much of the finite contribution is factorized defines what is called the \textit{factorization scheme}. This is completely analogous to the renormalization schemes introduced in the previous chapter.
For example, the redefinition in \eqref{PDF:redef} is called DIS scheme.
Observe that like for renormalization we had to introduce an unphysical scale $\mu_F$, such that physical observables do not depend on the choice of $\mu_F$.
Conceptually it is a different scale with respect to the renormalization scale $\mu_R$. However, being the two completely arbitrary, we can always choose $\mu_F=\mu_R=\mu$, as we will do from now on. Anyway, if we want to reinsert the $\mu_R$ dependence, we can expand $\alpha_s(\mu_F^2)$ in terms of $\alpha_s(\mu_R^2)$ using that 
\begin{equation}
    \alpha_s(\mu_F^2)= \alpha_s(\mu_R^2)-\beta_0 \log \frac{\mu_F^2}{\mu_R^2}\alpha_s^2(\mu_R^2) - \biggl(\frac{\beta_1}{4\pi}\log \frac{\mu_F^2}{\mu_R^2} - \beta_0^2\log^2 \frac{\mu_F^2}{\mu_R^2}\biggr)\alpha_s^3(\mu_R^2).
\end{equation}

All this derivation can be carried out in dimensional regularization too. In this case we have that
\begin{equation}
    \hat{F}_{2,q}(z,Q^2)\Bigl|_{\text{div}}=e_q^2z\alpha_s(\mu^2)P_{qq}^{(0)}(z)\Bigl(-\frac{1}{\epsilon}\Bigr),
\end{equation}
Now we can factorize the singular term in the PDF using the $\overline{\mbox{MS}}$ scheme, i.e. we use the redefinition in \eqref{muMSbar} so we have to subtract only the poles in $\epsilon$.
Hence we get that
\begin{equation}
    F_{2,q}(x,Q^2)=\sum_{q,\bar{q}}e_q^2x\int_x^1\frac{dz}{z}\,q(z,\mu^2)\biggl[\delta\Bigr(1-\frac{x}{z}\Bigr)+\alpha_s(\mu^2)C_{2,q}^{(1)}\Bigl(\frac{x}{z},\frac{\mu^2}{Q^2}\Bigr) \biggr],
\end{equation}
with
\begin{align}
    C_{2,q}^{(1)}\Bigl(z,\frac{\mu^2}{Q^2}\Bigr)&=\frac{2C_F}{4\pi}\biggl[2\Bigl(\frac{\log(1-z)}{1-z}\Bigr)_+-\frac{3}{2}\Bigl(\frac{1}{1-z}\Bigr)_+-(1+z)\log(1-z)-\Bigl(\frac{1+z^2}{1-z}\Bigr)\log z \notag \\
    & \quad \quad \quad \quad  +3+2z-\Bigl(\frac{\pi^2}{3}+\frac{9}{2}\Bigr)\delta(1-z) - \Bigl( \frac{1+z^2}{1-z}\Bigr)_+ \log \Bigl(\frac{\mu^2}{Q^2}\Bigr)\biggr]. \label{Cq1}
\end{align}

So far we considered only the contribution of one quark to $F_2$. We can add the gluon contribution to the DIS: the first nonzero contribution is the $\ord{}$ that is given by the two diagrams shown in \figref{diagrams}.
These diagrams are divergent due to the zero mass of the quarks. Regularizing with a cutoff gives
\begin{equation}
    \hat{F}_{2,g}(z)=x\sum_{q,\bar{q}}e_q^2\alpha_s\Bigl(P_{qg}^{(0)}(z)\log\frac{Q^2}{\kappa^2}+c_{2,g}(z)\Bigr),
\end{equation}
with
\begin{equation}\label{Pqg}
    P_{qg}^{(0)}(z)=\frac{2T_F}{4\pi}\Bigl(z^2+(1-z)^2\Bigr).
\end{equation}
Adding this contribution to \eqref{hadr:F2} we find that the total structure function $F_2$ is
\begin{align}
   F_2(x,Q^2) ={}& \sum_{q,\bar{q}}e_q^2x\int_x^1\frac{dz}{z}\,q_0(z)\Biggl[\delta\Bigr(1-\frac{x}{z}\Bigr)+\alpha_s\biggl(P_{qq}^{(0)}\Bigl(\frac{x}{z}\Bigr)\log\frac{Q^2}{\kappa^2} + c_{2,q}\Bigl(\frac{x}{z}\Bigr)\biggr)\Biggr]  \notag \\
   &+\sum_{q,\bar{q}}e_q^2x\int_x^1\frac{dz}{z}\,g_0(z)\alpha_s\Biggl[P_{qg}^{(0)}\Bigl(\frac{x}{z}\Bigr)\log\frac{Q^2}{\kappa^2} + c_{2,g}\Bigl(\frac{x}{z}\Bigr)\Biggr].
\end{align}
In order to absorb the collinear singularities into the PDFs in DIS scheme, we make the following redefinition:
\begin{align}
    q_{_{DIS}}(x,\mu^2) ={}& \int_x^1\frac{dz}{z}\,q_0(z)\Biggl[\delta\Bigr(1-\frac{x}{z}\Bigr)+\alpha_s\biggl(P_{qq}^{(0)}\Bigl(\frac{x}{z}\Bigr)\log\frac{\mu^2}{\kappa^2} + c_{2,q}\Bigl(\frac{x}{z}\Bigr)\biggr)\Biggr]  \notag \\
   &+\int_x^1\frac{dz}{z}\,g_0(z)\alpha_s\Biggl[P_{qg}^{(0)}\Bigl(\frac{x}{z}\Bigr)\log\frac{\mu^2}{\kappa^2} + c_{2,g}\Bigl(\frac{x}{z}\Bigr)\Biggr], \\
   g_{_{DIS}}(x,\mu^2)={}&g_0(x),
\end{align}
so that we find
\begin{align}
     F_2(x,Q^2) ={}& \sum_{q,\bar{q}}e_q^2x\int_x^1\frac{dz}{z}\,q_{_{DIS}}(z, \mu^2)\biggl[\delta\Bigr(1-\frac{x}{z}\Bigr)+ \alpha_s P_{qq}^{(0)}\Bigl(\frac{x}{z}\Bigr)\log\frac{Q^2}{\mu^2} \biggr]  \notag \\
   &+\sum_{q,\bar{q}}e_q^2x\int_x^1\frac{dz}{z}\,g_{_{DIS}}(z,\mu^2)\alpha_sP_{qg}^{(0)}\Bigl(\frac{x}{z}\Bigr)\log\frac{Q^2}{\mu^2}.
\end{align}
A more common choice is to use the $\overline{\mbox{MS}}$ scheme: we find that 
\begin{equation}
    \hat{F}_{2,g}\Bigl|_{\text{div}}=ze_q^2\alpha_s(\mu^2)\sum_{q,\bar{q}}P_{qg}^{(0)}(z)\Bigl(-\frac{1}{\epsilon}\Bigr).
\end{equation}
Reabsorbing the collinear divergences in the the PDFs with the redefinition
\begin{align}
    q(x,\mu^2)={}&\int_x^1\frac{dz}{z}q_0(z)\biggl[\delta \Bigl(1 -\frac{x}{z}\Bigr) -\frac{1}{\epsilon} \alpha_s(\mu^2)P_{qq}^{(0)}\Bigl(\frac{x}{z}\Bigr)\biggr]  \notag \\
    &+\int_x^1\frac{dz}{z}g_0(z)\biggl[-\frac{1}{\epsilon}\alpha_s(\mu^2)P_{qg}^{(0)}\Bigl(\frac{x}{z}\Bigr)\biggr], \label{q:red}\\
    g(x,\mu^2)={}&g_0(x), \label{g:red}
\end{align}
we have that
\begin{align}
     F_2(x,Q^2) =& \,\sum_{q,\bar{q}}e_q^2x\int_x^1\frac{dz}{z}\,q(z, \mu^2)\biggl[\delta\Bigr(1-\frac{x}{z}\Bigr)+\alpha_s(\mu^2) C_{2,q}^{(1)}\Bigr(\frac{x}{z},\frac{\mu^2}{Q^2}\Bigr) \biggr]  \notag \\
   &\,\,\,\, +e_q^2x\int_x^1\frac{dz}{z}\,g(z,\mu^2) \alpha_s(\mu^2) C_{2,g}^{(1)}\Bigr(\frac{x}{z},\frac{\mu^2}{Q^2}\Bigr), \label{DIS:tot:fact}
\end{align}
with
\begin{align}
    C_{2,g}^{(1)}\Bigl(z,\frac{\mu^2}{Q^2}\Bigr)=\frac{4T_F}{4\pi}\biggl[&-8z^2+8z-1+\log\Bigl(\frac{1-z}{z}\Bigr)\bigl(z^2+(1-z)^2\bigr) \notag \\
   &-\bigl(z^2+(1-z^2)\bigr)\log \Bigl(\frac{\mu^2}{Q^2}\Bigr) \biggr]. \label{Cg1}
\end{align}
Notice that we have removed the sum over $q$ and $\bar{q}$ from the second line of \eqref{DIS:tot:fact} and this yields an additional factor of $2$ in \eqref{Cg1}.
\eqref{DIS:tot:fact} agrees with \eqref{coeff:func} with the only difference that in that case we choose $\mu=Q$.

The same computations can be performed for the longitudinal structure functions.
However, both $F_{L,q}$ and $F_{L,g}$ are finite at $\ord{}$, therefore at this order they are scheme independent and the redefinition of the PDFs in Eqs.~(\ref{q:red}) and (\ref{g:red}) don't change their expressions. Their form is the following
\begin{align}
    C_{L,q}^{(1)}(z)&=\frac{4C_F}{4\pi}z, \\
    C_{L,g}^{(1)}(z)&=\frac{16T_F}{4\pi}z(1-z).
\end{align}

\subsection{Computation of the gluon coefficient function}\label{massless:comp}

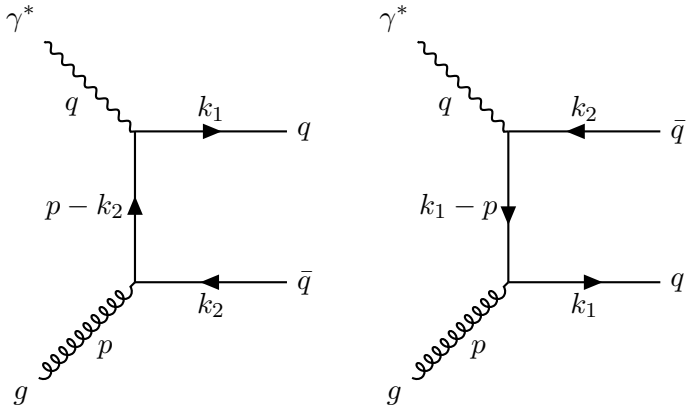
\begin{figure}[!b]
  \centering
  \vspace{1em}
\begin{tikzpicture} 
\begin{feynman}[large]
\vertex (i1) {\(\gamma^{*}\)};
\vertex[below=1.5cm of i1] (tmp1); 
\vertex[right=1.5cm of tmp1] (v1);
\vertex[below=2cm of v1] (v2);
\vertex[below=5cm of i1] (i2) {\(g\)};
\vertex[right=2cm of v1] (o1) {\(q\)};
\vertex[right=2cm of v2] (o2) {\(\bar{q}\)};
\diagram* {
(i1) -- [boson, edge label'=\(q\)] (v1),
(i2) -- [gluon, edge label'=\(p\)] (v2),
(o2) -- [fermion, edge label=\(k_2\)] (v2) -- [fermion, edge label=\(p-k_2\)] (v1) -- [fermion, edge label=\(k_1\)] (o1),
};
\end{feynman}
\end{tikzpicture}
\hspace{1em}
\begin{tikzpicture}
\begin{feynman}[large]
\vertex (i1) {\(\gamma^{*}\)};
\vertex[below=1.5cm of i1] (tmp1); 
\vertex[right=1.5cm of tmp1] (v1);
\vertex[below=2cm of v1] (v2);
\vertex[below=5cm of i1] (i2) {\(g\)};
\vertex[right=2cm of v1] (o1) {\(\bar{q}\)};
\vertex[right=2cm of v2] (o2) {\(q\)};
\diagram* {
(i1) -- [boson, edge label'=\(q\)] (v1),
(i2) -- [gluon, edge label'=\(p\)] (v2),
(o2) -- [anti fermion, edge label=\(k_1\)] (v2) -- [anti fermion, edge label=\(k_1-p\)] (v1) -- [anti fermion, edge label=\(k_2\)] (o1),
};
\end{feynman}
\end{tikzpicture} 
\vspace{1em}
\caption{Feynman diagrams that contribute to the partonic process $g+\gamma^* \rightarrow q+\bar{q}$ at tree level.}
\label{diagrams}
\end{figure}

In order to show how the collinear divergences appear in the coefficient functions when we have the emission of massless particles, we consider the deep inelastic scattering of a virtual photon with a real gluon that gives a quark-antiquark pair:
\begin{equation}\label{process}
  g(p)+\gamma^*(q)\rightarrow q(k_1)+\bar{q}(k_2),
\end{equation}
where $p^2=0$ and $q^2=-Q^2<0$ is the virtuality of the photon. 
As mentioned before, the quarks are considered massless and therefore we will have collinear divergences.
The two Feynman diagrams that contribute to this process at tree level are the ones shown in \figref{diagrams}.
The virtual photon is emitted by the electron participating in the DIS process. In the squared amplitude the part related to the initial electron and the one related to the partonic process factorize as
\begin{equation}
  |A|^2=\frac{1}{q^4}L^{\mu \nu}\hat{W}_{\mu \nu},
\end{equation}
where $L^{\mu \nu}$ is the part associated to the electron line and $\hat{W}_{\mu \nu}$ is the amplitude associated to the process in \eqref{process}. Therefore we will consider $\hat{W}_{\mu \nu}$ only. 
The most general expression of $\hat{W}_{\mu \nu}$ that satisfies current conservation $q^{\mu}\hat{W}_{\mu \nu}=q^{\nu}\hat{W}_{\mu \nu}=0$ is
\begin{equation} \label{partonic:tens}
  \hat{W}_{\mu \nu}(\hat{x},Q^2)=\Bigl(-g_{\mu \nu}+\frac{q^{\mu}q^{\nu}}{q^2}\Bigr)\hat{W}_1(\hat{x},Q^2)+\Bigl(p^{\mu}+\frac{q^{\mu}}{2\hat{x}}\Bigr)\Bigl(p^{\nu}+\frac{q^{\nu}}{2\hat{x}}\Bigr)\hat{W}_2(\hat{x},Q^2),
\end{equation}
where $\hat{x}$ is the partonic scaling variable and is defined as
\begin{equation}
\hat{x}=\frac{Q^2}{2q\cdot p}=\frac{Q^2}{2q\cdot zP}=\frac{x}{z},
\end{equation}
where $P$ is the momentum of the proton participating in the DIS.
What we want to compute are the partonic structure functions, defined as
%\begin{subequations}\label{strfunc}
\begin{align}
  \hat{F}_2(\hat{x},Q^2)&=\hat{\nu} \hat{W}_2(\hat{x}, Q^2), \label{strfunc:a}\\ 
  \hat{F}_L(\hat{x},Q^2)&=\hat{\nu} \hat{W}_2(\hat{x}, Q^2)- 2\hat{x}\hat{W}_1(\hat{x},Q^2), \label{strfunc:b}
\end{align}
%\end{subequations}
with $\hat{\nu}=q \cdot p$.
In order to extract them from $\hat{W}_{\mu \hat{\nu}}$ we have to define the 4-vectors $\bar{p}$ and $\bar{q}$ such that 
%\begin{subequations}\label{proj}
\begin{align}
  \bar{p}^2&=0, \quad \bar{p}\cdot p=1, \quad \bar{p}\cdot q=0, \label{proj:a}\\
  \bar{q}^2&=0, \quad \bar{q}\cdot p=0, \quad \bar{q}\cdot q=\hat{\nu}. \label{proj:b}
\end{align}
%\end{subequations}
With this definition they satisfy the relations
%\begin{subequations}\label{k1k2p}
\begin{gather}
\bar{p}\cdot k_1+\bar{p} \cdot k_2=1, \label{k1k2p:a} \\
\bar{q}\cdot k_1+\bar{q} \cdot k_2=\hat{\nu}, \label{k1k2p:b}
\end{gather}
%\end{subequations}
where we have used momentum conservation $q+p=k_1+k_2$.
Now we can extract the structure functions applying $\bar{p}$ and $\bar{q}$ to $\hat{W}^{\mu \nu}$ obtaining
%\begin{subequations}\label{strfunc2}
\begin{align}
  \hat{F}_2&=\hat{\nu} \bar{p}^{\mu} \bar{p}^{\nu} \hat{W}_{\mu \nu},\label{strfunc2:a} \\
  \hat{F}_L&=\frac{4\hat{x}^2}{\hat{\nu}}\bar{q}^{\mu}\bar{q}^{\nu}\hat{W}_{\mu \nu}.\label{strfunc2:b}
\end{align}
%\end{subequations}
Using Eqs.~(\ref{strfunc2:a}) and (\ref{strfunc2:b}) we can compute the partonic structure functions $\hat{F}_2$ and $\hat{F}_L$ from the diagrams in \figref{diagrams}.
Since massless particles give rise to IR divergences we will use the dimensional regularization in the $\overline{\mbox{MS}}$ scheme.
%, which means that we will perform the computation in dimension $d=4-2\epsilon$ and $\alpha_s$ becomes $\alpha_s(\mu^2)\tilde{\mu}^{2\epsilon}$ with $\tilde{\mu}^2=\bigl(\frac{\mu^2e^{\gamma}}{4\pi}\bigr)$ where $\alpha_s(\mu^2)$ is the dimensionless coupling. 
In this way such divergences are regularized and they will appear as poles of the form $\epsilon^{-1}$.

In the case of massless quarks we have that $k_1^2=k_2^2=0$ and therefore the amplitude of the two diagrams is
\begin{equation}\label{Mmu}
  M^{\mu}=e_qg_sT^a\bar{u}(k_1)\Biggl(\gamma^{\mu} \frac{\slashed{p}-\slashed{k_2}}{(p-k_2)^2}\slashed{\epsilon}(p)+\slashed{\epsilon}(p)\frac{\slashed{k_1}-\slashed{p}}{(k_1-p)^2}\gamma^{\mu}\Biggr)v(k_2),
\end{equation}
where $T^a$ are the Gell-Mann matrices and $\epsilon_\mu$ is the polarization vector of the gluon.
The tensor of \eqref{partonic:tens} is given by
\begin{equation}\label{Wmunu}
  \hat{W}^{\mu \nu}=\overline{\sum_{\text{pol}}} M^{\mu} M^{*\,\nu},
\end{equation}
where the symbol ${\overline{\sum}}_{\text{pol}}$ is the sum over the final polarizations and the average over the initial ones. When we sum over the colors of the final quarks and of the initial gluon we find a factor
\begin{equation}
\sum_{i,j,a}T^a_{ij}T^{a*}_{ij}=\frac{4}{3}\sum_{i,j}\delta_{ij}=4  ,  
\end{equation}
that averaged over the colors of the gluon becomes $T_F=1/2$. Since we are in $d$ dimensions, the polarizations of the gluon are $d-2=2(1-\epsilon)$. Putting all of these ingredients together we find
\begin{align} \label{hadr:ten}
  \hat{W}^{\mu \nu} = T_F\frac{e_q^2g_s^2}{2(1-\epsilon)}\tr\Biggl[&\slashed{k_1}\Biggl(\gamma^{\mu}\frac{\slashed{p}-\slashed{k_2}}{(p-k_2)^2}\slashed{\epsilon}(p)+\slashed{\epsilon}(p)\frac{\slashed{k_1}-\slashed{p}}{(k_1-p)^2}\gamma^{\mu}\Biggr) \slashed{k_2} \Biggl(\slashed{\epsilon^*}(p)\frac{\slashed{p}-\slashed{k_2}}{(p-k_2)^2}\gamma^{\nu}
  \notag \\
  &+\gamma^{\nu}\frac{\slashed{k_1}-\slashed{p}}{(k_1-p)^2}\slashed{\epsilon^*}(p)\Biggr)\Biggr].
\end{align}
It is convenient to define the Mandelstam variables
%\begin{subequations} \label{mand}
\begin{align}
  s&=(p+q)^2=-Q^2+2p\cdot q=(k_1+k_2)^2=2k_1\cdot k_2, \label{mand:a} \\
  t&=(q-k_1)^2=-Q^2-2k_1\cdot q=(k_2-p)^2=-2k_2\cdot p,\label{mand:b} \\
  u&=(q-k_2)^2=-Q^2-2k_2\cdot q=(k_1-p)^2=-2k_1\cdot p, \label{mand:c}
\end{align}
%\end{subequations}
that satisfy $s+t+u+Q^2=0$. 
Plugging these definitions in \eqref{hadr:ten} we find
\begin{equation}\label{wmunu}
  \hat{W}^{\mu \nu}=T_F\frac{e_q^2g_s^2}{2(1-\epsilon)}\Biggl(\frac{A^{\mu \nu}}{t^2}+\frac{B^{\mu \nu}}{u^2}+\frac{C^{\mu \nu}}{ut}\Biggr),
\end{equation}
where we have defined
\begin{align*}
  A^{\mu \nu}={}&\tr \Bigl(\slashed{k_1}\gamma^{\mu}(\slashed{p}-\slashed{k_2})\slashed{\epsilon}\slashed{k_2}\slashed{\epsilon}^*(\slashed{p}-\slashed{k_2})\gamma^{\nu}\Bigr), \notag \\
  B^{\mu \nu}={}&\tr \Bigl(\slashed{k_1}\slashed{\epsilon}(\slashed{k_1}-\slashed{p})\gamma^{\mu}\slashed{k_2}\gamma^{\nu}(\slashed{k_1}-\slashed{p})\slashed{\epsilon}^*\Bigr), \notag \\
  C^{\mu \nu}={}&\tr \Bigl(\slashed{k_1}\gamma^{\mu}(\slashed{p}-\slashed{k_2})\slashed{\epsilon}\slashed{k_2}\gamma^{\nu}(\slashed{k_1}-\slashed{p})\slashed{\epsilon}^*\Bigr) \notag \\
  &+\tr \Bigl(\slashed{k_1}\slashed{\epsilon}(\slashed{k_1}-\slashed{p})\gamma^{\mu}\slashed{k_2}\slashed{\epsilon}^*(\slashed{p}-\slashed{k_2})\gamma^{\nu}\Bigr) .
\end{align*}
Now we can compute these factors using that $\sum_{pol}\epsilon_{\mu}\epsilon^*_{\nu}=-g_{\mu \nu}$, using the definitions in Eqs.~(\ref{mand:a}), (\ref{mand:b}) and (\ref{mand:c}) and the fact that since we are in $d=4-2\epsilon$ dimensions we still have that $\{\gamma^{\mu},\gamma^{\nu}\}=2g^{\mu \nu}$ and $\tr(\gamma^{\mu}\gamma^{\nu})=4g^{\mu \nu}$ but the following relations hold
\begin{align}
  &g^{\mu \nu}g_{\mu \nu}=d , \label{diracmatrices:a}\\
  &\gamma^{\mu}\gamma_{\mu}=d, \label{diracmatrices:b}\\
  &\gamma^{\mu}\gamma^{\rho}\gamma_{\mu}=-(d-2)\gamma^{\rho}, \label{diracmatrices:c}\\
  &\gamma^{\mu}\gamma^{\rho}\gamma^{\sigma}\gamma_{\mu}=4g^{\rho \sigma}-2\epsilon\gamma^{\rho}\gamma^{\sigma} ,\label{diracmatrices:d}\\
  &\gamma^{\mu}\gamma^{\rho}\gamma^{\sigma}\gamma^{\nu}\gamma_{\mu}=-2\gamma^{\nu}\gamma^{\sigma}\gamma^{\rho}+2\epsilon\gamma^{\rho}\gamma^{\sigma}\gamma^{\nu}.\label{diracmatrices:e}
\end{align}
What we find is
\begin{align}
  A^{\mu \nu}={}&-8t(1-\epsilon)\Bigl(k_1^{\mu}p^{\nu}+k_1^{\nu}p^{\mu}+\frac{u}{2}g^{\mu \nu}\Bigr), \label{A}\\
  B^{\mu \nu}={}&-8u(1-\epsilon)\Bigl(p^{\mu}k_2^{\nu}+p^{\nu}k_2^{\mu}+\frac{t}{2}g^{\mu \nu}\Bigr), \label{B}\\
  C^{\mu \nu}={}&8g^{\mu \nu}\Bigl(\epsilon t u -s^2-s(u+t)\Bigr)+8p^{\nu}\Bigl(2 \epsilon s p^{\mu}+\epsilon u k_2^{\mu}-sk_2^{\mu}+\epsilon t k_1^{\mu} - sk_1^{\mu} \Bigr) \notag\\ 
  &+8k_1^{\nu} \Bigl(\epsilon t p^{\mu}-s p^{\mu} - 2tk_1^{\mu} +k_2^{\mu}(u+t+2s)\Bigr) \notag \\
  &+8k_2^{\nu}\Bigr(\epsilon u p^{\mu} -sp^{\mu}-2uk_2^{\mu}+k_1^{\mu}(u+t+2s)\Bigr). \label{C}
\end{align}
\subsubsection*{Computation of $\hat{F}_2$}
Now that we have computed $\hat{W}^{\mu \nu}$ we can apply \eqref{strfunc2:a} to compute $\hat{F}_2$. Remembering the properties in Eqs.~(\ref{proj:a}) and (\ref{proj:b}) we find that
\begin{align*}
  \hat{F}_2={}&T_F \hat{\nu} \frac{e_q^2g_s^2}{2(1-\epsilon)}\Biggl[-\frac{16}{t}(1-\epsilon)(k_1\cdot \bar{p})-\frac{16}{u}(1-\epsilon)(k_2\cdot \bar{p})\\
  &+\frac{8}{ut}(k_1\cdot \bar{p})\Bigl(\epsilon t -s -2t(k_1\cdot \bar{p})+(k_2\cdot \bar{p})(s-Q^2)\Bigr) \\
  &+\frac{8}{ut}(k_2 \cdot \bar{p})\Bigl(\epsilon u -s -2 u(k_2 \cdot \bar{p})+(k_1\cdot \bar{p})(s-Q^2)\Bigr)\\
  &+\frac{8}{ut}\Bigl(2\epsilon s +\epsilon u(k_2 \cdot \bar{p})-s(k_2\cdot \bar{p})+\epsilon t (k_1\cdot \bar{p}) -s(k_1\cdot \bar{p})\Bigr)\Biggr].
\end{align*}
Using Eqs.~(\ref{k1k2p:a}) and (\ref{k1k2p:b}) to write $k_2 \cdot \bar{p}=1-k_1 \cdot \bar{p}$ and using $\hat{\nu}=Q^2/2\hat{x}$ we find that
\begin{equation}\label{F2}
  \hat{F}_2=\frac{4T_F e_q^2g_s^2}{\hat{x}(1-\epsilon)ut}Q^4\biggl(2(k_1\cdot \bar{p})^2-2(k_1\cdot \bar{p})+1-\epsilon\biggr).
\end{equation}
In order to compute $\hat{F}_2$ we move to the center-of-mass frame of the real gluon and virtual photon and we rotate the spatial axis such that their 3-momenta lie on the $z$ axis. Since the gluon is on-shell and the photon is off-shell we can write
\begin{equation*}
  p=(p^0,0,0,p^0),\quad q=(q^0,0,0,-p^0)\quad p^0,\,q^0>0.
\end{equation*}
If we define the two light-like 4-vectors
\begin{equation*}
  n=(1,0,0,1), \quad \bar{n}=(1,0,0,-1),
\end{equation*}
we can write
\begin{equation}\label{p;q}
  p=p^0n,\quad q=\alpha n +(p^0+\alpha)\bar{n}=(2\alpha+p^0,0,0,-p^0).
\end{equation}
The constants $\alpha$ and $p^0$ can be computed in terms of $s$ and $Q^2$ from the relations $s=(q+p)^2$ and $q^2=-Q^2$, finding
\begin{equation}
  p^0=\frac{s+Q^2}{2\sqrt{s}}, \quad \alpha=-\frac{Q^2}{2\sqrt{s}}.
\end{equation}
In order to compute $(k_1\cdot \bar{p})$ in this frame we parametrize $k_1$, $k_2$ and $\bar{p}$ in the following way
\begin{align*}
  \bar{p}&=A_{\bar{p}}n+B_{\bar{p}}\bar{n}+\bar{p}_t ,\\
  k_1&=A_{1}n+B_1\bar{n}+k_t, \\
  k_2&=A_{2}n+B_2\bar{n}-k_t,
\end{align*}
where $\bar{p}_t$ and $k_t$ have only the $x$ and $y$ components and the opposite signs of $k_t$ come from the momentum conservation in the transverse plane. If we impose the relations in Eqs.~(\ref{proj:a}) we can compute $A_{\bar{p}}$, $B_{\bar{p}}$ and $\bar{p}_t^2$, while imposing Eqs.~(\ref{mand:a}), (\ref{mand:b}) and (\ref{mand:c}) and $k_1^2=k_2^2=0$ we can compute $A_1$, $A_2$, $B_1$, $B_2$ and $k_t^2$. What we find is
%\begin{subequations}\label{pbar;k1;k2}
\begin{align}
  \bar{p}={}&\frac{Q^2}{\sqrt{s}(s+Q^2)}n+\frac{\sqrt{s}}{s+Q^2}\bar{n}+\bar{p}_t, \label{pbar;k1;k2:a} \\
  k_1={}&-\frac{t}{2\sqrt{s}}(1-\hat{x})n-\frac{u\sqrt{s}}{2(s+Q^2)}\bar{n}+k_t, \label{pbar;k1;k2:b} \\
  k_2={}& -\frac{u\sqrt{s}}{2(s+Q^2)}n-\frac{t\sqrt{s}}{2(s+Q^2)}\bar{n}-k_t, \label{pbar;k1;k2:c}
\end{align}
%\end{subequations}
with
\begin{equation}
  \bar{p}_t^2=-|\vec{\bar{p}}_t|^2=-\frac{4\hat{x}^2}{Q^2}, \quad k_t^2=-|\vec{k_t}|^2=-\frac{tu}{s+Q^2}(1-\hat{x}).
\end{equation}
In conclusion we find, using the relations $s+Q^2=Q^2/\hat{x}$ and $Q^2+t\hat{x}=-uQ^2/(s+Q^2)$, that
\begin{equation}
  k_1 \cdot \bar{p}=\frac{\hat{x}}{Q^2}\Bigl(Q^2+t(2\hat{x}-1)\Bigr)+k_t\cdot \bar{p}_t. \label{k*p}
\end{equation}
$\bar{p}_t$ and $k_t$ are two space-like vectors in the $xy$ plane so $k_t \cdot \bar{p}_t=-|k_t||\bar{p}_t|\cos \phi$ where $\phi$ is the angle between the two vectors. Since the integration of the phase space of the final particles includes an integration in $d\phi$, observing that
\begin{equation}\label{integral}
  \frac{1}{2\pi}\int_0^{2\pi}\cos \phi\, d \phi=0, \quad \frac{1}{2\pi}\int_0^{2\pi}\cos^2\phi\, d \phi=\frac{1}{2},
\end{equation}
we have that the factor $k_t \cdot \bar{p}_t$ contributes only in the quadratic form. So we can write
\begin{equation}\label{ktpt}
  \frac{1}{2\pi}\int_0^{2\pi}d\phi\,(k_t \cdot \bar{p}_t)^2=-\frac{2\hat{x}^2}{Q^2}\frac{t(1-\hat{x})(Q^2+t\hat{x})}{Q^2}.
\end{equation}

Now we can plug the results of \eqref{k*p} and \eqref{ktpt} into \eqref{F2} and integrate the $d$-dimensional phase space that has the expression
\begin{equation}
  d\phi_2=\frac{1}{8\pi}\frac{(4\pi)^{\epsilon}}{\Gamma(1-\epsilon)}\Bigl(\frac{1-\hat{x}}{\hat{x}}\Bigr)^{-\epsilon}Q^{-2\epsilon}y^{-\epsilon}(1-y)^{-\epsilon}dy, \label{PS}
\end{equation}
with 
\begin{equation}
y=\frac{1-\cos \theta}{2},
\end{equation}
where $\theta$ is the angle between $\vec{k}_1$ and the $z$ axis.
In \eqref{PS} the angle $\phi$ (or its analogous in $d$ dimensions) is already integrated assuming independence from $\phi$. This gives a factor that for $\epsilon \rightarrow 0$ goes to $2\pi$. Therefore, since we want to integrate a function that has a dependence from $\phi$ we have divided a factor of $2\pi$ in the \eqref{integral} that cancels the $2\pi$ present in \eqref{PS}. Using
\begin{equation*}
  t=-2k_2\cdot p=-2(k_2^0p^0-|\vec{k_2}||\vec{p}\,|\cos \theta)=-2k_2^0p^0(1-\cos \theta),
\end{equation*}
we can show that 
\begin{equation}\label{y}
t=-\frac{Q^2y}{\hat{x}},
\end{equation}
from which we find
\begin{equation}\label{u}
 u=-\frac{Q^2(1-y)}{\hat{x}}.
\end{equation}  
Now we can plug Eqs.~(\ref{y}) and (\ref{u}) into \eqref{F2} and integrate in the phase space given \eqref{PS}. Using the $\overline{\mbox{MS}}$ scheme, so that the coupling becomes $\alpha_s(\mu^2)\bigl(\frac{\mu^2e^{\gamma}}{4\pi}\bigr)^{\epsilon}$ with $\alpha_s(\mu^2)$ dimensionless, we arrive to the expression
\begin{align}
  \hat{F}_2&=\hat{x}\frac{2T_F e_q^2\alpha_s(\mu^2)e^{\gamma \epsilon}}{(1-\epsilon)\Gamma(1-\epsilon)}\Bigl(\frac{\mu^2}{Q^2}\Bigr)^\epsilon \Bigl(\frac{1-\hat{x}}{\hat{x}}\Bigr)^{-\epsilon}\times \notag \\
  &\quad\int_0^1\frac{2\hat{x}(\hat{x}-1)+(1-\epsilon)-2(6\hat{x}^2-6\hat{x}+1)y(1-y)}{y^{\epsilon+1}(1-y)^{\epsilon+1}}\,dy.
\end{align}
The integral can be evaluated using the properties of the gamma function:
\begin{align*}
  &\int_0^1\frac{2\hat{x}(\hat{x}-1)+(1-\epsilon)-2(6\hat{x}^2-6\hat{x}+1)y(1-y)}{y^{\epsilon+1}(1-y)^{\epsilon+1}}\,dy \\
  &=\, (2\hat{x}(\hat{x}-1)+1-\epsilon)\int_0^1y^{-\epsilon-1}(1-y)^{-\epsilon-1}\,dy-2(6\hat{x}^2-6\hat{x}+1)\int_0^1y^{-\epsilon}(1-y)^{-\epsilon}\,dy  \\
 &=\, (2\hat{x}(\hat{x}-1)+1-\epsilon)\frac{\Gamma(-\epsilon)\Gamma(-\epsilon)}{\Gamma(-2\epsilon)}-2(6\hat{x}^2-6\hat{x}+1)\frac{\Gamma(1-\epsilon)\Gamma(1-\epsilon)}{\Gamma(2-2\epsilon)},
\end{align*}
where we used that
\begin{equation}\label{gamma}
  \int_0^1y^{a-1}(1-y)^{a-1}\,dy=\frac{\Gamma(a)\Gamma(a)}{\Gamma(2a)}.
\end{equation}

Finally we can expand in $\epsilon$ using that
\begin{align*}
 \Gamma(\epsilon)&=\frac{1}{\epsilon}-\gamma+\mathcal{O}(\epsilon), \\
 e^{\gamma \epsilon}& =1+\gamma \epsilon +\mathcal{O}(\epsilon^2), \\
 \Bigl(\frac{1-\hat{x}}{\hat{x}}\Bigr)^{-\epsilon} &=1-\epsilon \log \Bigl(\frac{1-\hat{x}}{\hat{x}}\Bigr)+\mathcal{O}(\epsilon^2), \\
 \Bigl(\frac{\mu^2}{Q^2}\Bigr)^\epsilon&=1+\epsilon \log \Bigl(\frac{\mu^2}{Q^2}\Bigr)  +\mathcal{O}(\epsilon^2)
\end{align*}
In conclusion the expression for $\hat{F}_2$ is, neglecting terms linear in $\epsilon$ and inserting the ``flux factor'' $1/4\pi$
\begin{align}
  \hat{F}_2=\hat{x}e_q^2\alpha_s(\mu^2)\frac{4T_F}{4\pi}\biggl[&-8\hat{x}^2+8\hat{x}-1+\log\Bigl(\frac{1-\hat{x}}{\hat{x}}\Bigr)\bigl(\hat{x}^2+(1-\hat{x})^2\bigr)-\frac{1}{\epsilon}\bigl(\hat{x}^2+(1-\hat{x})^2\bigr)\notag \\
  &-\bigl(\hat{x}^2+(1-\hat{x})^2\bigr) \log \Bigl(\frac{\mu^2}{Q^2}\Bigr)\biggr].
\end{align}
and therefore, using Eqs.~(\ref{q:red}) and (\ref{g:red}), we find the expression in \eqref{Cg1},
that is the final result.
\subsubsection*{Computation of $\hat{F}_L$}
Now we can apply \eqref{strfunc2:b} in order to compute $\hat{F}_L$. Remembering the relations in Eqs.~(\ref{wmunu}), (\ref{A}), (\ref{B}) and (\ref{C}) and in Eqs.~(\ref{proj:a}) and (\ref{proj:b}) we find
\begin{align}
  \hat{F}_L=\frac{4\hat{x}^2}{\hat{\nu}}{e_q^2g_s^2}{4(1-\epsilon)ut}\biggl[&8k_1\cdot\bar{q}\Bigl(-2tk_1\cdot\bar{q}+k_2\cdot\bar{q}(u+t+2s)\Bigr)\notag \\
  &+8k_2\cdot\bar{q}\Bigl(-2uk_2\cdot\bar{q}+k_1\cdot\bar{q}(u+t+2s)\Bigr)\biggr].
\end{align}
Using Eqs.~(\ref{k1k2p:a}) and (\ref{k1k2p:b}) to write $k_2\cdot\bar{q}=\hat{\nu}-k_1\cdot\bar{q}$ we can find
\begin{equation}\label{FL}
  \hat{F}_L=\frac{8T_F \hat{x}^2}{\hat{\nu}}\frac{2e_q^2g_s^2Q^2}{(1-\epsilon)ut}\biggl[4(k_1\cdot\bar{q})^2+(k_1\cdot\bar{q})\Bigl(\frac{2u+s-Q^2}{\hat{x}}\Bigr)-\frac{uQ^2}{2\hat{x}^2}\biggr].
\end{equation}
In order to compute $k_1\cdot\bar{q}$ we have to parametrize $\bar{q}$ as we did for $\bar{p}$ in the previous section. So we write
\begin{equation}
  \bar{q}=A_{\bar{q}}n+B_{\bar{q}}\bar{n}+\bar{q}_t.
\end{equation}
Imposing \eqref{proj:b} we find
\begin{equation*}
  A_{\bar{q}}=\frac{Q^2}{2\hat{x}\sqrt{s}}, \quad B_{\bar{q}}=0, \quad \bar{q}_t^2=0.
\end{equation*}
Since $\bar{q}_t$ is a spatial vector $\bar{q}_t^2=0$ implies $\bar{q}_t=0$. So we've found
\begin{equation}\label{qbar}
  \bar{q}=\frac{Q^2}{2\hat{x}\sqrt{s}}n.
\end{equation}
Using this parametrization we can compute
\begin{equation}
  k_1\cdot\bar{q}=-\frac{Q^2}{2\hat{x}}\frac{u}{s+Q^2}.
\end{equation}
Using this expression in \eqref{FL} we find after some algebra
\begin{equation}
  \hat{F}_L=32T_F\frac{e_q^2g_s^2}{1-\epsilon}\hat{x}^2(1-\hat{x})
\end{equation}
Now we integrate the phase space in \eqref{PS} finding
\begin{equation*}
  \hat{F}_L=32T_F\frac{e_q^2\alpha_s(\mu^2) e^{\gamma \epsilon}}{(1-\epsilon)\Gamma(1-\epsilon)}\Bigl(\frac{\mu^2}{Q^2}\Bigr)^\epsilon \Bigl(\frac{1-\hat{x}}{\hat{x}}\Bigr)^{-\epsilon}\hat{x}^2(1-\hat{x})\int_0^1y^{-\epsilon}(1-y)^{-\epsilon}\,dy,
\end{equation*}
and using the integral in \eqref{gamma} we get
\begin{equation*}
  \int_0^1y^{-\epsilon}(1-y)^{-\epsilon}\,dy=\frac{\Gamma(1-\epsilon)\Gamma(1-\epsilon)}{\Gamma(2-2\epsilon)}.
  \end{equation*}
In conclusion, neglecting terms of $\mathcal{O}(\epsilon)$ and inserting the same factor $1/4\pi$,  we have that
\begin{equation}
  \hat{F}_L=\hat{x} e_q^2\alpha_s(\mu^2)\frac{16T_F}{4\pi}\hat{x}(1-\hat{x}),
\end{equation}
that is finite, as we said previously.

\subsection{Discussion to all orders}

So far we have done a discussion at leading order in perturbation theory: we considered the expansion of $C_{2,q}$ and $C_{2,g}$ up to $\ord{}$ and we factorized their divergences in the PDFs order by order. Moreover we considered only one flavour of quarks. Obviously this discussion can be generalized beyond leading order. The most general expression of a generic structure functions (i.e.\ either $F_2$ or $F_L$) is, suppressing all $z$ and $Q^2$ dependence, and considering all the light flavours
\begin{equation}\label{bare:str:func}
    F=x\sum_{i=q,\bar{q},g} C_{0,i}\otimes f_{0,i}, \qquad q=u,d,s,
\end{equation}
where $C_{0,i}$ and $f_{0,i}$ are respectively the bare coefficient functions and the bare PDF. It means that $C_{0,i}$ contains the collinear singularities. One can show that the bare coefficient functions always factorize as
\begin{equation}\label{div:fact}
    C_{0,i}=\sum_{j=q,\bar{q},g}C_{j}(\mu_F^2)\otimes \Gamma_{ji}(\mu_F^2),
\end{equation}
where $C_j$ is no longer singular and therefore $\Gamma_{ji}$ contains the collinear divergences. As we did in the previous discussion, we introduced the factorization scale $\mu_F$. In fact we have that, using the $\MSbar$ scheme with $\mu_R=\mu_F=\mu$,
\begin{equation}
    \Gamma_{ji}(z, \mu^2)=\delta_{ji}\delta(1-z) - \frac{1}{\epsilon} \alpha_s (\mu^2) P_{ji}^{(0)}(z)+\ord{2}.
\end{equation}
The objects $P_{ji}^{(0)}$ will be given in the next section.
Plugging \eqref{div:fact} into \eqref{bare:str:func} we find
\begin{equation}\label{bare:str:func:fact}
    F=x\sum_{i,j=q,\bar{q},g} C_{j}(\mu^2)\otimes \Gamma_{ji}(\mu^2)\otimes f_{0,i} .
\end{equation}
Now we can absorb the collinear divergences into a redefinition of the PDFs. Hence we define
\begin{equation}
    f_j(\mu^2)=\sum_{i=q,\bar{q},g}\Gamma_{ji}(\mu^2)\otimes f_{0,i},
\end{equation}
so that \eqref{bare:str:func:fact} becomes
\begin{equation}\label{finite:str:func}
    F=x\sum_{j=q,\bar{q},g} C_{j}(\mu^2)\otimes f_{j}(\mu^2),
\end{equation}
that is written in terms of finite quantities.
Expanding these equations at $\ord{}$ we get exactly what we found in the previous sections.

\section{DGLAP equations}
\label{DGLAP:sec}

Imposing the independence of the bare PDFs from $\mu^2$ we can derive the so-called Dokshitzer-Gribov-Lipatov-Altarelli-Parisi (DGLAP) equations:
\begin{equation}\label{DGLAP}
\mu^2\frac{d}{d\mu^2}f_i(x,\mu^2)=\sum_{j=q,\bar{q},g}\int_x^1 \frac{dz}{z} P_{ij} \Bigl( \frac{x}{z},  \alpha_s(\mu^2) \Bigr) f_j(z,\mu^2), \qquad i=q,\bar{q},g,
\end{equation}
where $P_{ij}$ are called \textit{splitting functions} and are computed in perturbation theory, i.e.\
\begin{equation}\label{DGLAP:exp}
P_{ij}(z)=\alpha_sP_{ij}^{(0)}(z)+\alpha_s^2P_{ij}^{(1)}(z)+ \ord{3}.    
\end{equation}
DGLAP equations are a system of $2n_f+1$ integro-differential equations whose solution gives the running of the PDFs with the factorization scale. They are the analogous of the RGE, \eqref{RGE}, for the running of $\alpha_s$. 
It means that, solving \eqref{DGLAP} with a certain initial condition at the scale $\mu_0$, we can find the PDFs at every scale $\mu$. Therefore
\begin{equation}
    f_i(x,\mu^2)=\sum_{j=q,\bar{q},g}U_{ij}(\mu, \mu_0)f_j(x,\mu_0^2), \qquad i=q,\bar{q},g,
\end{equation}
where $U_{ij}(\mu,\mu_0)$ are the DGLAP evolution operators from the scale $\mu_0$ to the scale $\mu$. They resum logarithms of the
form $\log\bigl(\mu/\mu_0\bigr)$ to all orders in $\alpha_s$. In fact we can write
\begin{equation}
    U_{ij}(\mu,\mu_0)=U^{LL}_{ij}(\alpha_s L)+ \alpha_sU_{ij}^{NLL}(\alpha_s L)+ \alpha_s^2U_{ij}^{NNLL}(\alpha_s L)+\ord{3},
\end{equation}
where $L=\log(\mu/\mu_0)$ and $U^{N^kLL}_{ij}(\alpha_s L)$ are functions to all orders in $\alpha_s L$. The PDFs evolved at N$^k$LL are usually called N$^k$LO PDFs.

We report the other $\ord{}$ splitting functions in addition to the ones given in Eqs.~(\ref{Pqq}) and (\ref{Pqg}):
\begingroup
\allowdisplaybreaks
\begin{align}
    &P_{qq}^{(0)}(z)=P_{\bar{q}\bar{q}}^{(0)}(z)=\frac{C_F}{2\pi}\biggl(\frac{1+z^2}{1-z}\biggr)_+, \\
    &P_{qg}^{(0)}(z)=P_{\bar{q}g}^{(0)}(z)=\frac{T_F}{2\pi}\bigl(z^2+(1-z)^2\bigr), \\
    &P_{gq}^{(0)}(z)=P_{g\bar{q}}^{(0)}(z)=\frac{C_F}{2\pi}\frac{1+(1-z)^2}{z}, \\
    &P_{gg}^{(0)}(z)=\frac{2C_F}{2\pi}\biggl[z\Bigl(\frac{1}{1-z}\Bigr)_++\frac{1-z}{z}+z(1-z)+\Bigl(\frac{11}{12}-\frac{n_f}{2C_A}\Bigr)\delta(1-z)\biggr].
\end{align}
\endgroup
The higher orders can be found in Ref.~\cite{Moch_2004, Vogt_2004}. Starting from the $\ord{2}$, the components $P_{q_i q_j}$ for $i\neq j$ and $P_{q_i \bar{q}_j}$ for any $i$ or $j$ arise.
Using the definition in \eqref{conv} we can write \eqref{DGLAP} as
\begin{equation}
    \mu^2\frac{d}{d\mu^2}f_i(\mu^2)=\sum_{j=q,\bar{q},g}P_{ij} \otimes f_j(\mu^2),
\end{equation}
where we suppressed the $x$ and $z$ dependence.

In order to solve DGLAP equations we have to write it in a simplified form. First of all we consider
\begin{equation}
    \mu^2\frac{d}{d\mu^2}\bigl(f_i(\mu^2)-f_j(\mu^2)\bigl)=\sum_{k=q,\bar{q},g}\bigl(P_{ik}\otimes  f_k(\mu^2)-P_{jk}\otimes f_k(\mu^2)\bigr) ,
\end{equation}
with $i,j=q,\bar{q}$. The contribution for $k=g$ exactly cancels among the two terms in parenthesis. At leading order, using that $P_{q_iq_j}^{(0)}=0$ for $i\neq j$ and $P_{\bar{q}_iq_j}^{(0)}=0$ for any $i$ or $j$, we find that
\begin{equation}\label{nonsingletDGLAP}
    \mu^2\frac{d}{d\mu^2}\bigl(f_i(\mu^2)-f_j(\mu^2)\bigl)=P_{qq} \otimes \bigl(f_i(\mu^2)-f_j(\mu^2)\bigr).
\end{equation}
Therefore, if we have $n_f$ light flavours, there are $2n_f-1$ independent combinations of PDFs that evolve independently from each other. They are called \textit{non-singlet} components. Beyond LO the non-singlet components can be still diagonalized but in a more complex way, see Ref.~\cite{ellis_stirling_webber_1996} for a complete description.
Then we consider
\begin{align}
    \mu^2\frac{d}{d\mu^2}\sum_{i=q,\bar{q} }f_i(\mu^2)&=\sum_{\substack{i=q,\bar{q} \\ k=q,\bar{q},g}}P_{ik} \otimes f_k(\mu^2)=\sum_{i,k=q,\bar{q}}P_{ik}\otimes f_k(\mu^2)+\sum_{i=q,\bar{q}}P_{ig}\otimes f_g(\mu^2)\notag \\
    &=P_{qq}\otimes \sum_{k=q,\bar{q}}f_k(\mu^2)+2 n_f P_{qg} \otimes f_g(\mu^2).
\end{align}
In the last step we used again that at leading order $P_{q_i q_j}$ is diagonal. This form is found also beyond LO, but with a different definition of $P_{qq}$ \cite{ellis_stirling_webber_1996}.
Instead, for the gluon evolution we have that
\begin{equation}
    \mu^2\frac{d}{d\mu^2}f_g(\mu^2)=\sum_{i=q,\bar{q},g}P_{gi} \otimes f_i(\mu^2) = \sum_{i=q,\bar{q}}P_{gi} \otimes f_i(\mu^2) + P_{gg} \otimes f_g(\mu^2).
\end{equation}
Therefore, defining 
\begin{equation}
    S(\mu^2)=\sum_{i=q,\bar{q}}f_i(\mu^2),
\end{equation}
that is called \textit{singlet} component, we get that
\begin{align}
    \mu^2\frac{d}{d\mu^2}S(\mu^2)&=P_{qq}\otimes S(\mu^2) + 2n_fP_{qg} \otimes f_g(\mu^2), \label{singletDGLAP} \\
    \mu^2\frac{d}{d\mu^2}f_g(\mu^2)&=P_{gq}\otimes S(\mu^2) + P_{gg} \otimes f_g(\mu^2). \label{gluonDGLAP}
\end{align}
It means that, while the non-singlet components evolve independently, the singlet component mixes with the gluon density in its evolution.

After that we have almost diagonalized \eqref{DGLAP}, in order to solve it we move to Mellin space: in this way DGLAP equations become
\begin{equation}\label{DGLAP:mellin}
    \mu^2\frac{d}{d\mu^2}f_i(N,\mu^2)=\sum_{i,j}\gamma_{ij}\bigl(N,\alpha_s(\mu^2)\bigr)f_j(N,\mu^2),
\end{equation}
where $f_i(N,\mu^2)$ and $\gamma_{ij}(N,\alpha_s)$ are the Mellin transform of $f_i(x,\mu^2)$ and $P_{ij}(x,\alpha_s)$ and are given by
\begin{align}
    f_i(N,\mu^2)&=\int_0^1dz\,z^{N-1}f_i(z,\mu^2), \\
    \gamma_{ij}(N,\alpha_s)&=\int_0^1dz\,z^{N-1}P_{ij}(z,\alpha_s).
\end{align}
The functions $\gamma_{ij}$ are called \textit{anomalous dimensions}. In this way we get an ordinary differential equation that is easier to solve than \eqref{DGLAP}.
Obviously, the same procedure can be applied to Eqs.~(\ref{nonsingletDGLAP}), (\ref{gluonDGLAP}) and (\ref{singletDGLAP}) giving analogous results.
After that we have solved \eqref{DGLAP:mellin} in Mellin space we have to transform the result back to $z$ space, finding the final solution of \eqref{DGLAP}.

%%%%%%%%%%%%%%%%%%%%%%%%%%%%%%%%%%%
\section{Factorization of large logarithms}
\label{big:logs}
\subsection{Three flavour scheme}
In the previous sections we have said that sending the quark mass to zero gives rise to other IR divergences, in addition to those coming from the zero mass of the gluon.
However, if we are treating heavy quarks, for example in heavy quark pair production, considering them as massless would be a rude approximation. Therefore we will treat the light quark as massless, where light is referred with respect to $\Lambda$, and we will keep the mass dependence of the heavy quarks. It means that the top and bottom quarks will be considered as massive, while up, down and strange will be considered as massless. The charm is a little bit more complicated since its mass is of the same order of $\Lambda$. In order to avoid complications we will treat the charm as heavy, as we did in the previous chapter. 

When we consider the mass of the quarks, it acts as a regulator to collinear singularities. Hence, the coefficient functions will be finite.
In particular, if we choose $\mu=Q$, the cross section will be of the form
\begin{align}
    \sigma  \propto{} & \qquad\,\,\,\,  a_0 \notag \\
    &+\alpha_s\Bigl(b_0 + b_1 \log \frac{Q^2}{m^2} \Bigr) \notag \\
   &+ \alpha_s^2 \Bigl( c_0 + c_1 \log\frac{Q^2}{m^2} + c_2\log^2\frac{Q^2}{m^2}\Bigr) \notag \\
    &+ \alpha_s^3 \Bigl( d_0 + d_1 \log\frac{Q^2}{m^2} + d_2\log^2\frac{Q^2}{m^2} + d_3\log^3\frac{Q^2}{m^2}\Bigr) \notag \\
    &+\dots,
\end{align}
where $m$ is the mass of the quark and the logarithms come from the integration in the collinear region.
It means that while for massless quarks we are forced to perform factorization, for massive quarks there is no need to factorize the logarithms of $Q^2/m^2$. However, also in this case some problems arise: first of all, treating the massive quarks is way more complicated than treating the massless ones.
Second, the appearance of logarithms of the form $\log \frac{Q^2}{m^2}$ is problematic:
when $Q^2 \sim m^2$ the logarithms are small and therefore every term in the perturbative expansion is smaller then the previous ones thanks to the extra powers of $\alpha_s$. It means that we can safely approximate the cross section taking all the contributions up to a certain order in $\alpha_s$ and neglecting all the higher orders, because we are throwing away small terms. 
Instead, when $Q^2 \gg m^2$ the factors $\log \frac{Q^2}{m^2}$ become large and we have that $\alpha_s(Q^2)\log Q^2/m^2 \sim 1$: it means that every order in perturbation theory is not negligible with respect to the previous ones but rather it is comparable with them. This means that truncating the perturbative expansion up to a certain order is no longer a good approximation. Therefore we have to resum the series to all orders if we want to have reliable predictions.

The \textit{three flavour scheme} (3FS) uses the $\MSbar$ scheme for the three light quarks, it keeps the mass logarithms of the heavy quarks and it uses the DS for the heavy quarks loops. It means that the coefficient function will contain unresummed logarithms, that can potentially be large.
A generic structure function for DIS will be of the form 
\begin{equation}\label{3FS}
    F(x,Q^2)=x\sum_{i=q,\bar{q},g} C^{[3]}_{i}\Bigl(\frac{m^2}{Q^2},\alpha_s^{[3]}(Q^2)\Bigr)\otimes f^{[3]}_{i}(Q^2),
\end{equation}
where $C^{[3]}$ is the coefficient function in the 3FS. In this scheme the PDFs are labeled as $f^{[3]}$ to remember that they follow DGLAP evolution with three active flavours and therefore they satisfy
\begin{equation}\label{DGLAP3FS}
    f_i^{[3]}(Q^2)=\sum_{j=q,\bar{q},g}U^{[3]}_{ij}(Q^2,Q_0^2) \otimes f_j^{[3]}(Q_0^2), \quad i=q,\bar{q},g.
\end{equation}
Notice that \eqref{3FS} is written in terms of $\alpha_s^{[3]}$ because we have used the DS for the heavy quarks and therefore the beta function is written in terms of $n_f=3$.
In the case of heavy quark pair production, since the coefficient functions contain the exact heavy quark mass dependence, they will contain also kinematical constraints like a term $\theta(s-4m^2)$, that comes from imposing momentum conservation of the partonic process.
In fact, the heavy quark can be produced in the final state only if $s>4m^2$.
Using that $s=Q^2(1/z-1)$ the kinematical constraint becomes $\theta(z_{\rm max}-z)$ with $z_{\rm max}=1/(1+4m^2/Q^2)$.
If $s<4m^2$ the heavy quarks doesn't appears in the final state. It means that, since the adoption of the DS assures the decoupling of heavy quark loops, the heavy quark completely decouples below its threshold and thus at scales $Q^2$ such that $Q^2 \ll m^2$. 
Therefore, if $Q^2 \ll m^2$ the heavy quark mass dependence completely disappears from \eqref{3FS}.

So far we focused on the charm quark. However, these considerations apply equivalently to the case of bottom and top. Therefore, the top quark can be ignored if we work at energies below its threshold. This is why in \eqref{3FS} we considered only the charm.

\subsection{Computation of the massive gluon coefficient function}

The fact that keeping the heavy quark mass dependence regulates the IR divergences and gives the kinematical constraint $\theta(z_\text{max}-z)$ can be shown for example computing the process in \eqref{process} without sending $m\rightarrow 0$.
It means that now we consider the case $k_i^2=m^2\neq 0$, where $m$ is the mass of the quark produced in the process that we are considering. In this case the mass of the quark behaves as a IR regulator, so the amplitudes will be finite: it means that we will not need dimensional regularization and therefore there will not be poles in $\epsilon$. However we still expect to find logarithms of the form $\log Q^2/m^2$ that are large for $Q^2\gg m^2$. In this case \eqref{Mmu} becomes
\begin{equation}\label{Mmu2}
  M^{\mu}=e_q g_s T^a\bar{u}(k_1)\Biggl(\gamma^{\mu} \frac{\slashed{p}-\slashed{k_2}+m}{(p-k_2)^2-m^2}\slashed{\epsilon}(p)+\slashed{\epsilon}(p)\frac{\slashed{k_1}-\slashed{p}+m}{(k_1-p)^2-m^2}\gamma^{\mu}\Biggr)v(k_2),
\end{equation}
and \eqref{hadr:ten} becomes
\begin{align}
  \hat{W}^{\mu \nu}&=T_F\frac{e_q^2g_s^2}{2}\tr\Biggl[(\slashed{k_1}+m)\Biggl(\gamma^{\mu}\frac{\slashed{p}-\slashed{k_2}+m}{(p-k_2)^2-m^2}\slashed{\epsilon}(p)+\slashed{\epsilon}(p)\frac{\slashed{k_1}-\slashed{p}+m}{(k_1-p)^2-m^2}\gamma^{\mu}\Biggr)\,\,\times \notag\\
   &\quad\,\,(\slashed{k_2}-m) \Biggl(\slashed{\epsilon^*}(p)\frac{\slashed{p}-\slashed{k_2}+m}{(p-k_2)^2-m^2}\gamma^{\nu}+\gamma^{\nu}\frac{\slashed{k_1}-\slashed{p}+m}{(k_1-p)^2-m^2}\slashed{\epsilon^*}(p)\Biggr)\Biggr]. \label{hadr:ten2}
\end{align}
Now the Mandelstam variables will be
\begin{align}
s&=(p+q)^2=-Q^2+2p\cdot q=(k_1+k_2)^2=2m^2+2k_1\cdot k_2, \label{mand2:a} \\
t&=(q-k_1)^2=-Q^2+m^2-2k_1\cdot q=(k_2-p)^2=m^2-2k_2\cdot p, \label{mand2:b}\\
u&=(q-k_2)^2=-Q^2+m^2-2k_2\cdot q=(k_1-p)^2=m^2-2k_1\cdot p. \label{mand2:c}
\end{align}
Using these definitions we can write \eqref{hadr:ten2} in the form
\begin{equation}\label{wmunu2}
\hat{W}^{\mu \nu}=T_F\frac{e_q^2g_s^2}{2}\Biggl(\frac{A^{\mu \nu}}{(t-m^2)^2}+\frac{B^{\mu \nu}}{(u-m^2)^2}+\frac{C^{\mu \nu}}{(u-m^2)(t-m^2)}\Biggr),
\end{equation}
where
\begin{align*}
A^{\mu \nu}={}&\tr \Bigl((\slashed{k_1}+m)\gamma^{\mu}(\slashed{p}-\slashed{k_2}+m)\slashed{\epsilon}(\slashed{k_2}-m)\slashed{\epsilon}^*(\slashed{p}-\slashed{k_2}+m)\gamma^{\nu}\Bigr), \notag \\
B^{\mu \nu}={}&\tr \Bigl((\slashed{k_1}+m)\slashed{\epsilon}(\slashed{k_1}-\slashed{p}+m)\gamma^{\mu}(\slashed{k_2}-m)\gamma^{\nu}(\slashed{k_1}-\slashed{p}+m)\slashed{\epsilon}^*\Bigr), \notag \\
C^{\mu \nu}={}&\tr \Bigl((\slashed{k_1}+m)\gamma^{\mu}(\slashed{p}-\slashed{k_2}+m)\slashed{\epsilon}(\slashed{k_2}-m)\gamma^{\nu}(\slashed{k_1}-\slashed{p}+m)\slashed{\epsilon}^*\Bigr) \notag \\
&+\tr \Bigl((\slashed{k_1}+m)\slashed{\epsilon}(\slashed{k_1}-\slashed{p}+m)\gamma^{\mu}(\slashed{k_2}-m)\slashed{\epsilon}^*(\slashed{p}-\slashed{k_2}+m)\gamma^{\nu}\Bigr).
\end{align*}
Computing these factors with the properties of the Dirac matrices in $d=4$ dimensions (it is sufficient to use the relations in Eqs.~(\ref{diracmatrices:a}-\ref{diracmatrices:e}) with $\epsilon=0$) we find
\begin{align*}
A^{\mu \nu}={}&8m^2(3p^{\mu}k_1^{\nu}+3k_1^{\mu}p^{\nu}-2k_1^{\mu}k_2^{\nu}-2k_2^{\mu}k_1^{\nu})-8t(p^{\mu}k_1^{\nu}+k_1^{\mu}p^{\nu}),  \\
B^{\mu \nu}={}& 8m^2(3p^{\mu} k_2^{\nu} +3k_2^{\mu} p^{\nu} -2k_1^{\mu} k_2^{\nu} -2k_2^{\mu }k_1^{\nu}) -8u(p^{\mu}k_2^{\nu}+k_2^{\mu}p^{\nu}), \\
C^{\mu \nu}={}& 8m^2(2k_1^{\mu}k_1^\nu+2k_2^\mu k_2^\nu-6k_1^\mu k_2^\nu-6k_2^\mu k_1^\nu +2p^\mu k_1^\nu+2p^\mu k_2^\nu+2k_1^\mu p^\nu+2k_2^\mu p^\nu-4p^\mu p^\nu) \\
&+8t(k_1^\mu k_2^\nu+k_2^\mu k_1^\nu-2k_1^\mu k_1^\nu)+8u(k_1^\mu k_2^\nu+k_2^\mu k_1^\nu-2k_2^\mu k_2^\nu)+8s(2k_1^\mu k_2^\nu+2k_2^\mu k_1^\nu-p^\mu k_1^\nu  \\
&-p^\mu k_2^\nu-k_1^\mu p^\nu-k_2^\mu p^\nu).
\end{align*}
where we have omitted terms proportional to $g^{\mu \nu}$ since they are irrelevant for the computation of $\hat{F}_2$ and $\hat{F}_L$ thanks to the relations in Eqs.~(\ref{proj:a}) and (\ref{proj:b}).

\subsubsection*{Computation of $\hat{F}_2$}
We can apply \eqref{strfunc2:a} to compute $\hat{F}_2$. What we find is
\begingroup
\allowdisplaybreaks
\begin{align}
\hat{F}_2={}&8T_F\hat{\nu}\frac{e_q^2g_s^2}{2}\Biggl\{\frac{1}{(t-m^2)^2}\biggl[m^2\Bigl(6(k_1\cdot \bar{p})-4(k_1\cdot \bar{p})(k_2\cdot \bar{p}) \Bigr)-2t(k_1\cdot \bar{p})\biggr] \notag \\
&+\frac{1}{(u-m^2)^2}\biggl[m^2\Bigl(6(k_2\cdot \bar{p})-4(k_1\cdot \bar{p})(k_2\cdot \bar{p}) \Bigr)-2u(k_2\cdot \bar{p})\biggr] \notag \\
&+\frac{1}{(t-m^2)(u-m^2)}\biggl[ m^2\Bigl( 2(k_1\cdot \bar{p})^2+2(k_2\cdot \bar{p})^2 -12(k_1\cdot \bar{p})(k_2\cdot \bar{p})+4(k_1\cdot \bar{p})  \notag \\
&+4(k_2\cdot \bar{p})-4 \Bigr)+t\Bigl( 2(k_1\cdot \bar{p})(k_2\cdot \bar{p})-2(k_1\cdot \bar{p})^2\Bigr)+u\Bigl( 2(k_1\cdot \bar{p})(k_2\cdot \bar{p})-2(k_2\cdot \bar{p})^2\Bigr) \notag \\
&+s\Bigl(4(k_1\cdot \bar{p})(k_2\cdot \bar{p})-2(k_1\cdot \bar{p})-2(k_2\cdot \bar{p})\Bigr)\biggr]\Biggr\},
\end{align}
\endgroup
and using $ k_2 \cdot \bar{p}=1-k_1\cdot \bar{p}$ we find
\begingroup
\allowdisplaybreaks
\begin{align} 
\hat{F}_2={}&\frac{4T_F\hat{\nu} e_q^2g_s^2}{(t-m^2)^2(u-m^2)^2} \Biggl\{(k_1\cdot \bar{p})^2\Biggl[4m^2\Bigl((u-m^2)^2+(t-m^2)^2\Bigr) \notag \\
&+(8m^2+4Q^2)(u-m^2)(t-m^2)\Biggr] +(k_1 \cdot \bar{p})\Biggl[2(m^2-t)(u-m^2)^2\notag \\
&+(-10m^2+2u)(t-m^2)^2+ (-12m^2-2Q^2+4u+2s)(u-m^2)(t-m^2)\Biggr]\notag \\
&+(6m^2-2u)(t-m^2)^2+2(m^2-u-s)(u-m^2)(t-m^2)\Biggr\}. \label{F2last}
\end{align}
\endgroup
Now we have to parametrize the vectors $q$, $p$, $\bar{p}$, $k_1$, $k_2$ like we did in the massless case in order to compute $k_1\cdot \bar{p}$. Following the steps of the massless computation we move to the center-of-mass frame of the virtual photon and the real gluon (that is also the center-of-mass frame of the two final quarks) and we align the 3-momenta of these two initial particles with the $z$ axis. Notice that the expressions that we found for $q$, $p$, $\bar{p}$ (\eqref{p;q} and \eqref{pbar;k1;k2:a}) are still valid because the equations that we imposed to derive them still hold. Computing $k_1$ and $k_2$ in the same way we did in the massless case we find
\begin{align}
&k_1=\frac{1}{2\sqrt{s}}\Bigl(m^2-Q^2-t+\hat{x}(m^2-u)\Bigr)n+\frac{\sqrt{s}}{2(s+Q^2)}(m^2-u)\bar{n}+k_t, \label{param2:a} \\
&k_2=\frac{1}{2\sqrt{s}}\Bigl(m^2(1+\hat{x})-u-Q^2-t\hat{x}\Bigr)n+\frac{\hat{x}\sqrt{s}}{2Q^2}(m^2-t)\bar{n}-k_t ,\label{param2:b}
\end{align}
with
\begin{equation}
k_t^2=-|\vec{k_t}|^2=m^2-\frac{\hat{x}}{Q^2}(m^2-u)\Bigl(m^2-Q^2-t+\hat{x}(m^2-u)\Bigr).
\end{equation}
From the expression of
\begin{equation*}
t=m^2-2k_2\cdot p=m^2-2(k_2^0p^0-|\vec{k_2}||\vec{p}\,|\cos \theta)=m^2-2k_2^0p^0\biggl(1-\frac{|\vec{k_2}|}{k_2^0}\cos \theta\biggr),
\end{equation*}
we can define $t-m^2=-\frac{Q^2y}{\hat{x}}$, where
\begin{equation}\label{y:beta}
y=\frac{1}{2}\Biggl(1-\sqrt{1-\frac{m^2}{Q^2}\frac{4\hat{x}}{(1-\hat{x})}}\cos \theta\Biggl)=\frac{1}{2}(1-\beta \cos \theta),
\end{equation}
where the value of $k_2^0$ has been computed using the parametrization in Eqs.~(\ref{param2:a}) and (\ref{param2:b}) (and since we are in the center-of-mass frame of two particles with the same mass we find $k_2^0=\sqrt{s}/2$). Observe that $\beta$ is exactly the velocity of the final quarks in the partonic center-of-mass frame. Using the definition of $y$ and the results in Eqs.~(\ref{param2:a}) and (\ref{param2:b}) we can show that
\begin{equation}
k_1 \cdot \bar{p}=k_t\cdot \bar{p}_t+y(1-\hat{x})+\hat{x}(1-y),
\end{equation}
where we have also used that $u-m^2=-\frac{Q^2}{\hat{x}}(1-y)$.
Also in this case we have that $k_t \cdot \bar{p}_t=-|k_t||\bar{p}_t|\cos \phi$ and thanks to the integrals in \eqref{integral} only the terms proportional $(k_t\cdot \bar{p})^2$ survive. In conclusion we find that
\begin{equation}
\frac{1}{2\pi}\int_0^{2\pi}(k_t \cdot \bar{p_t})^2 = \frac{1}{2} |k_t|^2|\bar{p}_t|^2=\frac{1}{2}k_t^2\bar{p_t}^2=2\hat{x}y(1-\hat{x})(1-y)-\frac{2\hat{x}^2m^2}{Q^2},
\end{equation}
where the factor $1/2$ comes from the integration over $\phi$. 

Putting everything inside \eqref{F2last} what we find after some algebra is
\begin{align}
\hat{F}_2={}&\frac{4T_F \hat{x}}{Q^2}\frac{e_q^2g_s^2}{y^2(1-y)^2}\biggl[-\frac{4m^4}{Q^2}\hat{x}^2+2\hat{x}m^2\Bigl(\hat{x}\bigl(y^2+(1-y)^2\bigr)-2y(1-y)(3\hat{x}-1)\Bigr)\notag \\
&+Q^2y(1-y)\Bigl((2\hat{x}^2-2\hat{x}+1)-2y(1-y)(6\hat{x}^2-6\hat{x}+1)\Bigr)\biggr]. \label{F2final}
\end{align}
Now we can integrate the phase space that has the form
\begin{equation}\label{PS:CM}
d\phi_2=\frac{\lambda^{\frac{1}{2}}(s,m^2,m^2)}{16\pi s}d\cos\theta=\frac{2}{\beta}\frac{\lambda^{\frac{1}{2}}(s,m^2,m^2)}{16\pi s}dy\,\, , \quad\,\,\, \frac{1-\beta}{2}\leq y\leq\frac{1+\beta}{2},
\end{equation}
where $\lambda(x,y,z)=x^2+y^2+z^2-2xy-2xz-2yz$ is the triangular function and in particular $\lambda^{\frac{1}{2}}(s,m^2,m^2)=\sqrt{s}\sqrt{s-4m^2}$. Integrating \eqref{F2final} we find
\begin{align}
\hat{F}_2={}&\frac{2T_F}{4\pi}\hat{x}\frac{e_q^2\alpha_s}{Q^2\beta}\frac{\sqrt{s-4m^2}}{\sqrt{s}}\Biggl[-4\hat{x}^2\frac{m^4}{Q^2}\biggl(\frac{8\beta}{(1+\beta)(1-\beta)}+4\log\Bigl(\frac{1+\beta}{1-\beta}\Bigr)\biggr)+ \frac{16\hat{x}^2m^2\beta}{(1+\beta)(1-\beta)} \notag \\
&-8\hat{x}m^2(3\hat{x}-1)\log\Bigl(\frac{1+\beta}{1-\beta}\Bigr)+2Q^2(2\hat{x}^2-2\hat{x}+1)\log\Bigl(\frac{1+\beta}{1-\beta}\Bigr)-2\beta Q^2(6\hat{x}^2-6\hat{x}+1)\Biggr] \notag \\
={}&\hat{x}e_q^2 \alpha_s  \frac{4T_F}{4\pi}  \Biggl[ \log\Bigl(\frac{1+\beta}{1-\beta}\Bigr)\biggl(-8\hat{x}^2\frac{m^4}{Q^4}-4\hat{x}\frac{m^2}{Q^2}(3\hat{x}-1)+(2\hat{x}^2-2\hat{x}+1)\biggr)\notag\\
&+\beta\biggl(4\frac{m^2}{Q^2}\hat{x}(\hat{x}-1)-(8\hat{x}^2-8\hat{x}+1)\biggr)\Biggr],
\end{align}
where we used that $\sqrt{1-4m^2/s}=\beta$ and we have again inserted the flux factor. The term $\log\frac{1+\beta}{1-\beta}$ is the large logarithm that we expected to find since it explodes for $Q^2\gg m^2$ ($\beta \rightarrow 1$ for $\frac{m^2}{Q^2}\rightarrow 0$).
Moreover, from the expression for $\beta$ we find that $\hat{x}<(1+4m^2/Q^2)^{-1}$, that is exactly the kinematic condition that we mentioned before. Using the definition in \eqref{coeff:func} we obtain
\begin{align}
   C_{2,g}^{(1)}(\hat{x}) = \frac{4T_F}{4\pi}  \Biggl[& \log\Bigl(\frac{1+\beta}{1-\beta}\Bigr)\biggl(-8\hat{x}^2\frac{m^4}{Q^4}-4\hat{x}\frac{m^2}{Q^2}(3\hat{x}-1)+(2\hat{x}^2-2\hat{x}+1)\biggr)\notag\\
&+\beta\biggl(4\frac{m^2}{Q^2}\hat{x}(\hat{x}-1)-(8\hat{x}^2-8\hat{x}+1)\biggr)\Biggr]. \label{C2g1}
\end{align}

\subsubsection*{Computation of $\hat{F}_L$}
Applying \eqref{strfunc2:b} to compute $\hat{F}_L$ we find that
\begingroup
\allowdisplaybreaks
\begin{align}
\hat{F}_L={}&T_F\frac{4\hat{x}^2}{\hat{\nu}}\frac{e_q^2g_s^2}{2}8\Biggl\{\frac{-4m^2}{(t-m^2)^2}(k_1 \cdot \bar{q})(k_2 \cdot \bar{q})+\frac{-4m^2}{(u-m^2)^2}(k_1 \cdot \bar{q})(k_2 \cdot \bar{q}) \notag \\
&+\frac{1}{(t-m^2)(u-m^2)}\Biggl[2m^2\biggl((k_1 \cdot \bar{q})^2+(k_1 \cdot \bar{q})^2-6(k_1 \cdot \bar{q})(k_2 \cdot \bar{q})\biggr)\notag \\
&+2t\biggl((k_1 \cdot \bar{q})(k_2 \cdot \bar{q})-(k_1 \cdot \bar{q})^2\biggr)+2u\biggl((k_1 \cdot \bar{q})(k_2 \cdot \bar{q})-(k_2 \cdot \bar{q})^2\biggr) \notag \\
&+4s(k_1 \cdot \bar{q})(k_2 \cdot \bar{q})\Biggr]\Biggr\},
\end{align}
\endgroup
and using that $(k_2 \cdot \bar{q})=\hat{\nu}-(k_1 \cdot \bar{q})$ we get
\begin{align}
\hat{F}_L={}&16T_F\frac{\hat{x}^2e_q^2g_s^2}{\hat{\nu}}\Biggl\{(k_1 \cdot \bar{q})^2\Biggl[\frac{8m^2+4Q^2}{(u-m^2)(t-m^2)}+4m^2\biggl(\frac{1}{(t-m^2)^2}+\frac{1}{(u-m^2)^2}\biggr)\Biggr]\notag \\
&+\hat{\nu}(k_1 \cdot \bar{q})\Biggl[\frac{4u+2s-12m^2-2Q^2}{(t-m^2)(u-m^2)}-4m^2\biggl(\frac{1}{(u-m^2)^2}+\frac{1}{(t-m^2)^2}\biggr)\Biggr]-\frac{2\nu^2}{(t-m^2)}\Biggr\}.
\end{align}
Computing this factor, using that the expression of $\bar{q}$ in the massive case is the same of the massless case, \eqref{qbar}, we find
\begin{equation}
\hat{F}_L=32T_F\frac{\hat{x}^3e_q^2g_s^2}{Q^2}\biggl[Q^2\Bigl(\frac{1-\hat{x}}{\hat{x}}\Bigr)-\frac{m^2}{y(1-y)}\biggr],
\end{equation}
where $y$ is the same we used in the computation of $F_2$.

Integrating this expression in the phase space in \eqref{PS:CM}, observing that this time we have no $\phi$ dependence, we get
\begin{equation}
\hat{F}_L=\hat{x}e_q^2\alpha_s \frac{16T_F}{4\pi}\biggl[ \hat{x}(1-\hat{x}) \beta- 2\hat{x}^2 \frac{m^2}{Q^2}\log\Bigl(\frac{1+\beta}{1-\beta}\Bigr)\biggr].
\end{equation}
that gives
\begin{equation}\label{CLg1}
    C_{L,g}^{(1)}(\hat{x})=\frac{16T_F}{4\pi}\biggl[ \hat{x}(1-\hat{x}) \beta- 2\hat{x}^2 \frac{m^2}{Q^2}\log\Bigl(\frac{1+\beta}{1-\beta}\Bigr)\biggr].
\end{equation}
This time the potentially large logarithm is multiplied by a factor that goes to zero as a power term for $Q^2 \gg m^2$ and therefore gives no problem. Indeed, in the massless computation of $\hat{F}_L$ we had no pole in $\epsilon$, so the insertion of the mass dependence has no divergence to regularize.
We have again the kinematical constraint $\hat{x}<(1+4m^2/Q^2)^{-1}$ that comes from considering heavy flavours pair production with the exact quark mass dependence.

\subsection{Four flavour scheme}
As mentioned before, if $Q^2\gg m^2$ the logarithms become big and must be resummed.
This can be achieved by factorizing the large logarithms into a redefinition of the PDFs in the same way we factorized the collinear divergences, and then performing the DGLAP evolution with four active flavours.
Thanks to Collins theorem \cite{Collins_1998} we can write 
\begin{equation}\label{mass:fact}
    C^{[3]}_{i}\Bigl(\frac{m^2}{Q^2},\alpha_s^{[4]}(Q^2)\Bigr)=\sum_{j=q,\bar{q},g,c,\bar{c}} C^{[4]}_{j}\Bigl(\frac{m^2}{Q^2},\alpha_s^{[4]}(Q^2)\Bigr) \otimes K_{ji}\Bigl(\frac{m^2}{Q^2},\alpha_s^{[4]}(Q^2)\Bigr).
\end{equation}
The functions $K_{ji}$ exactly contain the mass logarithms so that $C^{[4]}_{j}$ are free from large logarithms. They are called \textit{matching conditions}.
Observe that now the 3FS coefficient functions have been written in terms of $\alpha_s^{[4]}(Q^2)$: this can be easily done order by order in perturbation theory by re-expanding $\alpha_s^{[3]}(Q^2)$ in terms of $\alpha_s^{[4]}(Q^2)$ using \eqref{alpha:matching}. Obviously the two expansions are equivalent to all orders. 
With this definition \eqref{3FS} becomes
\begin{equation}
     F(x,Q^2)=x\sum_{\substack{i=q,\bar{q},g \\ j=q,\bar{q},g,c,\bar{c}}} C^{[4]}_{j}\Bigl(\frac{m^2}{Q^2},\alpha_s^{[4]}(Q^2)\Bigr) \otimes K_{ji}\Bigl(\frac{m^2}{Q^2},\alpha_s^{[4]}(Q^2)\Bigr)\otimes f^{[3]}_{i}(Q^2).
\end{equation}
If now we redefine the PDFs as 
\begin{equation}\label{matching:PDF}
    f_j^{[4]}(Q^2)=\sum_{i=q,\bar{q},g} K_{ji}\Bigl(\frac{m^2}{Q^2},\alpha_s^{[4]}(Q^2)\Bigr) \otimes f^{[3]}_{i}(Q^2),\qquad j=q,\bar{q},g,c,\bar{c}\,,
\end{equation}
we find that 
\begin{equation}\label{4FS}
     F(x,Q^2)=x\sum_{j=q,\bar{q},g,c,\bar{c}} C^{[4]}_{j}\Bigl(\frac{m^2}{Q^2},\alpha_s^{[4]}(Q^2)\Bigr) \otimes f^{[4]}_{j}(Q^2).
\end{equation}
\eqref{4FS} is free from large logarithms and therefore we can approximate it truncating the perturbative expansion in $\alpha_s^{[4]}$ up to a certain order.
This is called \textit{four flavour scheme} (4FS).
Observe that since we have removed the large logarithms we can take the massless limit (or equivalently the limit $Q^2 \gg m^2$). The result that we get will be exactly the same we would have if we had considered a theory with four massless quark and we would had factorized the collinear divergences coming from the fourth quark too.
It means that we have that
\begin{equation}\label{massless:limit}
    C_i^{[4]}\Bigl(\frac{m^2}{Q^2}, \alpha_s^{[4]}\Bigr) \xrightarrow[]{Q^2\gg m^2} C^{[4]}_i\bigl(0, \alpha_s^{[4]}\bigr) + \mathcal{O}\Bigl(\frac{m^2}{Q^2}\Bigr),
\end{equation}
where $C^{[4]}_i\bigl(0, \alpha_s^{[4]}\bigr)$ is the coefficient function computed neglecting the heavy quark mass, i.e.\ the one computed in Sec.~\ref{massless:comp}.
Therefore the DGLAP evolution will be written in terms of four active flavours as
\begin{equation} \label{DGLAP4FS}
    f_i^{[4]}(Q^2)=\sum_{j=q,\bar{q},g,c,\bar{c}}U^{[4]}_{ij}(Q^2,Q_0^2) \otimes f_j^{[4]}(Q_0^2),  \quad i=q,\bar{q},g,c,\bar{c}.
\end{equation}
It means that the splitting functions will depend from $n_f=4$ flavours. Observe that in the 4FS we have the appearance of a heavy quark PDF. 
The evolution in \eqref{DGLAP4FS} is resumming collinear logarithms to all orders in perturbation theory.

The matching conditions $K_{ij}$ are computed from the comparison of calculations of the DIS coefficients functions in the 3FS and in the 4FS and using \eqref{massless:limit}. They have the following perturbative expansion
\begin{equation}
    K_{ij}\Bigl(z,\frac{m^2}{\mu^2}\Bigr)=\delta(1-z)\delta_{ij}+\sum_{k=1}^\infty \bigl(\alpha_s^{[4]}(\mu^2)\bigr)^k K_{ij}^{(k)}\Bigl(z,\frac{m^2}{\mu^2}\Bigr),
\end{equation}
where $\mu$ is the factorization scale (previously we assumed $\mu=Q$).
The non-diagonal components involving a gluon and a heavy quark, i.e.\ $K_{cg}$, $K_{\bar{c}g}$, $K_{gc}$ and $K_{g\bar{c}}$ start being nonzero at $\ord{}$, while the off-diagonal components that involve a light quark start contributing only at $\ord{2}$. All the diagonal quantities are of the form $K_{ii} = 1 + \ord{}$ except for $K_{qq}$ which starts at $\ord{2}$.

\section{Variable flavour number scheme}
\label{VFNS}

We have seen that if the DIS scale $Q^2$ is smaller than the heavy quark production threshold, then the heavy quark decouples and cannot be produced. As this threshold is crossed we start producing it and we have a scheme (the 3FS) in which we have unresummed mass logarithms. Increasing again $Q^2$ these logarithms become large and spoil perturbation theory and therefore must be resummed with the 4FS. When we cross the production threshold of the next heavy quark analogous considerations hold.
Therefore we need a scheme such that for $Q^2\ll m^2$ the heavy quark decouples, for $Q^2 \sim m^2$ the fixed order result is used, while for $Q^2 \gg m^2$ we resum large logarithms and the heavy quark can be considered massless. Since the number of active flavours changes with the scale $Q$, such scheme is called variable flavour number (factorization) scheme (VFNS). 

In the literature various schemes are available, such as ACOT \cite{Aivazis_1994,Aivazis_1994_2}, S-ACOT \cite{PhysRevD.57.3051, Kr_mer_2000}, TR and TR' \cite{Thorne_1998, Thorne_2006} and FONLL \cite{Buza_1998, Cacciari_1998, Forte_2010}.
All of them are equivalent to all
orders, because physical observables like cross sections must be scheme-independent,
but differ in the way the ingredients are combined at finite order. 

If we have the PDFs at the scale $Q_0\lesssim m$, we evolve them in the 3FS using \eqref{DGLAP3FS}. As $Q$ approaches $m$ we have to perform the scheme change.
Thus, we have to use \eqref{DGLAP4FS} to evolve the PDFs towards higher energies. It means that the evolution of the PDFs will be described by
\begingroup
\allowdisplaybreaks
\begin{align}
    f_i^{[4]}(\mu^2)\,\,\,\,&\stackrel{\mathclap{\mu > \mu_c}}{=} \sum_{j=q,\bar{q},g,c,\bar{c}}U_{ij}^{[4]}(\mu,\mu_c)f_j^{[4]}(\mu_c^2) \notag \\
   &= \sum_{j=q,\bar{q},g,c,\bar{c}}U_{ij}^{[4]}(\mu,\mu_c)\sum_{k=q,\bar{q},g}K_{jk}\Bigl(\frac{m^2}{\mu_c^2},\alpha_s^{[4]}(\mu_c^2)\Bigr)f_h^{[3]}(\mu_c^2) \notag \\
    &=\sum_{j=q,\bar{q},g,c,\bar{c}}U_{ij}^{[4]}(\mu,\mu_c)\sum_{k=q,\bar{q},g}K_{jk}\Bigl(\frac{m^2}{\mu_c^2},\alpha_s^{[4]}(\mu_c^2)\Bigr)\sum_{h=q,\bar{q},g}U_{kh}^{[3]}(\mu_c,\mu_0)f_h^{[3]}(\mu_0^2) \notag \\
   &=\sum_{h=q,\bar{q},g}T_{ih}(\mu,\mu_0,m)f_h^{[3]}(\mu_0^2),
\end{align}
\endgroup
where $i=q,\bar{q},g,c,\bar{c}$ and $\mu_c$ is the matching scale, i.e.\ the scale at which we perform the scheme change. Obviously we have to choose $\mu_c\sim \mathcal{O}(m)$, otherwise in the matching condition in \eqref{matching:PDF} we would have unresummed large logarithms. It means that for $\mu < \mu_c$ we use the 3FS, while for $\mu > \mu_c$ we use the 4FS.
We defined the evolution matrix $T_{ih}$ as
\begin{equation}
    T_{ih}(\mu,\mu_0,m)=\sum_{\substack{j=q,\bar{q},g,c,\bar{c} \\ k=q,\bar{q},g}}U_{ij}^{[4]}(\mu,\mu_c)K_{jk}\Bigl(\frac{m^2}{\mu_c^2},\alpha_s^{[4]}(\mu_c^2)\Bigr)U_{kh}^{[3]}(\mu_c,\mu_0),
\end{equation}
that is resumming large logarithms of the form $\log \mu^2/\mu_0^2$. 
Observe that in the evolution matrix $T_{ih}$ the dependence from the matching scale $\mu_c$ disappears. In fact, being it a completely arbitrary scale, the physical quantities cannot depend from it. Observing \eqref{4FS}, since the coefficient function in the 4FS does not depend from $\mu_c$, if the PDFs in the 4FS depended from it then the structure function would depend from $\mu_c$ as well. For this reason the PDFs cannot depend on the choice of $\mu_c$ and it means that $T_{ih}$ is independent from $\mu_c$ at any given order.
%When the evolution matrix $T_{ih}$ is expanded at fixed order the dependence from $\mu_c$ disappears. 
This procedure applies in the same way when we cross the threshold for production of the bottom and top quarks: for $\mu>\mu_b$ we use the 5FS and for $\mu>\mu_t$ we use the 6FS where we have respectively five and six active flavours.

In conclusion we have constructed a factorization scheme in which the number of active flavours, i.e. that take part to DGLAP evolution, varies with the scale $\mu$. This is completely analogous to what we did in Sec.~\ref{VFNS:ren}.

%%%%%%%%%%%%%%%%%%%%%%%%%%%%%%%%%%%%%%%%%%%%%%%%%%%%%%%%%%%%%%%%%%%%%%%%%%%%%%%%%%%%%%%%%%%%%%%%

\chapter{Approximate coefficient function at \texorpdfstring{N$^3$LO}{N3LO}}\label{approx}

As we explained in Chapter~\ref{intro}, in order to increase the accuracy of PDF determination we need the N$^3$LO exact massive coefficient functions of both quark and gluon. Unfortunately, such functions are not fully known at present time.
In fact, only the $\mu$-dependent terms are currently available in the literature, since the are computed numerically due to renormalization-group arguments \cite{Laenen_1999, van_Neerven_2000} using the NNLO coefficient functions and the NLO splitting functions \cite{Laenen_1999,Alekhin_2011}.
It means that the only unknown terms are the $\mu$-independent ones.
	
From now on we will consider heavy quark pair production in DIS, considering the charm quark as the heavy quark. This process is slightly different from the totally inclusive one, because it requires the presence of a heavy quark pair in the final state.
It follows that the light quark coefficient function starts at $\ord{2}$ because in such processes two strong vertices are needed to produce a heavy quark pair.
Moreover, we will take into consideration only the structure function $F_2$ which is dominant with respect to $F_L$.
In the end, even if both the gluon and the quark coefficient functions in principle contribute, being the gluon PDF dominant a small-$x$ with respect to the quark PDFs, the gluon initiated processes are much more important than the quark ones. For this reason we will focus on the gluon exact massive coefficient function at N$^3$LO. However, all the following considerations can be applied to the quark coefficient function too.
%	A consequence of considering heavy quark pair production is that the quark coefficient function starts at $\ord{2}$.
	
In every coefficient function for heavy quark production, assuming to be in the 3FS, the dependence from $\mu$ always factorizes as 
\begin{equation}\label{Ci:mu}
C_i\Bigl(z,\frac{m^2}{Q^2}, \frac{m^2}{\mu^2}\Bigr)=\sum_{k=1}^\infty \alpha_s^k \sum_{j=0}^{k-1} C_i^{(k,j)}\Bigl(z,\frac{m^2}{Q^2}\Bigr)\log^j\Bigl(\frac{\mu^2}{m^2}\Bigr),
\end{equation}
where $k$ starts from 1 because in heavy quark pair production the zeroth order of both the quark and the gluon processes are absent.
For this reason the first order expansion of the gluon coefficient function is sometimes called LO, the second order is called NLO and so on. In this thesis we will follow the convention in which the zero order is called LO and is absent, so that the gluon coefficient function starts at NLO.
From \eqref{Ci:mu} it follows that the $\ord{3}$ perturbative expansion of the quark and gluon coefficient functions can be written as
\begin{equation}\label{mu:dependence}
C_i^{(3)}\Bigl(z,\frac{m^2}{Q^2}, \frac{m^2}{\mu^2}\Bigr)=C_i^{(3,0)}\Bigl(z,\frac{m^2}{Q^2}\Bigr)+C_i^{(3,1)}\Bigl(z,\frac{m^2}{Q^2}\Bigr)\log \Bigl(\frac{\mu^2}{m^2}\Bigr)+C_i^{(3,2)}\Bigl(z,\frac{m^2}{Q^2}\Bigr)\log^2\Bigl(\frac{\mu^2}{m^2}\Bigr).
\end{equation}
As we said before, the terms $C_i^{(3,1)}$ and $C_i^{(3,2)}$ in \eqref{mu:dependence} are exactly known. For example, for the $F_2$ gluon coefficient function their expression in the $\MSbar$ scheme is \cite{Laenen_1999, van_Neerven_2000, Alekhin_2011}
\begin{align}
C_{2,g}^{(3,1)}&=C_{2,g}^{(1)} \otimes \Bigl(P_{gg}^{(1)}-\beta_1\Bigr)+C_{2,q}^{(2,0)} \otimes P_{qg}^{(0)}+C_{2,g}^{(2,0)} \otimes \Bigl(P_{gg}^{(0)}-2\beta_0\Bigr), \label{C31}\\
C_{2,g}^{(3,2)}&=C_{2,g}^{(1)} \otimes \biggl(\frac{1}{2}P_{gg}^{(0)}\otimes P_{gg}^{(0)} +\frac{1}{2} P_{gq}^{(0)} \otimes P_{qg}^{(0)}- \frac{3}{2}\beta_0 P_{gg}^{(0)} + \beta_0^2\biggr). \label{C32}
\end{align}
$C_{2,g}^{(1)}$ is given in \eqref{C2g1}. The NNLO quark and gluon massive coefficient functions are exactly known \cite{LAENEN1993162}, even if not in an analytical form. Parametrizations for these functions are given in Ref.~\cite{Riemersma_1995}, with minor corrections provided in Ref.~\cite{Harris_1995}.
Observe that in Eqs.~(\ref{C31}) and (\ref{C32}) the splitting functions are expanded as in \eqref{DGLAP:exp}, i.e.\ in terms of $\alpha_s$ and not in terms of $\alpha_s/4\pi$ as they are often presented.
Analogous expressions hold for the $\mu$-dependent terms of the N$^3$LO quark coefficient function.
	
Although the complete expression of the $\mu$-independent part of the N$^3$LO massive coefficient function of the gluon is still unknown, its expansions in various kinematic limits are available in the literature. In particular the \textit{high-scale} limit (i.e.\ $Q^2 \gg m^2$), the \textit{high-energy} limit (i.e.\ $s \rightarrow \infty$, or equivalently $z\rightarrow 0$) and the \textit{threshold} limit (i.e.\ $s \rightarrow 4m^2$, or equivalently $z\rightarrow z_{\rm max}$) are all known. 
These limits are good approximations of the exact curve only in some kinematic regions but not for all the values of $z$. Therefore, if we want to construct an (hopefully good) approximation of the N$^3$LO missing term that is valid in the whole $z$ range, we have to combine such limits in a proper way, as it has already been done in the literature \cite{Kawamura_2012}. 

In this chapter we will construct an approximation for the $\mu$-independent part of the $\ord{3}$ exact massive DIS gluon coefficient function for $F_2$ in heavy quark pair production. In order to do so, we will combine the aforementioned limits of the exact function, but in a different way with respect to the other approximations that can be found in the literature.
Then, adding the exact $\mu$-dependent parts to our approximation of the $\mu$-independent one, we will find the full (approximate) expression of the N$^3$LO exact massive gluon coefficient function.
Being the exact result unknown, in order to test the accuracy of our approximation and to tune its parameters, we will apply the same approximation to the $\ord{}$ and $\ord{2}$ coefficient functions, that are exactly known. In this way we will verify the accuracy of our procedure by comparing such approximate curves with the exact ones.
For this reason, we will give the expressions of the various limits and of the final approximation also at NLO and at NNLO.
Obviously the exact $\ord{3}$ unknown term will be a different function with respect to the $\ord{}$ and $\ord{2}$. It means that, even if our approximation will give precise results when applied to the NLO and to the NNLO, we cannot be sure that it will be equally accurate for the N$^3$LO.
Therefore, we will not have to make a construction that is too specific for the NLO and for the NNLO, but rather we will have to make an approximation that is general enough to be applied to the N$^3$LO $\mu$-independent term. In this way we expect accurate results even if we constructed and tested our approximation on curves that can be in principle very different from the one we are interested in.
	
The logical footsteps of our procedure are the following:
in Sec.~\ref{asy} we will construct the asymptotic limit, i.e.\ the small-$z$ approximation of the exact function. It will be done using the $Q^2 \gg m^2$ limit at which we will reinsert the neglected power terms in the $z\rightarrow 0$ limit. In this way we will assure that the asymptotic limit approaches the exact curve for $z\rightarrow 0$.
In Sec.~\ref{thresh} we will present the threshold limit that has already been derived in Ref.~\cite{Kawamura_2012}. 
These two limits will be combined in Sec.~\ref{comb} using two damping functions in such a way that the final result overlaps the asymptotic limit for $z\rightarrow 0$ and the threshold limit for $z\rightarrow z_{\rm max}$ and interpolates the two curves for the intermediate values of $z$. 	
		
\section{Asymptotic limit}\label{asy}
In this section we will construct an approximation for the N$^3$LO exact massive gluon coefficient function in the limit $z \rightarrow 0$, where $z$ is the argument of the coefficient function. 
%Being $s=Q^2(1/z-1)$, where $s$ is the partonic center-of-mass energy, the small-$z$ limit corresponds to $s \rightarrow \infty$.
In order to construct such approximation we will consider the high-scale limit and the high-energy limit, that at $\ord{3}$ are both known with the exception of some contribution for which we will give approximate results. 
The high-scale limit is a good approximation of the exact massive coefficient function for $Q^2 \gg m^2$, but not for values of $Q^2$ such that $Q^2\sim m^2$. Moreover, it does not approaches the exact coefficient function in the small-$z$ region.
The high-energy limit is the expansion for small-$z$ with $Q$ fixed. It means that for any $Q$ it gives the limit of the exact curve for $z \rightarrow 0$. 
In order to find an approximation that describes the exact coefficient function in the small-$z$ limit for every values of $Q$ and therefore that approaches it for $z\rightarrow 0$, we have to reinsert in the high-scale limit the power terms that have been neglected, at least in the small-$z$ limit.
Such power terms will be estimated comparing the high-energy expansion with its $Q^2 \gg m^2$ limit.
In fact, the difference between the two will give exactly the power terms in the small-$z$ limit.
In this way the result will approach the exact coefficient functions for $z\rightarrow 0$ for every value of $Q$. Moreover for $Q^2 \gg m^2$ it will be a good approximation of the exact curve in the whole range of $z$, with the exception of the region $z\simeq z_{\rm max}$. This will be our small-$z$ approximation of the exact massive coefficient function and it will be called asymptotic limit.

\subsection{High-scale limit}
The first ingredient of our approximation is the so-called high-scale limit. In the previous chapter we have said that in the 3FS, the exact massive coefficient functions have unresummed mass logarithms. In particular, using $\mu=Q$, it can be written as
\begin{equation} \label{3FS:expansion}
C^{[3]}_i\Bigl(z,\frac{m^2}{Q^2}, \alpha^{[3]}_s\bigl(Q^2\bigr)\Bigr)=\sum_{k=0}^\infty \Bigl(\alpha_s^{[3]}\bigl(Q^2\bigr)\Bigr)^k\sum_{j=0}^{k}A_{i,k,j}\Bigl(z,\frac{m^2}{Q^2}\Bigr) \log^j\frac{Q^2}{m^2}.
\end{equation}
The high-scale limit is obtained sending $Q^2 \gg m^2$. It is equivalent to neglecting the power terms $m^2/Q^2$ while keeping the logarithmic ones in \eqref{3FS:expansion}. Therefore, in this limit, \eqref{3FS:expansion} can be written as
\begin{equation}\label{high:scale:expansion}
C^{[3,0]}_i\Bigl(z,\frac{m^2}{Q^2}, \alpha^{[3]}_s\bigl(Q^2\bigr)\Bigr)=\sum_{k=0}^\infty \Bigl(\alpha_s^{[3]}\bigl(Q^2\bigr)\Bigr)^k\sum_{j=0}^{k}A_{i,k,j}(z,0) \log^j\frac{Q^2}{m^2},
\end{equation}
where $C^{[3,0]}_i$ is the high-scale limit of the massive coefficient function in the 3FS.
For example, it is easy to show that the $Q^2 \gg m^2$ limit of the gluon coefficient function at $\ord{}$, \eqref{C2g1}, is
\begin{align}\label{1ord:m0:explicit}
C_{2,g}^{[3,0](1)}\Bigl(z,\frac{m^2}{Q^2}\Bigr)=\frac{4T_F}{4\pi} \biggl[ &( 1 - 2 z + 2  z^2 )  \log \Bigl( \frac{1 - z }{ z }\Bigr) - 1 + 8 z - 8  z^2 \notag \\
&+ ( 1 - 2 z + 2 z^2 )\log\Bigl(\frac{Q^2}{m^2}\Bigr)\biggr].
\end{align}

The high-scale limit can be written in a convenient form taking the $Q^2 \gg m^2$ limit of \eqref{mass:fact}. In this way we get that, suppressing the $z$ dependence for ease of notation,
\begin{equation}\label{high:scale1}
C^{[3,0]}_{i}\Bigl(\frac{m^2}{Q^2},\alpha_s^{[3]}(Q^2)\Bigr)=\sum_{j=q,\bar{q},g,c,\bar{c}} C^{[4]}_{j}\Bigl(0,\alpha_s^{[3]}(Q^2)\Bigr) \otimes K_{ji}\Bigl(\frac{m^2}{Q^2},\alpha_s^{[3]}(Q^2)\Bigr).
\end{equation}
Where $C^{[4]}_{j}(0)$ are the massless coefficient functions computed for $\mu=Q$.
This formula can be inverted finding
\begin{equation}\label{high:scale2}
\sum_{i=q,\bar{q},g,c,\bar{c}}C_i^{[3]}\Bigl(\frac{m^2}{Q^2}, \alpha_s^{[3]}(Q^2)\Bigr) \otimes K^{-1}_{ij}\Bigl(\frac{m^2}{Q^2}, \alpha_s^{[3]}(Q^2)\Bigr)\xrightarrow[]{Q^2\gg m^2}C^{[4]}_j\bigl(0,\alpha_s^{[3]}(Q^2)\bigr),
\end{equation}
which shows that $K_{ij}^{-1}(m^2/Q^2)$ factors out the mass logarithms from the massive coefficient functions in the 3FS so that we can safely take the massless limit.	
	
\eqref{high:scale1} holds in the case $\mu=Q$. Instead, in the general case $\mu \neq Q$, $\mu \neq m$ we have that
\begin{equation}\label{high:scale:mu}
C^{[3,0]}_{i}\Bigl(\frac{m^2}{Q^2},\frac{m^2}{\mu^2},\alpha_s^{[3]}(\mu^2)\Bigr)=\sum_{j=q,\bar{q},g,c,\bar{c}} C^{[4]}_{j}\Bigl(\frac{Q^2}{\mu^2},\alpha_s^{[3]}(\mu^2)\Bigr) \otimes K_{ji}\Bigl(\frac{m^2}{\mu^2},\alpha_s^{[3]}(\mu^2)\Bigr),
\end{equation}
where $C^{[4]}_{j}\bigl(Q^2/\mu^2\bigr)$ are the massless coefficient functions in which we keep the dependence from $\mu$. 
%Such dependence is realized through $\log Q^2/\mu^2=\log Q^2/m^2 - \log \mu^2/m^2$. 
Observe that Eqs.~(\ref{high:scale1}), (\ref{high:scale2}) and (\ref{high:scale:mu}), in contrast to \eqref{mass:fact}, are written in terms of $\alpha_s^{[3]}$ since the high-scale coefficient functions belong to the 3FS. 
	
In order to write the high-scale limit at N$^3$LO we have to expand every term in \eqref{high:scale:mu} and then we have to isolate the $\ord{3}$ contribution. Observe that the massless coefficient functions and the matching conditions are usually given as an expansion in terms $\alpha_s^{[4]}$. Since we need the expansion of the high-scale limit in terms of $\alpha_s^{[3]}$, we have to re-expand $\alpha_s^{[4]}$ in terms of $\alpha_s^{[3]}$ using \eqref{alpha:matching}. In the following we will call $C^{(k)}$ the $k$-th order expansion of the coefficient functions in terms of $\alpha_s^{[3]}$ and $D^{(k)}$ the one in terms of $\alpha_s^{[4]}$. It means that
\begin{align}
C_i&=C_i^{(0)}+\alpha_s^{[3]}C_i^{(1)}+\bigl(\alpha_s^{[3]}\bigr)^2C_i^{(2)}+\bigl(\alpha_s^{[3]}\bigr)^3C_i^{(3)}+\mathcal{O}\bigl((\alpha_s^{[3]})^4\bigr), \label{expansion:C}\\
C_i&=D_i^{(0)}+\alpha_s^{[4]}D_i^{(1)}+\bigl(\alpha_s^{[4]}\bigr)^2 D_i^{(2)}+\bigl(\alpha_s^{[4]}\bigr)^3D_i^{(3)}+\mathcal{O}\bigl((\alpha_s^{[4]})^4\bigr). \label{expansion:D}
\end{align}
Obviously Eqs.~(\ref{expansion:C}) and (\ref{expansion:D}) are equal to all orders but differ at fixed order in perturbation theory. The relation between $C^{(k)}$ and $D^{(k)}$ is obtained using \eqref{alpha:matching} to re-expand $\alpha_s^{[3]}$ in \eqref{expansion:C} in terms of $\alpha_s^{[4]}$ and then comparing it with with \eqref{expansion:D} (or vice-versa).

The expansion at NLO in terms of $\alpha_s^{[3]}$ of \eqref{high:scale:mu} for the gluon coefficient function for $F_2$ gives
\begin{equation}\label{1ord:m0}
C_{2,g}^{[3,0](1)}\Bigl(\frac{m^2}{Q^2}\Bigr)=D_{2,g}^{[4](1)}\Bigl(\frac{Q^2}{\mu^2}\Bigr)+2D_{2,c}^{[4](0)}\Bigl(\frac{Q^2}{\mu^2}\Bigr) \otimes K^{(1)}_{cg}\Bigl(\frac{m^2}{\mu^2}\Bigr).
\end{equation}
At this order the exact massive coefficient function is independent from $\mu$, therefore in \eqref{1ord:m0} the $\mu$ dependence cancels, giving the result in \eqref{1ord:m0:explicit}. The factor of 2 in front of the second term of \eqref{1ord:m0} comes from summing both on the charm and on the anti-charm.
At NNLO we have that
\begin{align}
C_{2,g}^{[3,0](2)}\Bigl(\frac{m^2}{Q^2},\frac{m^2}{\mu^2}\Bigr)={}& D_{2,g}^{[4](2)}\Bigl(\frac{Q^2}{\mu^2}\Bigr)+D_{2,g}^{[4](1)}\Bigl(\frac{Q^2}{\mu^2}\Bigr)\frac{1}{6\pi}\log\Bigl(\frac{\mu^2}{m^2}\Bigr) \notag \\ 
&+D_{2,g}^{[4](1)}\Bigl(\frac{Q^2}{\mu^2}\Bigr) \otimes K^{(1)}_{gg}\Bigl(\frac{m^2}{\mu^2}\Bigr)+2D_{2,c}^{[4](0)}\Bigl(\frac{Q^2}{\mu^2}\Bigr) \otimes K^{(2)}_{cg}\Bigl(\frac{m^2}{\mu^2}\Bigr) \notag \\
&+2D_{2,c}^{[4](0)}\Bigl(\frac{Q^2}{\mu^2}\Bigr) \otimes K^{(1)}_{cg}\Bigl(\frac{m^2}{\mu^2}\Bigr)\frac{1}{6\pi}\log\Bigl(\frac{\mu^2}{m^2}\Bigr)\notag \\
&+2D_{2,c}^{[4](1)}\Bigl(\frac{Q^2}{\mu^2}\Bigr) \otimes K^{(1)}_{cg}\Bigl(\frac{m^2}{\mu^2}\Bigr), \label{2ord:m0}
\end{align}
where the two terms multiplying $\log(\mu^2/m^2)/6\pi$ come from the NLO expansion of $\alpha_s^{[4]}$ in terms of $\alpha_s^{[3]}$, i.e.\ the first two terms of \eqref{alpha:matching}. Observe that the second and the third terms of \eqref{2ord:m0} exactly cancel (however it is only a coincidence). \eqref{2ord:m0} can be written as
\begin{equation}
C_{2,g}^{[3,0](2)}\Bigl(\frac{m^2}{Q^2},\frac{m^2}{\mu^2}\Bigr)=D_{2,g}^{[3,0](2)}\Bigl(\frac{m^2}{Q^2},\frac{m^2}{\mu^2}\Bigr)+\frac{1}{6\pi}\log \Bigl(\frac{\mu^2}{m^2}\Bigr)D_{2,g}^{[3,0](1)}\Bigl(\frac{m^2}{Q^2}\Bigr),
\end{equation}
where
\begin{align}
D_{2,g}^{[3,0](1)}\Bigl(\frac{m^2}{Q^2}\Bigr)={}&D_{2,g}^{[4](1)}\Bigl(\frac{Q^2}{\mu^2}\Bigr)+2D_{2,c}^{[4](0)}\Bigl(\frac{Q^2}{\mu^2}\Bigr) \otimes K^{(1)}_{cg}\Bigl(\frac{m^2}{\mu^2}\Bigr), \label{Cm0:1ord}\\
D_{2,g}^{[3,0](2)}\Bigl(\frac{m^2}{Q^2},\frac{m^2}{\mu^2}\Bigr) ={}& D_{2,g}^{[4](2)}\Bigl(\frac{Q^2}{\mu^2}\Bigr)+ D_{2,g}^{[4](1)}\Bigl(\frac{Q^2}{\mu^2}\Bigr) \otimes K^{(1)}_{gg}\Bigl(\frac{m^2}{\mu^2}\Bigr)\notag \\ 
&+2D_{2,c}^{[4](0)}\Bigl(\frac{Q^2}{\mu^2}\Bigr) \otimes K^{(2)}_{cg}\Bigl(\frac{m^2}{\mu^2}\Bigr)+2D_{2,c}^{[4](1)}\Bigl(\frac{Q^2}{\mu^2}\Bigr) \otimes K^{(1)}_{cg}\Bigl(\frac{m^2}{\mu^2}\Bigr).
\end{align}
Observe that \eqref{Cm0:1ord} coincides with \eqref{1ord:m0} because at LO order we have that $\alpha_s^{[3]}=\alpha_s^{[4]}$.
Finally, the N$^3$LO expansion of the high-scale limit is
\begingroup
\allowdisplaybreaks
\begin{align}
C_{2,g}^{[3,0](3)}\Bigl(\frac{m^2}{Q^2},\frac{m^2}{\mu^2}\Bigr)={}&D_{2,g}^{[4](3)}\Bigl(\frac{Q^2}{\mu^2}\Bigr)+D_{2,g}^{[4](2)}\Bigl(\frac{Q^2}{\mu^2}\Bigr) \otimes K^{(1)}_{gg}\Bigl(\frac{m^2}{\mu^2}\Bigr) +D_{2,g}^{[4](1)}\Bigl(\frac{Q^2}{\mu^2}\Bigr) \otimes K^{(2)}_{gg}\Bigl(\frac{m^2}{\mu^2}\Bigr)\notag \\
&+2D_{2,c}^{[4](0)}\Bigl(\frac{Q^2}{\mu^2}\Bigr) \otimes K^{(3)}_{cg}\Bigl(\frac{m^2}{\mu^2}\Bigr)+2D_{2,c}^{[4](1)}\Bigl(\frac{Q^2}{\mu^2}\Bigr) \otimes K^{(2)}_{cg}\Bigl(\frac{m^2}{\mu^2}\Bigr) \notag \\
&+ 2D_{2,c}^{[4](2)}\Bigl(\frac{Q^2}{\mu^2}\Bigr) \otimes K^{(1)}_{cg}\Bigl(\frac{m^2}{\mu^2}\Bigr)+2\frac{1}{6\pi}\log\Bigl(\frac{\mu^2}{m^2}\Bigr)\biggl[D_{2,g}^{[4](2)}\Bigl(\frac{Q^2}{\mu^2}\Bigr) \notag\\
&+D_{2,g}^{[4](1)}\Bigl(\frac{Q^2}{\mu^2}\Bigr) \otimes K^{(1)}_{gg}\Bigl(\frac{m^2}{\mu^2}\Bigr)+2D_{2,c}^{[4](0)}\Bigl(\frac{Q^2}{\mu^2}\Bigr) \otimes K^{(2)}_{cg}\Bigl(\frac{m^2}{\mu^2}\Bigr)\notag \\
&+2D_{2,c}^{[4](1)}\Bigl(\frac{Q^2}{\mu^2}\Bigr) \otimes K^{(1)}_{cg}\Bigl(\frac{m^2}{\mu^2}\Bigr)\biggr]+\frac{1}{6\pi^2}\biggl[\frac{1}{6}\log^2\Bigl(\frac{\mu^2}{m^2}\Bigr) +\frac{19}{4}\log\Bigl(\frac{\mu^2}{m^2}\Bigr) \notag \\
&+\frac{7}{4}\biggr]\biggl[D_{2,g}^{[4](1)}\Bigl(\frac{Q^2}{\mu^2}\Bigr)+2D_{2,c}^{[4](0)}\Bigl(\frac{Q^2}{\mu^2}\Bigr) \otimes K^{(1)}_{cg}\Bigl(\frac{m^2}{\mu^2}\Bigr)\biggr],\label{3ord:m0}
\end{align}
\endgroup
that can be rewritten as
\begin{align}
C_{2,g}^{[3,0](3)}\Bigl(\frac{m^2}{Q^2},\frac{m^2}{\mu^2}\Bigr)={}&D_{2,g}^{[3,0](3)}\Bigl(\frac{m^2}{Q^2},\frac{m^2}{\mu^2}\Bigr)+\frac{1}{3\pi}\log\Bigl(\frac{\mu^2}{m^2}\Bigr)D_{2,g}^{[3,0](2)}\Bigl(\frac{m^2}{Q^2},\frac{m^2}{\mu^2}\Bigr) \notag \\
&+\frac{1}{6\pi^2}\biggl[\frac{1}{6}\log^2\Bigl(\frac{\mu^2}{m^2}\Bigr) +\frac{19}{4}\log\Bigl(\frac{\mu^2}{m^2}\Bigr) +\frac{7}{4}\biggr]D_{2,g}^{[3,0](1)}\Bigl(\frac{m^2}{Q^2}\Bigr),
\end{align}
where
\begin{align}
D_{2,g}^{[3,0](3)}\Bigl(\frac{m^2}{Q^2},\frac{m^2}{\mu^2}\Bigr)={}&D_{2,g}^{[4](3)}\Bigl(\frac{Q^2}{\mu^2}\Bigr)+D_{2,g}^{[4](2)}\Bigl(\frac{Q^2}{\mu^2}\Bigr) \otimes K^{(1)}_{gg}\Bigl(\frac{m^2}{\mu^2}\Bigr)  \notag \\
&+D_{2,g}^{[4](1)}\Bigl(\frac{Q^2}{\mu^2}\Bigr) \otimes K^{(2)}_{gg}\Bigl(\frac{m^2}{\mu^2}\Bigr)+2D_{2,c}^{[4](0)}\Bigl(\frac{Q^2}{\mu^2}\Bigr) \otimes K^{(3)}_{cg}\Bigl(\frac{m^2}{\mu^2}\Bigr)\notag \\
&+2D_{2,c}^{[4](1)}\Bigl(\frac{Q^2}{\mu^2}\Bigr) \otimes K^{(2)}_{cg}\Bigl(\frac{m^2}{\mu^2}\Bigr)+ 2D_{2,c}^{[4](2)}\Bigl(\frac{Q^2}{\mu^2}\Bigr) \otimes K^{(1)}_{cg}\Bigl(\frac{m^2}{\mu^2}\Bigr).
\end{align}
It is clear from \eqref{3ord:m0} that, in order to construct the high-scale limit at $\ord{3}$, we need $K_{gg}$ at $\ord{2}$ and $K_{cg}$ at $\ord{3}$. The matching conditions are exactly known at $\ord{2}$ from Ref.~\cite{Buza_1998}. This means that the high-scale limit is fully known at $\ord{2}$. Its expression is given for example in Ref.~\cite{Buza_1996}. Instead, the matching conditions $K_{cg}$ are not fully known at $\ord{3}$. In fact, although the $\mu$-dependent contribution of such functions is exactly known, the $\mu$-independent one is not. If we call $k^{(3)}_{cg}$ the $\mu$-independent term of $K^{(3)}_{cg}$, then it can be decomposed as
\begin{equation}
k^{(3)}_{cg}(z)=k^{(3)0}_{cg}(z)+n_fk^{(3)1}_{cg}(z).
\end{equation}
The term proportional to $n_f$ is known exactly from Ref.~\cite{Ablinger_2011}. The $n_f$-independent term, on the other hand, is not fully known yet. In fact, only some integer Mellin moments are currently known \cite{Bierenbaum_2009}.
Hence, it is the only missing term for the construction of $C_{2,g}^{[3,0](3)}$ through \eqref{3ord:m0}.
In Ref.~\cite{Kawamura_2012} it is presented the expression of the high-scale limit of the exact massive coefficient function at $\ord{3}$,
with $k_{cg}^{(3)0}(z)$ left unspecified.
In that expression the N$^3$LO massless coefficient function of the gluon computed for $\mu=Q$ appears.
Although such function is exactly known, being its complete expression unmanageable due to its length, we used the approximation presented in Ref.~\cite{Vermaseren_2005} that deviates from the exact result by less than one part in a thousand.
	
A comment on the massless coefficient functions given in Ref.~\cite{Vermaseren_2005} is now required. They are the completely inclusive coefficient functions, which means that they take in account every possible final state and not only heavy quark pair production. For this reason, starting from the N$^3$LO, the massless coefficient functions have new quark flavour topologies, denoted with $fl_{11}$ for the quark coefficient function and with $fl_{11}^g$ for the gluon. These coefficient functions are not identical to the the heavy quark production coefficient functions that we need, since they require a heavy quark in the final state. However, the information that is currently available is not sufficient to disentangle the heavy quark production term from the totally inclusive one. 
For this reason, the contribution proportional to $fl_{11}^g$, that is numerically small, in the gluon massless coefficient function that enters in the expression of the N$^3$LO high-scale limit has been omitted.

Finally, in order to get numerical results, it remains to find approximate expressions for the only missing term, i.e.\ $k_{cg}^{(3)0}(z)$ and to contruct an uncertainty band. Some possible approximations are given in Ref.~\cite{Kawamura_2012}, with minor corrections given in Ref.~\cite{Alekhin_2017}, that are based on the known Mellin moments, on the small-$z$ leading logarithm expansion and on the large-$z$ behavior. Their expressions are 
\begin{align}
k_{cg,A}^{(3)0}(z)={}&354.1002\log^3(1-z) + 479.3838\log^2(1-z) - 7856.784(2-z) \notag \\
&- 6233.530\log^2z + \frac{9416.621}{z} + \frac{1548.891}{z}\log z, \label{kcgA}\\
k_{cg,B}^{(3)0}(z)={}&226.3840\log^3(1-z) - 652.2045\log^2(1-z) - 2686.387\log(1-z) \notag \\
&- 7714.786(2-z) - 2841.851\log^2z + \frac{7721.120}{z} + \frac{1548.891}{z}\log z. \label{kcgB}
\end{align}	
Eqs.~(\ref{kcgA}) and (\ref{kcgB}) can be interpreted as the extremes of an error band, while the average of the two can be used as the central value of our approximate function.
	
\begin{figure}[!ht]
	\centering
	\includegraphics[width=\textwidth]{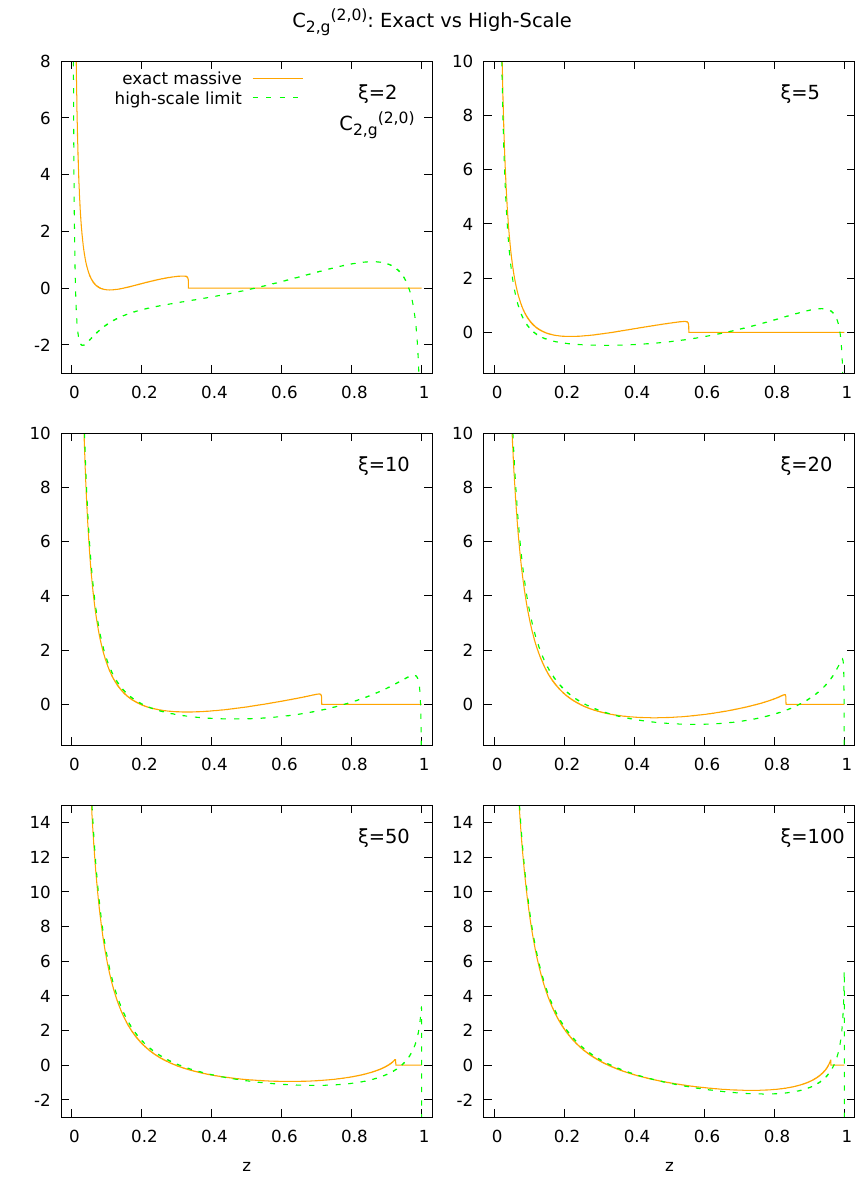}
	\caption{Comparison between the $\mu$-independent part of the NNLO exact massive coefficient function of the gluon for $F_2$ (solid orange) computed from the parametrization in Ref.~\cite{Riemersma_1995} and its high-scale limit (dashed green), i.e.\ \eqref{3ord:m0}. Six relevant values of $\xi=Q^2/m^2$ are shown. 
	%The $\mu$-independent parts are extracted computing the exact coefficient function and the high-scale limit for $\mu=m$.
	}
	\label{exact:highscale:x}
\end{figure}

\begin{figure}[!ht]
	\centering
	\includegraphics[width=\textwidth]{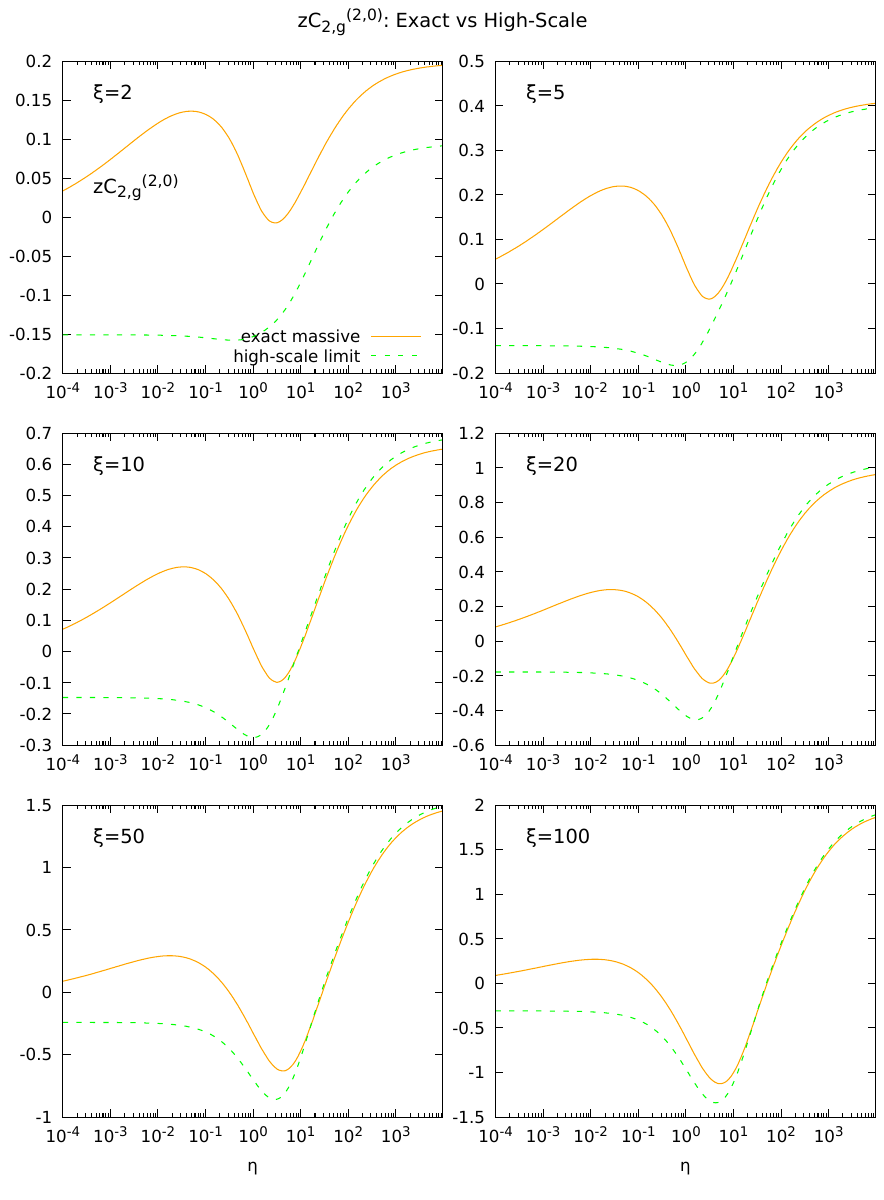}
	\caption{As \figref{exact:highscale:x} but the coefficient functions are multiplied by $z$ and the curves are plotted as a function of $\eta$, \eqref{eta}, instead of $z$, in order to stretch the threshold region.
	From this figure is even more clear that the high-scale limit does not approach the exact curve for $z\rightarrow 0$ and that even if $Q^2 \gg m^2$ it is not a good approximation for the threshold region.}
	\label{exact:highscale:eta}
\end{figure}

Now we have all the ingredients, at least in an approximate form, for the construction of the N$^3$LO high-scale limit of the exact massive gluon coefficient function for $F_2$ in heavy quark pair production.
As we said before, it is a good approximation of the exact curve in the limit $Q^2 \gg m^2$ but not for smaller values of $Q^2$. Moreover, comparing the $\ord{2}$ exact coefficient function with its high-scale limit, it can be easily verified that for values of $Q^2$ that are comparable with $m^2$, the approximation does not approaches the exact result for $z \rightarrow 0$. Figs.~\ref{exact:highscale:x} and \ref{exact:highscale:eta} show the comparison between the $\mu$-independent part of the NNLO exact massive coefficient function of the gluon for $F_2$ with its high-scale limit. 
In \figref{exact:highscale:eta} the curves are plotted as a function of the variable $\eta$, defined as
\begin{equation}\label{eta}
\eta=\frac{s}{4m^2}-1=\frac{Q^2}{4m^2}\Bigl(\frac{1-z}{z}\Bigr)-1,
\end{equation}
so that $\eta \rightarrow \infty$ corresponds to the $z\rightarrow 0$ limit, while $\eta \rightarrow 0$ corresponds to $z \rightarrow z_{\rm max}$ (or equivalently $s\rightarrow 4m^2$). 
For $\eta<0$ the heavy quark pair production is kinematically forbidden.
With such definition, plotting the coefficient functions as a function of $\eta$ we are stretching the threshold zone $z \rightarrow z_{\rm max}$, with respect to a plot as function of $z$.
\begin{figure}[!ht]
	\centering
	\includegraphics[width=\textwidth]{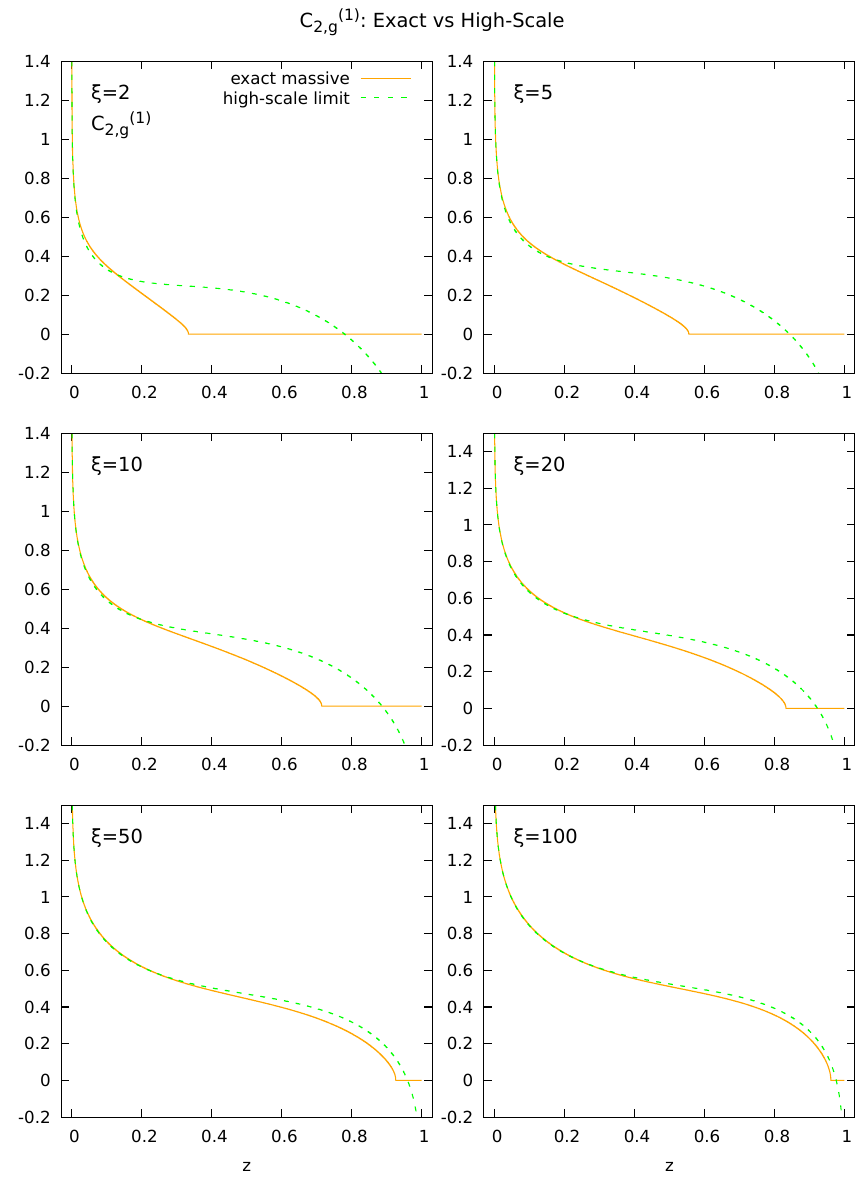}
	\caption{Comparison between the NLO exact massive coefficient function of the gluon for $F_2$ in \eqref{C2g1} (solid orange) and its high-scale limit in \eqref{1ord:m0:explicit} (dashed green). Six relevant values of $\xi=Q^2/m^2$ are shown.}
	\label{NLOexact:highscale:x}
\end{figure}
\begin{figure}[!ht]
	\centering
	\includegraphics[width=\textwidth]{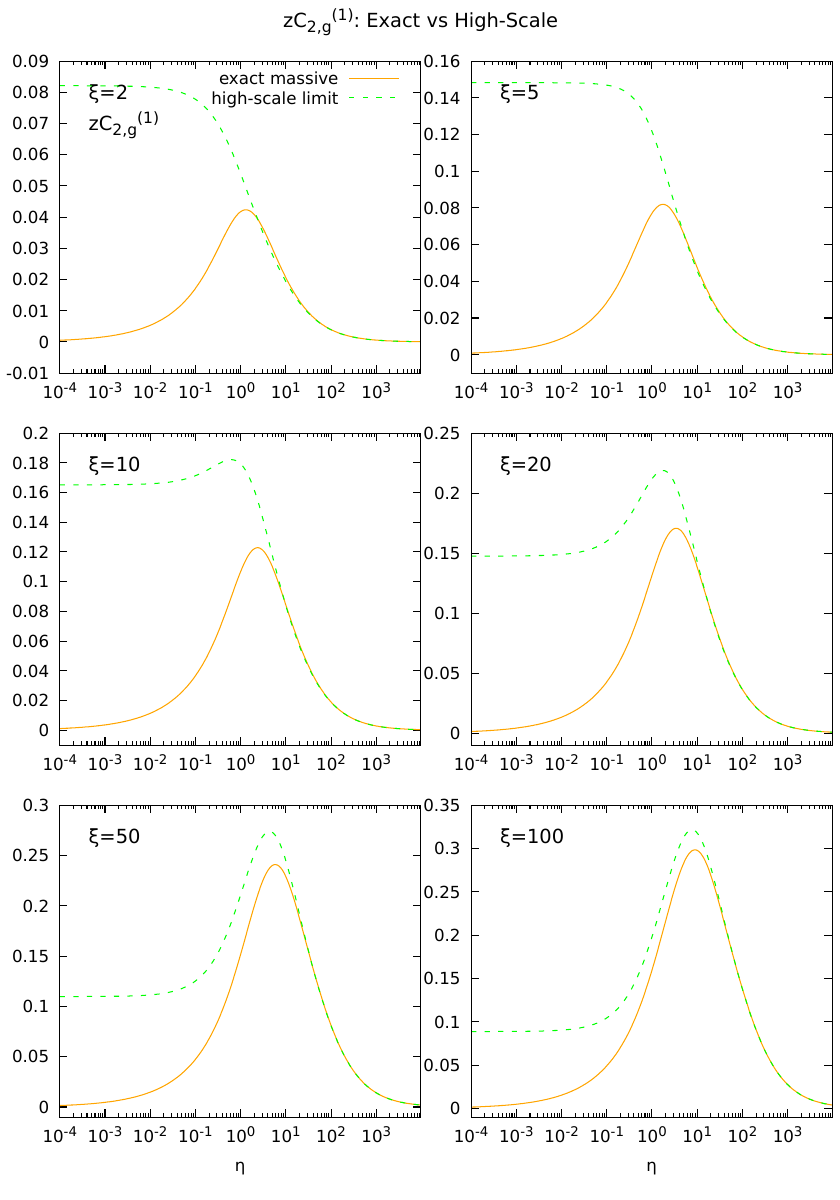}
	\caption{As \figref{NLOexact:highscale:x} but the coefficient functions are multiplied by $z$ and the curves are plotted as a function of $\eta$, \eqref{eta}, instead of $z$, in order to stretch the threshold region.}
	\label{NLOexact:highscale:eta}
\end{figure} We can observe that for small values of $\xi=Q^2/m^2$ the high-scale limit does not approximate the exact curve at all. Moreover, for $\xi \lesssim 100$, it does not have the correct behavior for $z\rightarrow 0$ (or equivalently $\eta\rightarrow \infty$) since it does not approaches the exact coefficient function. Starting from $\xi \sim 100$ the high-scale limit gives a good approximation of the exact curve for every value of $z$, except for those too close to the threshold point $z=z_{\rm max}$ (or $\eta=0$). 
For the NLO gluon coefficient function instead, we are in a lucky case since all the power terms that are neglected in the high-scale limit go to zero for $z \rightarrow 0$, as can be verified from \eqref{C2g1}. This means that the NLO high-scale limit in \eqref{1ord:m0:explicit} can be used as asymptotic limit since it approaches the exact massive coefficient function in the small-$z$ limit, for any value of $Q^2$.
This can be observed in Figs.~\ref{NLOexact:highscale:x} and \ref{NLOexact:highscale:eta}.

\subsection{Power terms}

So far in this section we have introduced the high-scale limit of the exact massive coefficient function and we have said that it is a good approximation of this one in the limit $Q^2 \gg m^2$ but not for $Q^2\sim m^2$. In fact in the second case the high-scale limit does not approaches the exact function for $z\rightarrow 0$ due to the absence of the power terms that for such values of $Q^2$ are not negligible.
Thus, in order to construct an asymptotic limit that approaches the exact massive coefficient function in the limit $z\rightarrow 0$, we need to reinsert such power terms, at least in the small-$z$ limit (if we could reinsert the power terms for every value of $z$ we would have the exact coefficient function and we would not need the approximate one). In order to do it we will use the small-$z$ expansion (with $Q$ fixed) of the exact massive coefficient function. 
Being $s=Q^2(1/z-1)$, where $\sqrt{s}$ is the partonic center-of-mass energy, the small-$z$ limit with $Q$ fixed corresponds to $s \rightarrow \infty$.
For this reason such limit is called high-energy limit.  
	
In Ref.~\cite{Kawamura_2012} the leading logarithm (LL) contributions of the high-energy limit of the gluon coefficient function are given at $\ord{2}$ and $\ord{3}$. They are computed performing small-$x$ resummation from the results of Ref.~\cite{CATANI1991135}. Their expressions are the following	
\begin{align}
C_{2,g}^{[3](2) \rm LL}={}&\frac{\xi}{z} \frac{C_A}{12\pi^2} \Biggl[ \frac{10}{3\xi} + \biggl(1-\frac{1}{\xi}\biggr)I(\xi) + \biggl(\frac{13}{6}-\frac{5}{3\xi}\biggr)J(\xi) +L_\mu\biggl( \frac{2}{\xi} + \Bigl(1-\frac{1}{\xi}\Bigr)J(\xi)\biggr) \Biggr], \label{CLL2}\\
C_{2,g}^{[3](3)\rm LL}={}&\xi \frac{\log z}{z}\frac{C_A^2}{8\pi^3}\Biggl[-\frac{184}{27\xi} - \frac{1}{3}\Bigl(1-\frac{1}{\xi}\Bigr)I(\xi) \log \Bigl( 1 + \frac{\xi}{4}\Bigr) - \frac{1}{9}\Bigl(13-\frac{10}{\xi}\Bigr)I(\xi) \notag \\
&- \frac{1}{27}\Bigl(71-\frac{92}{\xi}\Bigr)J(\xi) + \frac{1}{3}\Bigr(1-\frac{1}{\xi}\Bigl)K(\xi) +L_\mu\biggl(-\frac{20}{9\xi} - \frac{2}{3}\Bigl(1-\frac{1}{\xi}\Bigr)I(\xi)  \notag \\
&-\frac{1}{9}\Bigr(13-\frac{10}{\xi}\Bigr)J(\xi) \biggr)+L_\mu^2\biggl( -\frac{2}{3\xi} -\frac{1}{3}\Bigl(1-\frac{1}{\xi}\Bigr)J(\xi)\biggr)\Biggr],\label{CLL3}
\end{align}
where $\xi=Q^2/m^2$, $L_\mu=\log \bigl(m^2/\mu^2\bigr)$ and 
\begingroup
\allowdisplaybreaks
\begin{align}
I(\xi)&=\frac{4}{\xi}\sqrt{\frac{\xi}{\xi+4}}H\Bigl(+,-,\sqrt{\xi/(\xi+4)}\Bigr), \label{I}\\
J(\xi)&=\frac{4}{\xi}\sqrt{\frac{\xi}{\xi+4}}L\Bigl(\sqrt{\xi/(\xi+4)}\Bigr), \label{J}\\
K(\xi)&=\frac{4}{\xi}\sqrt{\frac{\xi}{\xi+4}}H\Bigl(-,+,-,\sqrt{\xi/(\xi+4)}\Bigr),\label{K}
\end{align}
\endgroup
with
\begingroup
\allowdisplaybreaks
\begin{align}
L(z)={}&\log\biggl(\frac{1+z}{1-z}\biggr), \label{Lz}\\
H\Bigl(+,-,z\Bigr)={}&H_{1,1}(z)+H_{1,-1}(z)-H_{-1,1}(z)-H_{-1,-1}(z) \label{Hpmz},\\
H\Bigl(-,+,-,z\Bigr)={}&H_{1,1,1}(z)-H_{1,1,-1}(z)+H_{1,-1,1}(z) - H_{1,-1,-1}(z) \notag \\
&- H_{-1,1,1}(z) + H_{-1,1,-1}(z) - H_{-1,-1,1}(z) + H_{-1,-1,-1}(z). \label{Hmpmz}
\end{align}
\endgroup
The functions $H_{i_1,i_2,\dots,i_n}(z)$ are the harmonic polylogarithms, and are defined in Ref.~\cite{REMIDDI_2000}. 
At $\ord{2}$, in the high-energy expansion, only the leading logarithm contribution is present.
However, at $\ord{3}$ also the next-to-leading logarithm (NLL) appears, and it is a factor proportional to $1/z$. 
Without this term the high-energy expansion would not give the correct small-$z$ limit of the N$^3$LO exact coefficient function since the exact result would not approach it for $z \rightarrow 0$.
Unfortunately, this contribution is still unknown.
In Appendix~\ref{high:energy:NLL} we will show how to compute an approximation of the next-to-leading logarithm small-$z$ expansion of the N$^3$LO coefficient function and how to estimate an uncertainty band associated to it. Therefore, the high-energy limits of the NNLO and N$^3$LO coefficient functions are given by
\begin{align}
    C_{2,g}^{[3](2)\rm h.e.}&=C_{2,g}^{[3](2)\rm LL}, \label{C:he2}\\
    C_{2,g}^{[3](3)\rm h.e.}&=C_{2,g}^{[3](3)\rm LL}+C_{2,g}^{[3](3)\rm NLL},\label{C:he3}
\end{align}
where $C_{2,g}^{[3](3) \rm NLL}$ is given in \eqref{CNLL}.

Now we can consider the $Q^2 \gg m^2$ limit of Eqs.~(\ref{C:he2}) and (\ref{C:he3}), that are
\begin{align}
    C_{2,g}^{[3,0](2)\rm h.e.}&=C_{2,g}^{[3,0](2)\rm LL}, \label{C:he:hs2}\\
    C_{2,g}^{[3,0](3)\rm h.e.}&=C_{2,g}^{[3,0](3)\rm LL}+C_{2,g}^{[3,0](3) \rm NLL},\label{C:he:hs3}
\end{align}
with 
\begin{align}
C_{2,g}^{[3,0](2)\rm LL}={}&\frac{1}{z}\frac{C_A}{16\pi^2} \biggl[\frac{8}{3}L_Q^2 - \frac{104}{9} L_Q + \frac{40}{9} - \frac{16}{3} \zeta_2 + \Bigl(-\frac{16}{3} L_Q + \frac{8}{3}\Bigr) L_\mu \biggr], \label{CLL2:highscale}\\
C_{2,g}^{[3,0](3) \rm LL}={}&-\frac{\log z}{z} \frac{C_A^2}{64\pi^3}\biggl[-\frac{32}{9}L_Q^3 + \frac{208}{9}L_Q^2 - \Bigl(\frac{2272}{27} - \frac{64}{3}\zeta_2\Bigr)L_Q + \frac{1472}{27}  \notag \\
&- \frac{416}{9}\zeta_2 +\frac{128}{3}\zeta_3 + \Bigl(\frac{32}{3}L_Q^2 - \frac{416}{9}L_Q + \frac{160}{9} - \frac{64}{3}\zeta_2\Bigr)L_\mu +  \notag \\
&+\Bigl(-\frac{32}{3}L_Q + \frac{16}{3}\Bigr)L_\mu^2\biggr], \label{CLL3:highscale}
\end{align}
where $L_Q=\log\bigl(m^2/Q^2\bigr)$ and $\zeta_i$ are the zeta functions, while $C_{2,g}^{[3,0](3)\rm NLL}$ is given in \eqref{CNLL:hs}.
Eqs.~(\ref{C:he:hs2}) and (\ref{C:he:hs3}) are the high-scale limits of the high-energy ones. If we assume the exact coefficient function to be smooth enough such that the two limits can be interchanged, we can state that Eqs.~(\ref{C:he:hs2}) and (\ref{C:he:hs3}) are the high-energy limits of the high-scale ones.
Therefore, while the exact coefficient function will tend to the high-energy limit for $z\rightarrow 0$, the high-scale limit will tend to the high-scale limit of the high-energy.
It means that we can estimate the power terms in the small-$z$ limit simply subtracting the high-energy limit with its high-scale limit. 
In fact, the high-energy limit of the exact coefficient function contains the power terms, while in its high-scale limit we have thrown away such power terms. Thus, the difference between the two gives exactly the power terms that we need.
In practice we have to subtract \eqref{C:he2} and \eqref{C:he:hs2} at $\ord{2}$, and \eqref{C:he3} and \eqref{C:he:hs3} at $\ord{3}$.
We conclude that 
\begin{align}
C_{2,g}^{[3](2)\rm p.t.\,h.e.}\Bigl(z,\frac{m^2}{Q^2}, \frac{m^2}{\mu^2}\Bigr)&=C_{2,g}^{[3](2)\rm h.e.}\Bigl(z,\frac{m^2}{Q^2}, \frac{m^2}{\mu^2}\Bigr)-C_{2,g}^{[3,0](2)\rm h.e.}\Bigl(z,\frac{m^2}{Q^2}, \frac{m^2}{\mu^2}\Bigr), \label{powerterms2}\\
C_{2,g}^{[3](3)\rm p.t.\,h.e.}\Bigl(z,\frac{m^2}{Q^2}, \frac{m^2}{\mu^2}\Bigr)&=C_{2,g}^{[3](3)\rm h.e.}\Bigl(z,\frac{m^2}{Q^2}, \frac{m^2}{\mu^2}\Bigr)-C_{2,g}^{[3,0](3)\rm h.e.}\Bigl(z,\frac{m^2}{Q^2}, \frac{m^2}{\mu^2}\Bigr),\label{powerterms3}
\end{align}
\begin{figure}[!t]
\centering
\includegraphics[width=\textwidth]{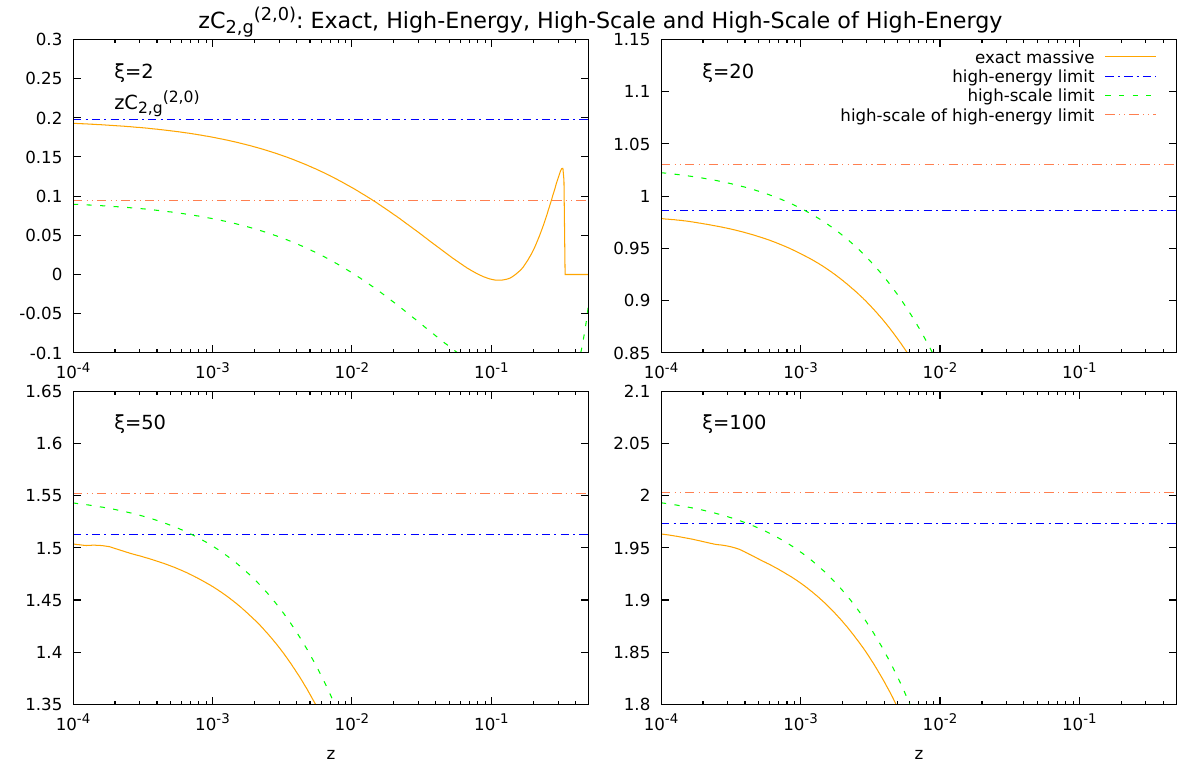}
\caption{Comparison between the $\mu$-independent part of the $\ord{2}$ exact massive (from Ref.~\cite{Riemersma_1995}), high-energy (\eqref{C:he2}), high-scale (\eqref{2ord:m0}) and high-scale of high-energy (\eqref{C:he:hs2}) gluon coefficient functions. Four different values of $\xi=Q^2/m^2$ are shown.}
\label{power:terms:2ord}
\hspace{1em}
\includegraphics[width=\textwidth]{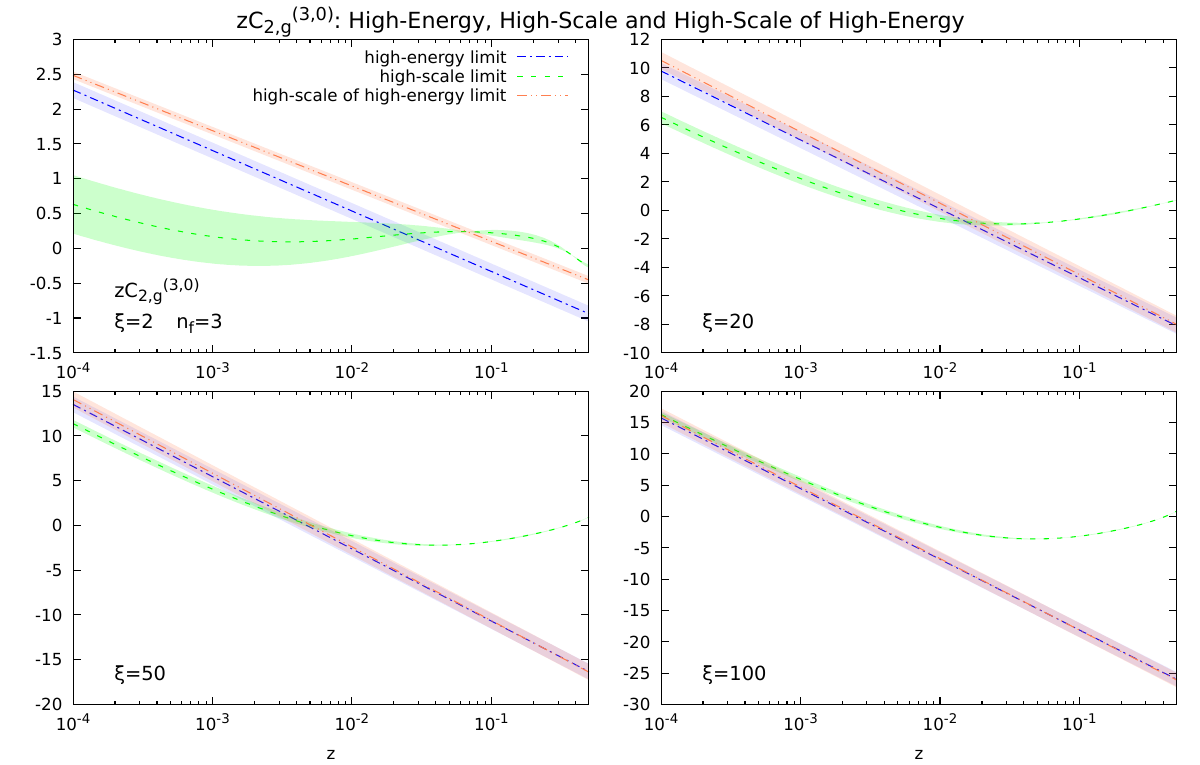}
\caption{Comparison between the $\mu$-independent part of the $\ord{3}$ high-energy (\eqref{C:he3}), high-scale (\eqref{3ord:m0}) and high-scale of high-energy (\eqref{C:he:hs3}) gluon coefficient functions for $n_f=3$. Four different values of $\xi=Q^2/m^2$ are shown.}
\label{power:terms:3ord}
\end{figure}\figref{power:terms:2ord} shows the comparison between the $\mu$-independent part of the $\ord{2}$ exact massive, high-energy, high-scale and high-scale of high-energy gluon coefficient functions. It is evident that the exact massive tends to the high-energy limit for $z\rightarrow 0$, while the high-scale tends to the $Q^2 \gg m^2$ limit of the high-energy.
Notice that the power terms correctly go to zero as $\xi$ increases.
\figref{power:terms:3ord} shows the comparison between the $\mu$-independent part of the $\ord{3}$ high-energy, high-scale and high-scale of high-energy gluon coefficient functions for $n_f=3$. Observe that the agreement for $z\rightarrow 0$ between high-scale limit and high-scale of high-energy limit is not perfect for small $\xi$. This may be due to the fact that the next-to-leading logarithm, that when multiplied for $z$ gives a constant factor, is an approximate expression, and therefore the final curve may be shifted by a constant. Instead, at high $\xi$ such difference goes to zero and the high-scale limit approaches the high-scale limit of the high-energy coefficient function. Moreover, for high values of $\xi$ the high-energy correctly tends to its high-scale limit, since the power terms go to zero. 
	
\begin{figure}[!ht]
	\centering
	\includegraphics[width=\textwidth]{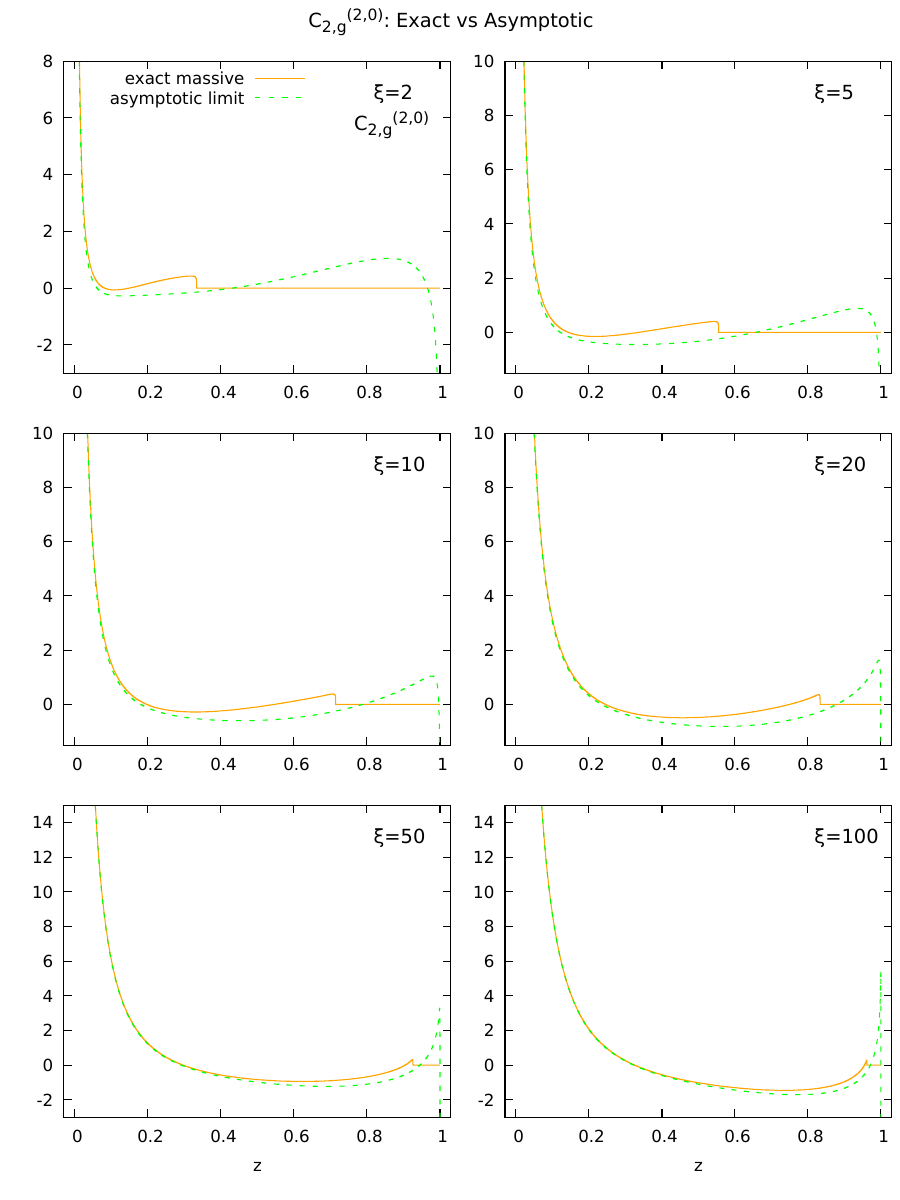}
	\caption{Comparison between the $\mu$-independent part of the NNLO exact massive coefficient function of the gluon for $F_2$ (solid orange) computed from the parametrization in Ref.~\cite{Riemersma_1995} and its asymptotic limit (dashed green), i.e.\ \eqref{asy:2ord}. Six relevant values of $\xi=Q^2/m^2$ are shown. 
	The $\mu$-independent parts are extracted computing the exact coefficient function and the high-scale limit for $\mu=m$.
	}
	\label{exact:asym:x}
\end{figure}

\begin{figure}[!ht]
	\centering
	\includegraphics[width=\textwidth]{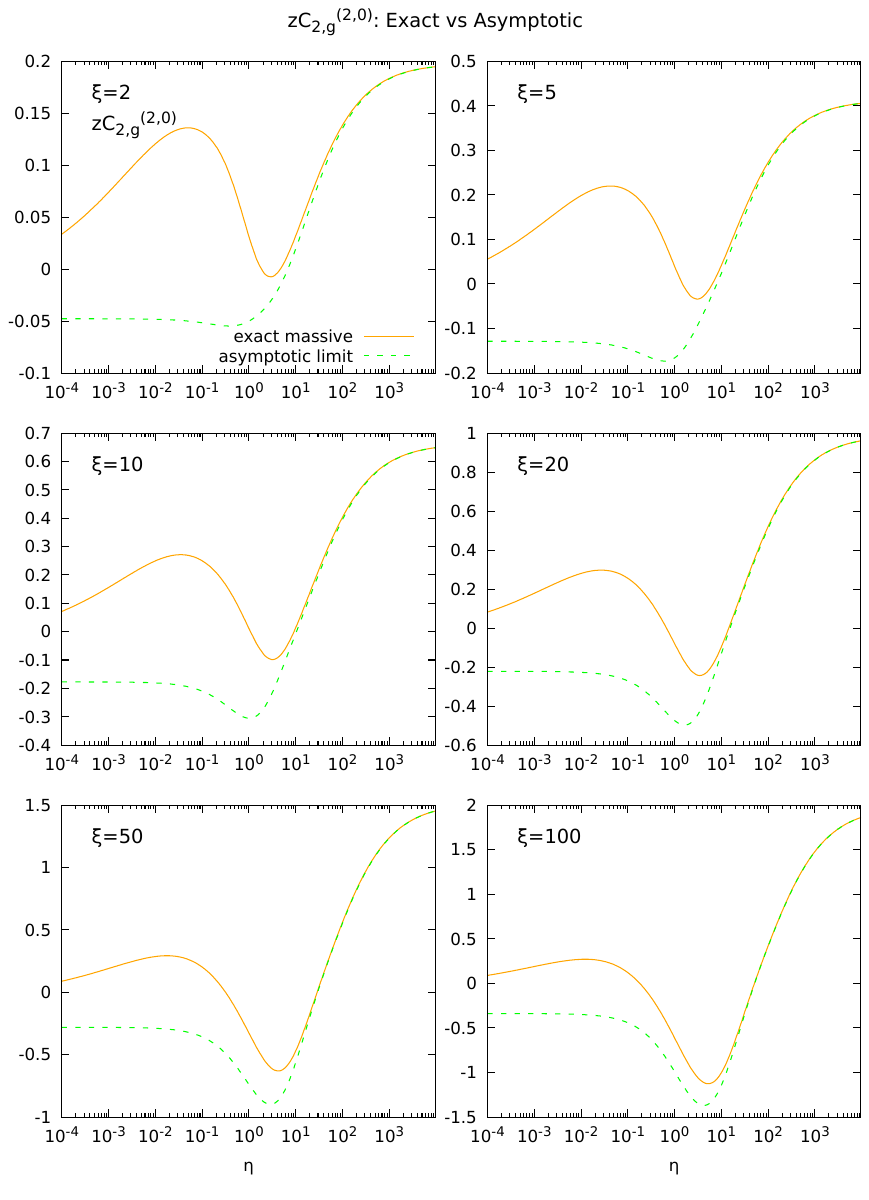}
	\caption{As \figref{exact:asym:x} but the coefficient functions are multiplied by $z$ and the curves are plotted as a function of $\eta$, \eqref{eta}, instead of $z$, in order to stretch the threshold region. From this figure is even more clear that the asymptotic limit approaches the exact curve for $z \rightarrow 0$, but even if $Q^ 2 \gg m^ 2$ it is not a good approximation for the threshold region.}
	\label{exact:asym:eta}
\end{figure}

Now that we have computed the power terms in the small-$z$ region we can add them to the high-scale to find our asymptotic limit. Therefore, what we get is
\begin{align}
C_{2,g}^{[3](2)\rm asy}\Bigl(z,\frac{m^2}{Q^2}, \frac{m^2}{\mu^2}\Bigr)&=C_{2,g}^{[3,0](2)}\Bigl(z,\frac{m^2}{Q^2}, \frac{m^2}{\mu^2}\Bigr)+C_{2,g}^{[3](2)\rm p.t.\,h.e.}\Bigl(z,\frac{m^2}{Q^2}, \frac{m^2}{\mu^2}\Bigr),  \label{asy:2ord} \\
%&=D_{2,g}^{[3,0](2)}\Bigl(z,\frac{m^2}{Q^2}, \frac{m^2}{\mu^2}\Bigr) + D_{2,g}^{(2)LL}\Bigl(z,\frac{m^2}{Q^2}, \frac{m^2}{\mu^2}\Bigr)-D_{2,g}^{[3,0](2)LL}\Bigl(z,\frac{m^2}{Q^2}, \frac{m^2}{\mu^2}\Bigr) ,\\
C_{2,g}^{[3](3)\rm asy}\Bigl(z,\frac{m^2}{Q^2}, \frac{m^2}{\mu^2}\Bigr)&=C_{2,g}^{[3,0](3)}\Bigl(z,\frac{m^2}{Q^2}, \frac{m^2}{\mu^2}\Bigr)+C_{2,g}^{[3](3)\rm p.t.\,h.e.}\Bigl(z,\frac{m^2}{Q^2}, \frac{m^2}{\mu^2}\Bigr).  \label{asy:3ord} 
%&=D_{2,g}^{[3,0](3)}\Bigl(z,\frac{m^2}{Q^2}, \frac{m^2}{\mu^2}\Bigr) + D_{2,g}^{(3)LL}\Bigl(z,\frac{m^2}{Q^2}, \frac{m^2}{\mu^2}\Bigr)-D_{2,g}^{[3,0](3)LL}\Bigl(z,\frac{m^2}{Q^2}, \frac{m^2}{\mu^2}\Bigr).
\end{align}
Eqs.~(\ref{asy:2ord}) and (\ref{asy:3ord}) are our final expressions of the asymptotic limit of the DIS exact massive gluon coefficient functions for $F_2$ in heavy quark production at $\ord{2}$ and $\ord{3}$.
As we have said previously, we want to find an approximation for the $\mu$-independent term of the N$^3$LO gluon coefficient function, i.e.\ $C_{2,g}^{[3](3,0)}$. Therefore, we have to extract the $\mu$-independent contribution of \eqref{asy:3ord}. This can be easily done computing it for $\mu=m$, so that all the logarithmic terms of the form $\log\bigl(\mu^2/m^2\bigr)$, see \eqref{mu:dependence}, go to zero and we are left only with the $\mu$-independent part.
In conclusion, the final approximation we are interested in is
\begin{align}
C_{2,g}^{[3](3,0)\rm asy}\Bigl(z,\frac{m^2}{Q^2}\Bigr)&=C_{2,g}^{[3,0](3,0)}\Bigl(z,\frac{m^2}{Q^2}\Bigr)+C_{2,g}^{[3](3,0)\rm p.t.\,h.e.}\Bigl(z,\frac{m^2}{Q^2}\Bigr)  \notag \\
&=C_{2,g}^{[3,0](3,0)}\Bigl(z,\frac{m^2}{Q^2}\Bigr) + C_{2,g}^{[3](3,0)\rm h.e.}\Bigl(z,\frac{m^2}{Q^2}\Bigr) -C_{2,g}^{[3,0](3,0)\rm h.e.}\Bigl(z,\frac{m^2}{Q^2}\Bigr).\label{asy:30ord}
\end{align}
Figs.~\ref{exact:asym:x} and \ref{exact:asym:eta} show the comparison of the $\mu$-independent part of the exact NNLO coefficient function with its asymptotic limit. 
We can observe that our asymptotic limit approaches the exact curve for $z\rightarrow 0$ for every value of $Q$. Moreover for $Q^2 \gg m^2$ it tends to the exact coefficient function for all the values of $z$ since in this case the high-scale approximates the exact curve while the power terms go to zero. As we have mentioned previously, the asymptotic limit does not describe correctly the threshold region even if $Q^2 \gg m^2$, as it can be clearly observed from \figref{exact:asym:eta}.

%%%%%%%%%%%%%%%%%%%%%%%%%%%%%%%%%%%%%%%%%%%%%%%%%%%%%

\section{Threshold limit}\label{thresh}

In the past section we have constructed the asymptotic limit of the exact massive gluon coefficient function at $\ord{2}$ and $\ord{3}$. Such limit approaches the exact curve for $z \rightarrow 0$. Moreover for $Q^2 \gg m^2$ it is a good approximation of the curve for every $z$, with the exception of those values that are too close to the threshold point $z=z_{\rm max}$. In fact, as can be observed from Figs.~\ref{exact:asym:x} and \ref{exact:asym:eta}, the asymptotic limit fails to describe the exact function in the threshold region even for $Q^2 \gg m^2$. 

In this section we will present the threshold limit of the exact massive coefficient function of the gluon. Such limit approaches the exact coefficient function for $z\rightarrow z_{\rm max}$ or equivalently $\eta \rightarrow 0$. 
This is the limit in which $\beta \rightarrow 0$, where $\beta$ is the center-of-mass velocity of the final heavy quarks and is given by
\begin{equation}\label{beta}
\beta = \sqrt{1- \frac{4m^2}{s}}=\sqrt{1-\frac{4m^2}{Q^2}\frac{z}{1-z}}.
\end{equation}
In this limit, corrections of the form $\alpha_s^n \log^m\beta$ with $m \leq 2n$ appear in the perturbative expansion. Such terms must be resummed performing the so-called threshold resummation \cite{STERMAN1987310, CATANI1989323, Contopanagos_1997, Kidonakis_1997,BONCIANI1998424}. Moreover, in heavy quark pair production, there are corrections of the form $\alpha_s^n \beta^{-m} \log^l \beta$ with $m \leq n$. They appear from diagrams in which the heavy quarks interact between each other through gluon exchanges. Also these contribution can be resummed \cite{Hoang:2000yr}.
Ref.~\cite{Kawamura_2012} provides the expressions of the threshold limit of the gluon coefficient function for $F_2$ up to $\ord{3}$.
At NLO the threshold limit is
\begin{equation}\label{thresh:ord1}
C_{2,g}^{[3](1)\rm thresh}\Bigl(z,\frac{m^2}{Q^2}\Bigr)=\frac{T_F}{4\pi}\frac{\xi}{z}\frac{\beta}{1+\frac{\xi}{4}} + \mathcal{O}(\beta^3).
\end{equation}
Beyond NLO, in the resummed contribution always factorizes the term $C_{2,g}^{[3](1)}$. So it is convenient not to expand $C_{2,g}^{[3](1)}$ for $\beta\rightarrow 0$, but instead to use the exact result. This will not change our results in the region in which the threshold limit is accurate, but only in the region where it is no more reliable.
Therefore, at NNLO we have that
\begin{align}
C_{2,g}^{[3](2)\rm thresh}\Bigl(z,\frac{m^2}{Q^2}, \frac{m^2}{\mu^2}\Bigr)={}&\frac{C_{2,g}^{[3](1)}\Bigl(z,\frac{m^2}{Q^2}\Bigr)}{4\pi}\bigg\{16 C_A \log^2 \beta + [ 48 C_A \log 2 - 40C_A ] \log \beta  \notag \\
&+(2 C_F  - C_A )\frac{\pi^2}{\beta}+ 8C_A \log \beta L_\mu + \Bigl[c_0(\xi) + 36C_A \log^2 2 \notag \\
&- 60 C_A \log 2+ L_\mu \big(8 C_A \log 2 - \bar{c}_0(\xi)\big)\Bigr] \bigg\} + {\mathcal{O}}(\beta^2), \label{thresh:ord2}
\end{align}
where $c_0(\xi)$ and $\bar{c}_0(\xi)$ are given in Eqs.~(\ref{c0}) and (\ref{c0bar}). 
Finally, at N$^3$LO the threshold limit is
%\begingroup
%\allowdisplaybreaks
\begin{align}
C_{2,g}^{[3](3)\rm thresh}&\Bigl(z,\frac{m^2}{Q^2},\frac{m^2}{\mu^2}\Bigr)=\frac{C_{2,g}^{[3](1)}\Bigl(z,\frac{m^2}{Q^2}\Bigr)}{(4\pi)^2}\Bigg\{128C_A^2 \log^4 \beta + \bigg[\bigg( 768 \log 2 - \frac{6464}{9}\bigg) C_A^2  \notag \\
&+ \frac{128}{9}C_A n_f + 128 C_A^2 L_\mu\bigg] \log^3 \beta + \bigg[\bigg( 1728 \log^2 2 - 3232 \log 2 -\frac{208}{3} \pi^2 \notag \\ 
&+ \frac{15520}{9}\bigg)C_A^2 + \bigg( 64 \log 2 - \frac{640}{9}\bigg) C_A n_f + 16C_A c_0(\xi) + 32 C_A \bigg(C_F - \frac{C_A}{2}\bigg) \frac{\pi^2}{\beta} \notag \\
&- \bigg\{ \bigg( -512 \log 2 + \frac{1136}{3}\bigg) C_A^2-\frac{32}{3} C_A n_f + 16 C_A \bar{c}_0(\xi) \bigg\} L_\mu+ 32C_A^2 L^2_\mu\bigg] \log^2 \beta  \notag \\
&+ \bigg[ \bigg( 1728 \log^3 2 - 4848 \log ^2 2 + \frac{15520}{3} \log 2 - 208 \pi^2 \log 2+ 936 \zeta_3 + \frac{608}{3} \pi^2  \notag \\
&- \frac{88856}{27} \bigg) C_A^2 + \bigg( 96 \log ^2 2 - \frac{640}{3} \log 2 - \frac{16}{3} \pi^2 +\frac{4592}{27} \bigg) C_A n_f-32 C_F \bigg(C_F - \frac{C_A}{2} \bigg) \pi^2 \notag \\
&+(48 \log 2 - 40) C_A c_0(\xi) + \bigg\{ \bigg(-\frac{92}{3} + 32 \log 2 \bigg) C_A + \frac{8}{3} n_f \bigg\} \bigg(C_F - \frac{C_A}{2}\bigg) \frac{\pi^2}{\beta}  \notag \\
&-\bigg\{\bigg( -672 \log^2 2 + 976 \log 2 + \frac{104}{3} \pi^2 - \frac{4160}{9} \bigg)C_A^2 + \bigg( - 32 \log 2 + \frac{320}{9} \bigg) C_A n_f \notag \\
&+ (48 \log 2 - 40)C_A \bar{c}_0(\xi) - 8 C_A c_0(\xi) - 16 C_A \bigg(C_F - \frac{C_A}{2} \bigg) \frac{\pi^2}{\beta} \bigg\} L_\mu \notag \\
&+ \bigg\{\bigg(64\log 2 - \frac{44}{3}\bigg) C_A^2 + \frac{8}{3}C_A n_f   - 8 C_A \bar{c}_0(\xi) \bigg\} L^2_\mu\bigg] \log \beta \notag \\
&+\bigg[ \bigg(8 \log ^2 2 - \frac{68}{3} \log 2 + \frac{8}{3} \pi^2 - \frac{658}{9} \bigg) C_A + \bigg(\frac{8}{3} \log 2 - \frac{20}{9} \bigg) n_f  + 2 c_0(\xi)  \notag \\
&+\bigg(\frac{26}{3} C_A + \frac{4}{3} n_f - 2 \bar{c}_0(\xi) \bigg) L_\mu \bigg] \bigg(C_F  - \frac{C_A}{2} \bigg) \frac{\pi^2}{\beta} + \frac{4}{3} \bigg(C_F - \frac{C_A}{2}\bigg)^2 \frac{\pi^4}{\beta^2} \notag \\
&+C_{2,g}^{[3](3) \rm const}\Bigl(z,\frac{m^2}{Q^2},\frac{m^2}{\mu^2}\Bigr) \Bigg\} + {\mathcal O}(\beta).\label{thresh:ord3}
\end{align}
%\endgroup
\begin{figure}[!ht]
	\centering
	\includegraphics[width=\textwidth]{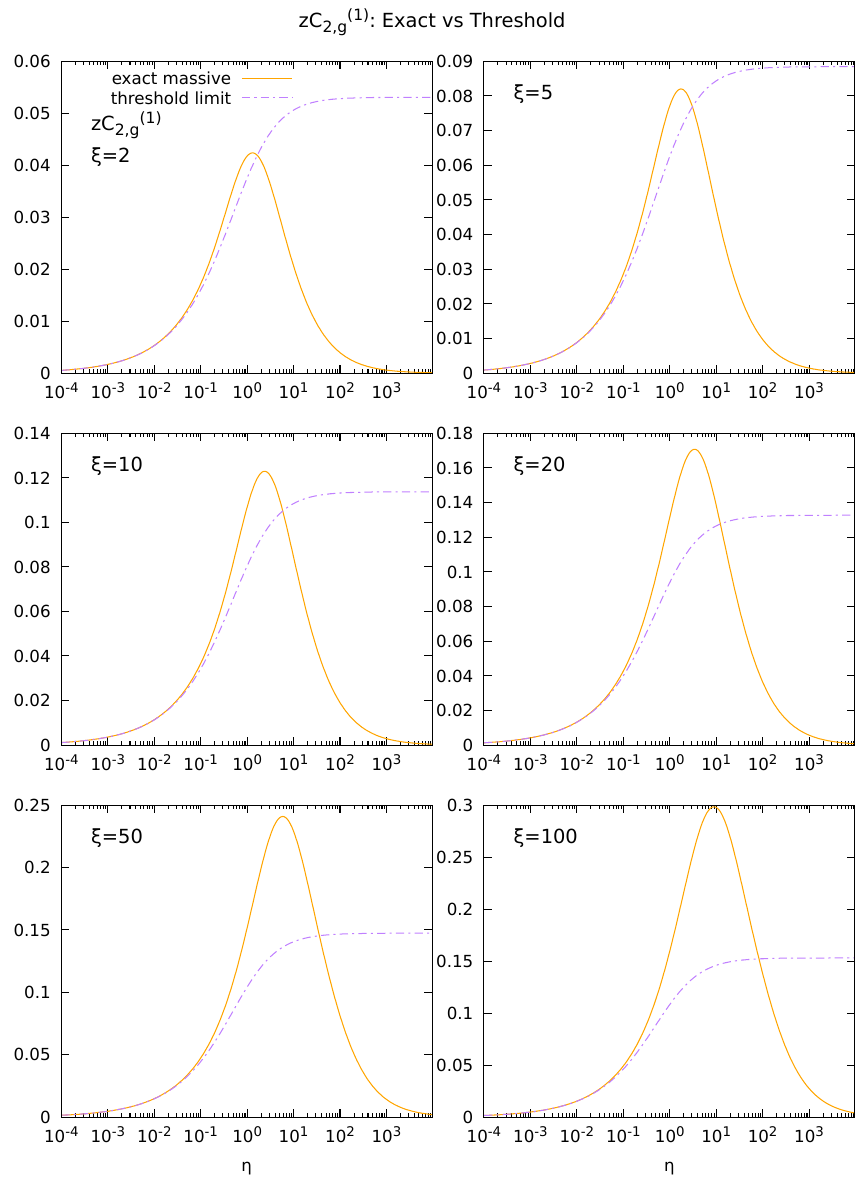}
	\caption{Comparison between the NLO exact massive coefficient function of the gluon for $F_2$ multiplied by $z$ (solid orange), i.e.\ \eqref{C2g1}, and its threshold limit (dashed purple), i.e.\ \eqref{thresh:ord1}, as a function of $\eta$. Six relevant values of $\xi=Q^2/m^2$ are shown.}
	\label{exact:threshold:eta:1ord}
\end{figure}
\begin{figure}[!ht]
	\centering
	\includegraphics[width=\textwidth]{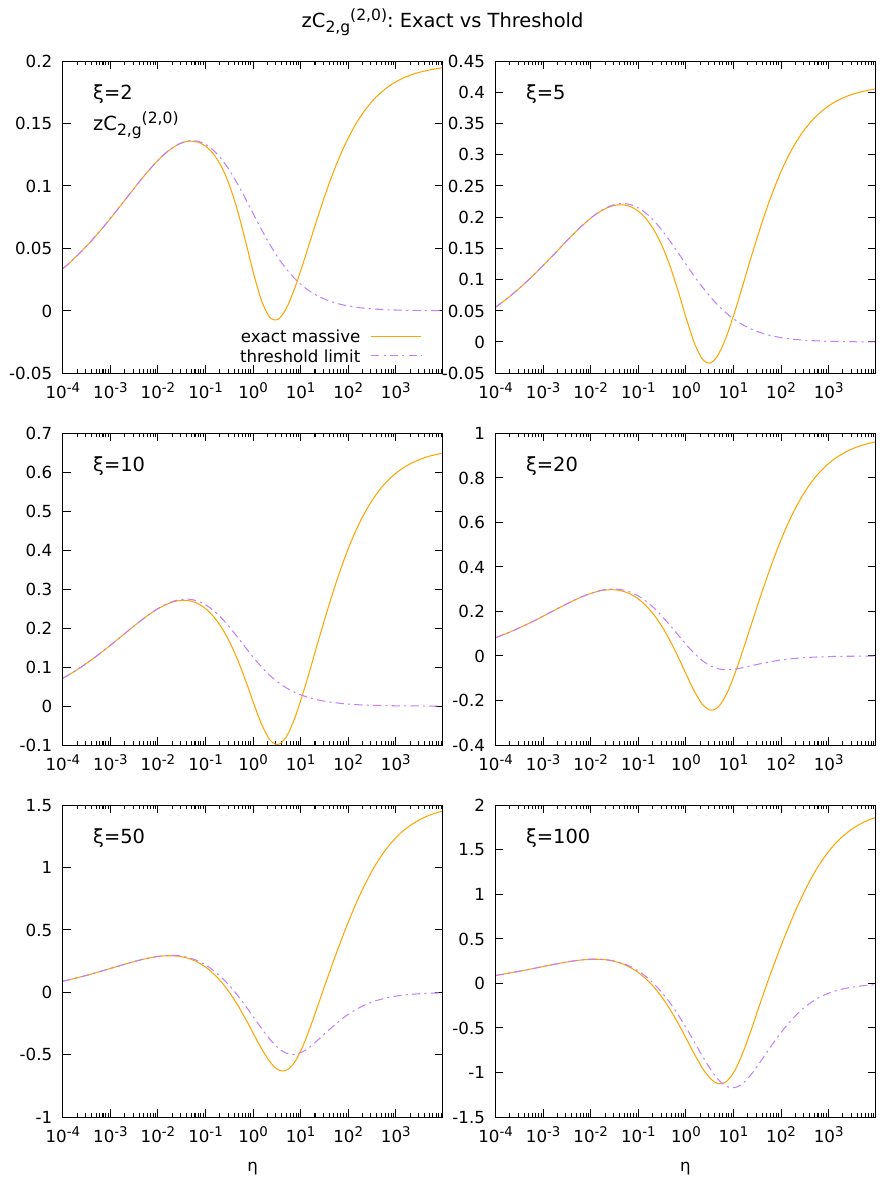}
	\caption{Comparison between the $\mu$-independent part of the NNLO exact massive coefficient function of the gluon for $F_2$ multiplied by $z$ (solid orange) computed from the parametrization in Ref.~\cite{Riemersma_1995} and its threshold limit (dashed purple), i.e.\ \eqref{thresh:ord3}, as a function of $\eta$. Six relevant values of $\xi=Q^2/m^2$ are shown. 
	The $\mu$-independent parts are extracted computing the exact coefficient function and the high-scale limit for $\mu=m$.}
	\label{exact:threshold:eta}
\end{figure}
$C_{2,g}^{[3](3) \rm const}$ is the term independent from $\beta$, that at this order is still unknown. An estimate is
\begin{align}
C_{2,g}^{[3](3) \rm const}\Bigl(z,\frac{m^2}{Q^2},\frac{m^2}{\mu^2}\Bigr)={}&\frac{C_{2,g}^{[3](1)}\Bigl(z,\frac{m^2}{Q^2}\Bigr)}{(4\pi)^2}\Bigl[c_0(\xi) + 36C_A\log^2 2 - 60 C_A\log 2 \notag \\
&+ L_\mu \bigr(8C_A \log 2 - \bar{c}_0(\xi)\bigl)\Bigr]^2\label{c-const}
\end{align}
\figref{exact:threshold:eta:1ord} shows the comparison between the NLO exact massive coefficient function and its threshold limit. It is evident that the two curves overlap in the region $\eta \rightarrow 0$.
\figref{exact:threshold:eta} shows the comparison between the $\mu$-independent part of the NNLO exact massive coefficient function for the gluon and its threshold limit. We can observe that it correctly approximates the exact result for $\eta \lesssim 10^{-1}$. Moreover the fact that in Eqs.~(\ref{thresh:ord2}) and (\ref{thresh:ord3}) there is the factor $C_{2,g}^{[3](1)}$ assures that $zC_{2,g}^{[3]\rm thresh}$ goes to zero for $\eta \rightarrow \infty$ both at NNLO and at N$^3$LO. This will be of great importance in the next chapter when we will combine threshold and asymptotic limits to get the final approximation. Notice that, as $\xi$ increases, the threshold limit gives a good approximation for higher values of $\eta$, but at the same time it tends more slowly to zero for $\eta \rightarrow \infty$.

\section{Combination of threshold and asymptotic limits}\label{comb}

In the previous sections we have constructed the asymptotic and the threshold limits of the exact coefficient function of the gluon. Such limits, obviously, do not describe the exact curve in the whole $z$ range, but only in specific kinematic regimes, i.e.\ respectively $z\rightarrow 0$ and $z\rightarrow z_{\rm max}$. 
In order to have a function that approximates the exact coefficient function for all the values of $z$, we have to combine these two limits. This is achieved using two damping functions: we multiply the threshold by a function $f_1(z)$ and the asymptotic by a function $f_2(z)$ such that
\begin{align}
    f_1(z)&\xrightarrow[]{z\rightarrow 0} 0, \quad f_1(z)\xrightarrow[]{z\rightarrow z_{\rm max}} 1, \label{f1}\\
    f_2(z)&\xrightarrow[]{z\rightarrow 0} 1, \quad f_2(z)\xrightarrow[]{z\rightarrow z_{\rm max}} 0. \label{f2}
\end{align}
In this way we make sure that for $z\rightarrow 0$ only the asymptotic limit contributes since the threshold has been suppressed, while for $z\rightarrow z_{\rm max}$ we have the opposite. Therefore, in these limits our approximation approaches the unknown exact massive coefficient function. Instead, in the intermediate values of $z$, the approximate curve will be an interpolation of the two limits.
The accuracy of the approximation for these values of $z$ will depend on the specific form that we choose for $f_1$ and $f_2$.
In the remaining of this section we will leave them unspecified, but in Chapter~\ref{results} we will choose their functional form and we will tune its parameters comparing our approximation applied to the NLO and to the NNLO with their known exact coefficient functions, in order to have the best agreement between the two.
%In conclusion, in Chapter \ref{results} such approximation will be tuned on the NLO and on the NNLO coefficient functions, that are known, and we will give the results at N$^3$LO.
Thus, our approximate coefficient function has the following form
\begin{equation}
C_{2,g}^{[3](k) \rm approx}\Bigl(z,\frac{m^2}{Q^2},\frac{m^2}{\mu^2}\Bigr)=C_{2,g}^{[3](k) \rm thresh}\Bigl(z,\frac{m^2}{Q^2},\frac{m^2}{\mu^2}\Bigr)f_1(z) + C_{2,g}^{[3](k) \rm asy}\Bigl(z,\frac{m^2}{Q^2},\frac{m^2}{\mu^2}\Bigr)f_2(z). \label{Dapprox}
\end{equation}
Since we want an approximation just for the $\mu$-independent part of the $\ord{3}$ gluon coefficient function, our final approximation is
\begin{equation}
    C_{2,g}^{[3](k,0)\rm approx}\Bigl(z,\frac{m^2}{Q^2}\Bigr)=C_{2,g}^{[3](k,0)\rm thresh}\Bigl(z,\frac{m^2}{Q^2}\Bigr)f_1(z) + C_{2,g}^{[3](k,0)\rm asy}\Bigl(z,\frac{m^2}{Q^2}\Bigr)f_2(z), \label{Dapprox:mu:indep}
\end{equation}
with $k=1,2,3$ since we are considering the NLO, the NNLO and the N$^3$LO. The $\mu$-independent part of \eqref{Dapprox} can be extracted simply computing it for $\mu=m$, as we have done in the previous sections.

Now a comment on \eqref{Dapprox:mu:indep} is required: in the construction of our approximation we have used two ingredients, i.e.\ the asymptotic and the threshold limit, that have been combined using the damping functions $f_1$ and $f_2$. This will be of great importance in the next chapter when we will tune the damping functions on the NLO and on the NNLO. In fact, constructing the approximation in the most simple way, we are confident that it will give accurate results also at the N$^3$LO. Instead, the other approximations that can be found in the literature construct their approximation using more ingredients and combining them in a more complicated  way, with the risk that, even if such approximation has been accurately tuned on the lower orders, it can fail when applied to the N$^3$LO, since it is a different function with respect to the NLO and NNLO. For this reason we expect our approximation to be more accurate than the ones that can be found in the literature.

Now that we have an approximation for the $\mu$-independent part of the N$^3$LO coefficient function of the gluon, we can add the exact $\mu$-dependent part to find the full function. What we get is
\begin{align}
    C_{2,g}^{[3](3) \rm approx}\Bigl(z,\frac{m^2}{Q^2},\frac{m^2}{\mu^2}\Bigr)={}&C_{2,g}^{[3](3,0) \rm approx}\Bigl(z,\frac{m^2}{Q^2}\Bigr)+ C_{2,g}^{[3](3,1)}\Bigl(z,\frac{m^2}{Q^2}\Bigr)\log\Bigl(\frac{\mu^2}{m^2}\Bigr)+\notag \\
    &+C_{2,g}^{[3](3,2)}\Bigl(z,\frac{m^2}{Q^2}\Bigr)\log^2\Bigl(\frac{\mu^2}{m^2}\Bigr),\label{D3full}
\end{align}
where $C_{2,g}^{[3](3,0)\rm approx}$ is given in \eqref{Dapprox:mu:indep} while $C_{2,g}^{[3](3,1)}$ and $C_{2,g}^{[3](3,2)}$ are given in Eqs.~(\ref{C31}) and (\ref{C32}).

%%%%%%%%%%%%%%%%%%%%%%%%%%%%%%%%%%%%%%%%%%%%%%%%%%%%%%%%%%%%%%%%%%%%

\chapter{Results}\label{results}

In Chapter~\ref{approx} we constructed our approximation for the $\mu$-independent part of the N$^3$LO gluon coefficient function, \eqref{Dapprox:mu:indep}, combining the threshold limit, \eqref{thresh:ord3}, and the asymptotic limit, \eqref{asy:30ord}, using two damping functions, i.e.\ $f_1(z)$ and $f_2(z)$.
In that discussion we left the functions $f_1(z)$ and $f_2(z)$ unspecified, only requiring that they had to satisfy Eqs.~(\ref{f1}) and (\ref{f2}). This provides the correct behavior in these two kinematic regimes for our approximation because in this way we make sure that for $z\rightarrow 0$ and $z\rightarrow z_{\rm max}$ the approximate function approaches the exact one.
However, as we have said before, for the intermediate values of $z$ our approximation interpolates the two limits. Therefore, the choice of the functions $f_1(z)$ and $f_2(z)$ will be crucial in order to make sure that our construction gives a good description of the exact curve for such values of $z$.

In this chapter we will choose the functional forms for the damping functions: it will be done applying our approximation to the NLO and NNLO gluon coefficient functions, i.e.\ using \eqref{Dapprox:mu:indep} for $k=1$ and $k=2$, and comparing them with their exact curves, that are known. In this way we will choose the functions $f_1(z)$ and $f_2(z)$ that provide the best agreement between the NLO and the NNLO exact coefficient functions and their approximations. 
After doing it, we will apply our construction to the $\mu$-independent part of the N$^3$LO gluon coefficient function for $F_2$, that is the one we are interested in. As we have said in Chapter~\ref{approx}, the N$^3$LO exact curve can in principle be very different from its lower orders. Therefore, even if we tested and tuned our approximation on such orders, we cannot be sure that at N$^3$LO it will give the same accuracy that it had for the NLO and the NNLO.
For this reason we have to construct our approximation always keeping in mind that in the end it will be applied on a different function with respect to the ones we tested it on. 
Thus, we will have to make an approximation that it is not too specific for the NLO and the NNLO, i.e.\ that is not ``order dependent'', but instead that is general enough to be applied to the unknown N$^3$LO still expecting accurate results. 

Another problem we have to consider is the fact that at N$^3$LO not every term of our approximation is exactly known. In fact the term $k_{cg}^{(3)0}$, appearing in the high-scale limit, \eqref{3ord:m0}, the NLL expansion for small-$z$ of the N$^3$LO coefficient function, \eqref{CNLL}, and the term independent of $\beta$ of the N$^3$LO threshold limit, \eqref{c-const}, are only known in an approximate form.
For this reason the final approximation will be associated with a bigger uncertainty that comes from the uncertainty associated to the approximate terms that we used.

This chapter is organized as follows: in Sec.~\ref{damp} we will choose the functional form for the damping functions $f_1(z)$ and $f_2(z)$, testing it on the known NLO and NNLO and using these functions to tune its various parameters.
In Sec.~\ref{N3LO} we will apply our construction to the unknown N$^3$LO gluon coefficient function, showing the plots of the curves that we will obtain and discussing their accuracy. 
In Sec.~\ref{band} we will construct the uncertainty band associated to our approximation.
In the end, in Sec.~\ref{comparing} we will compare the approximation that we propose with another approximation available in the literature, i.e.\ the one given in Ref.~\cite{Kawamura_2012}.

\section{Tuning the damping function}\label{damp}

In this section we will finally choose the functional form of the functions $f_1(z)$ and $f_2(z)$. In the following we will use the variable $\eta$ defined in \eqref{eta}, instead of $z$, since with such a variable the threshold region is stretched with respect to a plot as a function of $z$ and this allows to correctly check the agreement between our approximation and the exact curve in that kinematic region. Hence, the damping functions will depend on $z$ through the variable $\eta(z)$.
The functional form that we will adopt is 
\begin{align}
f_1(\eta)&=\frac{1}{1+\bigl(\frac{\eta}{h}\bigr)^k},\label{damping} \\
f_2(\eta)&=1-f_1(\eta).
\end{align}
It is trivial to show that these functions have the correct behaviors in the limits $\eta \rightarrow 0$ and $\eta \rightarrow \infty$ (that are respectively $z\rightarrow z_{\rm max}$ and $z\rightarrow 0$).
$h$ and $k$ are the two parameters that we want to tune because they control the ``form'' of the damping functions.
In particular, since $f_1(\eta=h)=f_2(\eta=h)=1/2$ for any value of $k$, $h$ controls the point in which the two dampings equal each others.
Increasing $h$, such point is shifted to bigger values of $\eta$ and vice-versa. 
The parameter $k$, instead, controls the slope of the function close to the turning point $\eta=h$. It means that it controls the ``speed'' with which we pass from asymptotic to threshold (or vice-versa). Therefore, changing it we change the size of the region in which \eqref{Dapprox:mu:indep} is dominated by the asymptotic or the threshold limit and the size of the interpolated region.

As we have observed in the previous chapter, as $\xi=Q^2/m^2$ increases, the asymptotic limit approximates well the exact coefficient function down to lower values of $\eta$, while the threshold limit approximates it well up to higher values of $\eta$. Therefore, increasing $\xi$ we have to reduce the interpolated region. It means that $k$ has to increase as $\xi$ increases. 
\begin{figure}[!ht]
	\centering
	\includegraphics[width=\textwidth]{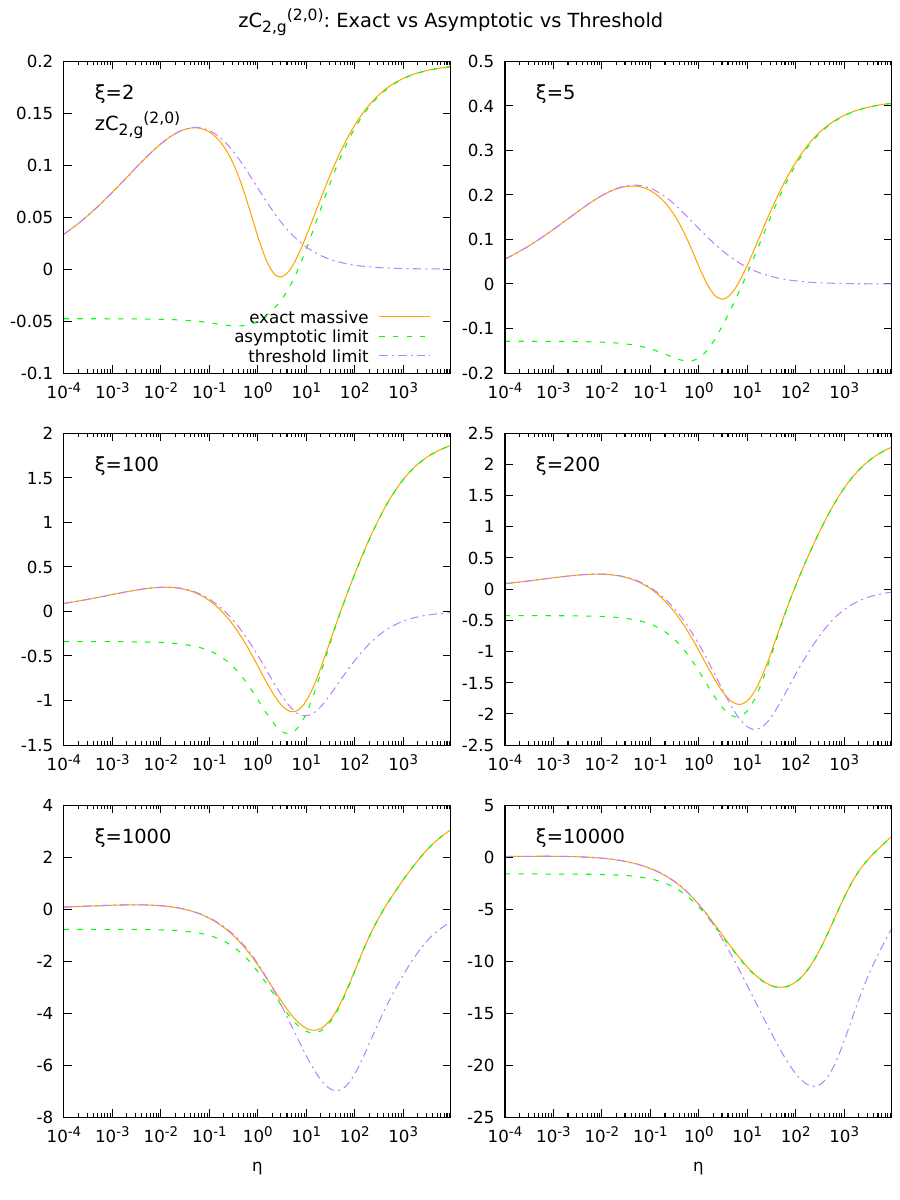}
	\caption{Comparison between the $\mu$-independent part of the NNLO exact massive coefficient function of the gluon for $F_2$ multiplied by $z$ (solid orange) computed from the parametrization in Ref.~\cite{Riemersma_1995} and its threshold (dashed purple) and asymptotic (dashed green) limits, i.e.\ Eqs.~(\ref{thresh:ord2}) and (\ref{asy:2ord}), as a function of $\eta$. Six relevant values of $\xi=Q^2/m^2$ are shown. 
	The $\mu$-independent parts are extracted computing the exact coefficient function and the high-scale limit for $\mu=m$.}
	\label{exact:threshold:asymp:eta}
\end{figure}
\figref{exact:threshold:asymp:eta} shows the comparison between the NNLO exact gluon coefficient function and its asymptotic and threshold limits. We can observe what we where discussing previously: for small values of $\xi$ the interpolation zone is quite large so $k$ must be small. As $\xi$ increases both the asymptotic and the threshold limits give a good approximation for a bigger range of values of $\eta$, therefore the transition from asymptotic to threshold must be quicker so $k$ must be large.
We can also observe that $h$ has to be a function of $\xi$ as well. 
It is because of different reasons: first of all, as we already observed in Sec.~\ref{thresh}, even if the fact that the threshold limit goes to zero for $\eta \rightarrow \infty$ helps the damping to send it to zero, for big values of $\xi$ the convergence to zero becomes slower. Thus, as $\eta \rightarrow \infty$ we want $f_1(z)$ to be practically zero, so that the threshold limit is totally suppressed. This is achieved both increasing the slope of the damping and shifting it leftwards. Second, we can observe from \figref{exact:threshold:asymp:eta} that the center of the interpolation zone, as $\xi$ becomes bigger, moves towards smaller values of $\eta$. For these reasons $h$ must decrease as $\xi$ increases.
In conclusion, we have shown that $h$ and $k$ must be functions of $\xi$. Once we have chosen these functions, we have all the ingredients for the construction of our approximation. 

The chosen functional forms for the parameters $h$ and $k$ are
\begin{align}
    h(\xi)&=A+\frac{B-A}{1+\exp{\bigl( a(\log \xi - b)\bigr)}}, \label{damp:h}\\
    k(\xi)&=C+\frac{D-C}{1+\exp{\bigl( c(\log \xi - d)\bigr)}}, \label{damp:k}
\end{align}
with $a,c>0$.
We chose these functional forms because for small $\xi$, we want $h$ and $k$ to be almost constant. As $\xi$ increases, these parameters have to start a slow transition from a value to another. Then, when $\xi$ becomes big, they have to settle on a constant value and do not change for all the higher $\xi$.
In fact, for $\xi \sim \mathcal{O}(1)$ the exponential is suppressed and we have, if $e^{-ab},e^{-cd} \ll 1$, that $h \simeq B$ and $k\simeq D$. For $\xi \gg 1$ the exponential diverges and thus we have that $h \simeq A$ and $k \simeq C$. We used $\log \xi$ instead of just $\xi$ in order to have a slower transition from the small values of $\xi$ to the big ones.
We choose this form since we want the growth of $k$ to stop at a certain point. In fact, due to the fact that the N$^3$LO coefficient function is unknown, having a too fast transition from threshold to asymptotic for large $\xi$, would probably fail to describe correctly the exact function since we do not know if the interchange between the two limits would be as good as for the NNLO in \figref{exact:threshold:asymp:eta}. This will be shown explicitly applying our approximation to the NLO.

All we have to do now is to tune the eight parameters $A$, $B$, $C$, $D$, $a$, $b$, $c$ and $d$.
Comparing the plots given by different values of these parameters we established that at small $\xi$ the choices $h\simeq 2.5$ and $k\simeq 1.2$ best describe the exact NNLO coefficient function. Instead, for large $\xi$, i.e.\ $\xi \gtrsim 10^{4}$, the best values of the parameters are $h \simeq 1.7$ and $k\simeq 2.5$. This value of $h$ has been extracted observing that the intersection point between the asymptotic and the threshold limits tends  to $h\simeq 1.7$ for high $\xi$.
Instead, the value of $k$ for large $\xi$ has been chosen in order to move quickly from threshold to asymptotic limit, see \figref{exact:threshold:asymp:eta}. 
%However, as we have said previously, this interchange between the two curves cannot be too fast since we cannot be sure that for the N$^3$LO they will intersect as they do for the NNLO. 
Therefore we choose
\begin{align}
    A&= 1.7, \qquad B=2.5, \label{ABNNLO}\\
    C&=2.5, \qquad D=1.2. \label{CDNNLO}
\end{align}
The remaining parameters are chosen in order to have a not too sharp transition between the small and high values of $\xi$. Thus, we made the following choice
\begin{equation*}
    a=c=2.5, \qquad b=d=5.
\end{equation*}
With such values the condition $e^{-ab},e^{-cd} \ll 1$ is satisfied.

Now we have all the ingredients for the construction of the approximation. 
Testing it on the NNLO gives promising results.
\begin{figure}[!ht]
\centering
\includegraphics[width=\textwidth]{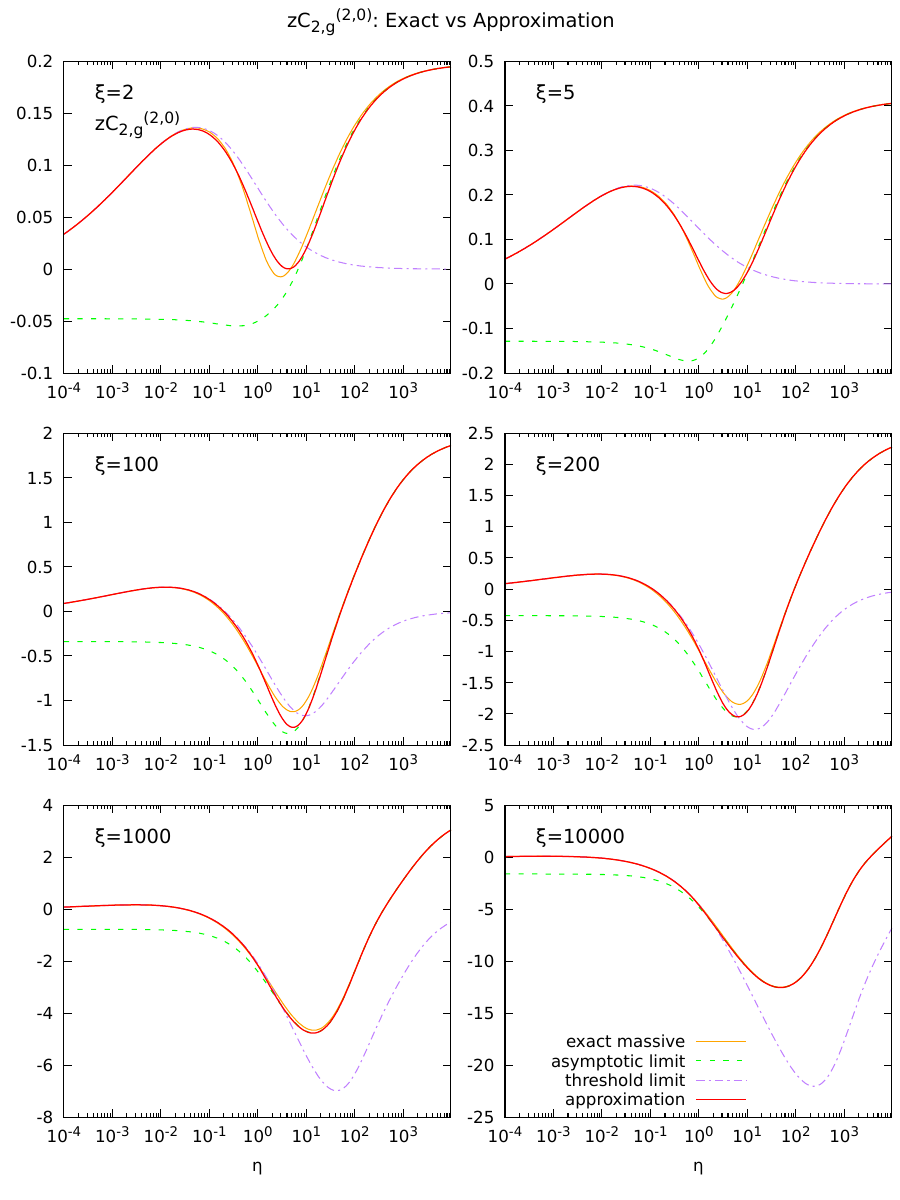}
\caption{Comparison between the $\mu$-independent part of the NNLO exact massive coefficient function of the gluon for $F_2$ multiplied by $z$ (solid orange) computed from the parametrization in Ref.~\cite{Riemersma_1995}, its threshold (dashed purple) and asymptotic (dashed green) limits, i.e.\ Eqs.~(\ref{thresh:ord3}) and (\ref{asy:2ord}), and the approximation that we propose (solid red) in \eqref{Dapprox:mu:indep}, as a function of $\eta$. Six relevant values of $\xi=Q^2/m^2$ are shown.}
\label{approxNNLO}
\end{figure}
\figref{approxNNLO} shows the comparison between the NNLO exact gluon coefficient function and the approximation that we propose. Different values of $\xi$ are shown. We can observe a good agreement between the two both at small and at high $\xi$. Moreover, for $\xi \gg 1$ our approximation tends to the exact coefficient function for all the values of $\eta$ and therefore in the whole $z$ range. 
Actually, in this limit the massless result is accurate. It means that we can use the massless computation, that at N$^3$LO is exactly known, and we don't really need the approximate massive coefficient function that we are constructing. 
However, our aim is to construct an approximation that gives precise results both for small $\xi$ and for $\xi \gg 1$. It means that for $\xi \gg 1$, in the computation of the structure functions via \eqref{coeff:func}, we can use both our approximation of the massive coefficient function and the massless limit, expecting with good approximation the same result.
Instead, if we constructed an approximation that deviates from the exact result as $\xi$ increases, we would have an intermediate range of values of $\xi$ that are not described neither from our approximation nor from the massless limit.
 
As a first check of the accuracy of the approximation procedure that we constructed from the NNLO, we can apply it to the NLO gluon coefficient function, \eqref{C2g1}. First of all we have to notice a small difference with respect to the NNLO and the N$^3$LO: the NLO threshold function doesn't approach zero for $\eta \rightarrow \infty$. For this reason, since we want to use exactly the same form for the damping function (but with a small difference in the parameters that we are going to discuss), the convergence to zero of the threshold limit times its damping will be slower. However, this problem affects only the small values of $\xi$. In fact, even for the NNLO and the N$^3$LO we have seen that for $\xi \gg 1$ the convergence to zero is pushed towards higher values of $\eta$.
For the approximation at NLO our approach consists in using the same damping function we used at NNLO, i.e.\ \eqref{damping}, and the same functional form for $h$ and $k$, i.e.\ Eqs.~(\ref{damp:h}) and (\ref{damp:k}). Regarding the parameters, the only one that will be changed is $A$, since the center of the interpolation region will approach a different value with respect to the case of the NNLO. 
The value of $A$ has been chosen in the following way:
we wanted the value of $h$ for small $\xi$ to be the same that we extracted from the NNLO. Hence, we have that $h\simeq 2.5$. Then, we studied the point of minimal distance between threshold and asymptotic limits for $\xi \gg 1$. We extracted the value $h\simeq 0.2$. It means that we have to set
\begin{equation}
    A=0.2, \qquad B=2.5.\label{ABNLO}
\end{equation}
All the others parameters will be left unchanged. Observe that this procedure of searching the point of minimal distance between threshold and asymptotic limit, or the point of intersection between the two, in the limit $Q^2 \gg m^2$, does not require the knowledge of the exact curve. It means that we can tune in the same way the parameters for the N$^3$LO approximation.

\begin{figure}[!ht]
	\centering
	\includegraphics[width=\textwidth]{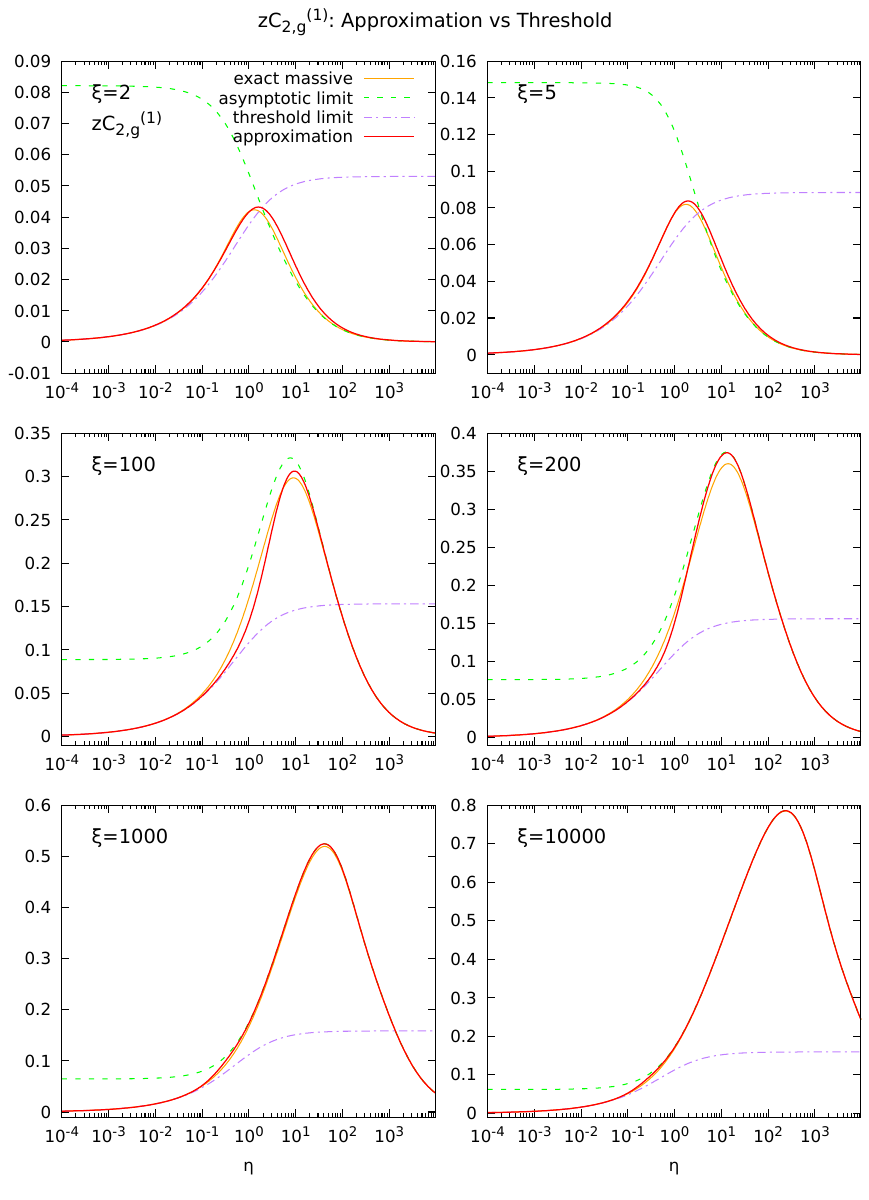}
	\caption{Comparison between the NLO exact massive coefficient function of the gluon for $F_2$ multiplied by $z$ (solid orange), \eqref{C2g1}, its threshold (dashed purple) and asymptotic (dashed green) limits, i.e.\ Eqs.~(\ref{thresh:ord1}) and (\ref{1ord:m0:explicit}), and the approximation that we propose in \eqref{Dapprox}, as a function of $\eta$. Six relevant values of $\xi=Q^2/m^2$ are shown.}
	\label{NLOexact:threshold:asymp:eta}
\end{figure}
Now we can apply our approximation to the NLO.
\figref{NLOexact:threshold:asymp:eta} shows the comparison between the exact curve and the approximate one. We can observe a good agreement between the two. Also in this case we have that the approximate coefficient function tends to the exact one for $\xi \gg 1$. From the plots for $\xi=1000$ and $\xi =10000$ we can observe that if we had chosen a value of $k$ that is too large, even if the accuracy of the approximation at NNLO would have been the same, it wouldn't have been so at NLO. In fact, in this case a too sharp transition from the threshold to the asymptotic limit would have spoiled the agreement between the approximation and the exact curve in the interpolation region.

In conclusion, tuning the damping functions $f_1$ and $f_2$ on the NNLO coefficient function, we have constructed an approximation that gives accurate results for small values of $\xi$ and tends to the exact function for $\xi \gg 1$. Since applying our approximation on the NLO still provides accurate results, we are confident that it will provide accurate results when we will apply it on the N$^3$LO too.

\section{Results at \texorpdfstring{N$^3$LO}{N3LO}}\label{N3LO}
Now that we have constructed our approximation and that we have tested it to the NLO and NNLO we can apply it to the $\mu$-independent part of the N$^3$LO coefficient function of the gluon. 
From the accuracy that such approximation had when we applied it to the lower orders, see Figs.~\ref{approxNNLO} and \ref{NLOexact:threshold:asymp:eta}, we expect good results also at N$^3$LO. Obviously, since at this order some ingredients are only approximate, the precision of the approximation will not be as good as the one of the NLO and NNLO coefficient functions.

\begin{figure}[!ht]
	\centering
	\includegraphics[width=\textwidth]{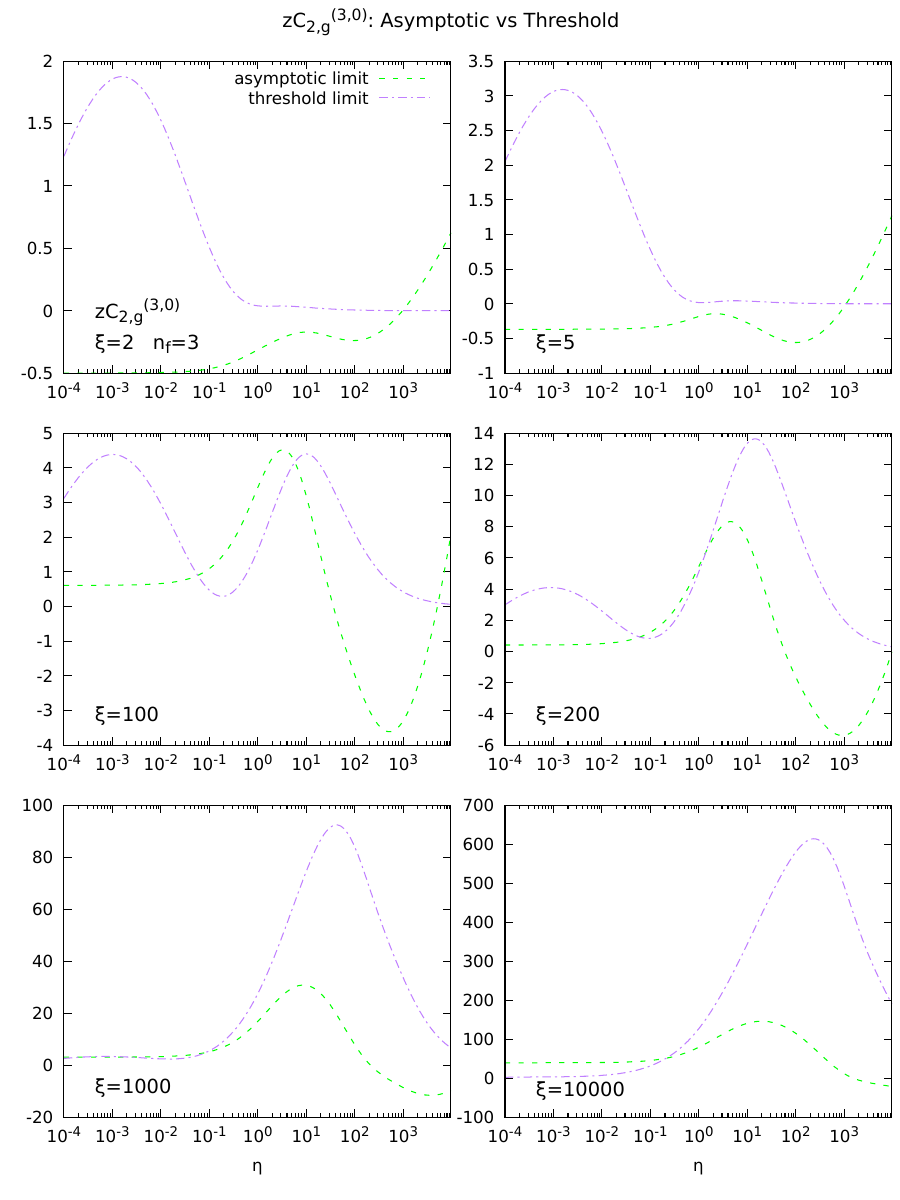}
	\caption{Comparison between the $\mu$-independent part of the N$^3$LO threshold (dashed purple) and asymptotic (dashed green) limits of coefficient function of the gluon for $F_2$ multiplied by $z$, i.e.\ Eqs.~(\ref{thresh:ord3}) and (\ref{asy:3ord}), as a function of $\eta$. Six relevant values of $\xi=Q^2/m^2$ are shown. 
	The $\mu$-independent parts are extracted computing the exact coefficient function and the high-scale limit for $\mu=m$.}
	\label{exact:threshold:asymp:eta3}
\end{figure}

First of all we have to set the parameter $A$ of the damping functions, see \eqref{damp:h}. In order to do it, we have to study the behavior of the asymptotic and threshold limits for $\xi \gg 1$, as we did for the lower orders. 
\figref{exact:threshold:asymp:eta3} shows the comparison between these two limits. For the high-scale limit we used the center of the band given by the two approximations of the term $k_{cg}^{(3)0}$ in Eqs.~(\ref{kcgA}) and (\ref{kcgB}).
As it happened for the NLO, for $\xi \gg 1$ we do not have the good transition from the asymptotic to the threshold limit that we had at NNLO. This may be due to the fact that we used approximate expressions for the construction of both the asymptotic and the threshold limits.
Therefore, the fact that the growth of $k(\xi)$ stops at a certain point helps to increase the accuracy of our approximation.
From the comparison of the two limits for the highest values of $\xi$ we extracted that the center of the interpolation region is approaching roughly the value of $\eta\simeq 0.3$. We conclude that $A=0.3$. Instead, at small $\xi$, as explained in the previous section, we still want that $h\simeq 2.5$.
Therefore, we set 
\begin{equation}
    A=0.3, \quad B=2.5\,.\label{ABN3LO}
\end{equation}

Now that we have all the parameters of the damping functions we can construct our approximation for the $\mu$-independent part of the $\ord{3}$ DIS massive coefficient function of the gluon for $F_2$ in heavy flavour production.
\begin{figure}[!ht]
\centering
\includegraphics[width=\textwidth]{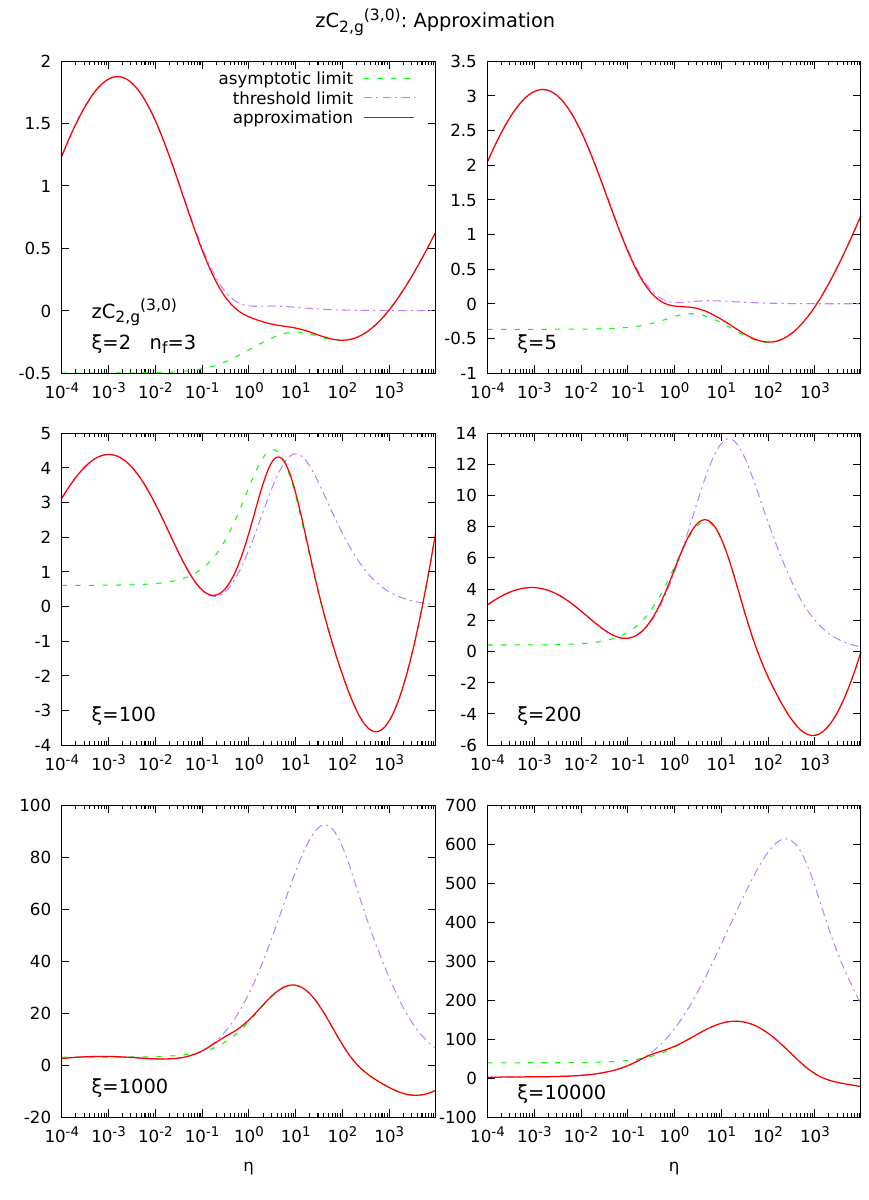}
\caption{Comparison between the $\mu$-independent part of the N$^3$LO threshold (dashed purple) and asymptotic (dashed green) limits of the coefficient function of the gluon for $F_2$ multiplied by $z$, i.e.\ Eqs.~(\ref{thresh:ord3}) and (\ref{asy:3ord}), and the approximation that we propose (solid red) in \eqref{Dapprox:mu:indep}, as a function of $\eta$. Six relevant values of $\xi=Q^2/m^2$ are shown.}
\label{approxN3LO}
\end{figure}
Fig.~\ref{approxN3LO} shows the results that we obtained. Six values of $\xi$ are shown and we used $n_f=3$ since we are considering the charm as heavy quark.
A comment on such results is now required: first of all, being the N$^3$LO next-to-leading logarithm expansion for small-$z$ and the $\mu$-independent term of the high-scale limit of the massive coefficient function only approximate expressions, we cannot be sure that in the small-$z$ region the asymptotic limit exactly approaches the (yet unknown) exact massive coefficient function. 
Second, also the threshold limit contains approximate ingredients: in fact in \eqref{thresh:ord3} the term constant in $\beta$ is not exact but it is an approximation \cite{Kawamura_2012}. For this reason we expect our approximation to approach the exact result both in the asymptotic and in the threshold regions only approximately. Third, as we mentioned before, at N$^3$LO the transition from asymptotic limit to threshold limit at high $\xi$ is not perfect as it was for the NNLO. Moreover, contrary to the lowest orders, we do not have the exact curve as a reference to help us to tune the point that the parameter $h(\xi)$ has to tend to. For these reasons, at high $\xi$ the accuracy of our N$^3$LO approximation will not be as good as the one of the NNLO and NLO. Last, we observe again that even if the application of the approximation to the NNLO and to the NLO gives promising results, the N$^3$LO will be a different function and therefore we cannot be sure that such approximation will be accurate in the same way.
Despite the sources of uncertainty that we have just mentioned, we expect our approximation to have a good level of precision. 
In order to check the accuracy of our approximate coefficient function, in Sec.~\ref{comparing} we will compare it with an other approximation that is available in the literature.
%In fact, our procedure gives accurate results for the NLO and NNLO approximations. Moreover, our construction is general enough that we are confident that it can give accurate results at N$^3$LO even if we tuned it on the NNLO and on the NLO coefficient functions. 
%%%%%%%%%%%%%%%%%%%%%%%%%%%%%%%%%%%%%%%%%%%%%%%%%%%%%%%%%%%%%%%%%%%%%%%%%%%%%%%%%%%
\section{Construction of the uncertainty band}\label{band}

In this section we will construct an uncertainty band associated to our approximation of the gluon coefficient function at $\ord{3}$. It will be done varying the parameters $A$, $B$, $C$ and $D$ around their central value and varying the approximate contributions that we used in our construction between the two extremes of their uncertainty estimate. The parameters $a$, $b$, $c$ and $d$ will not be varied since the final approximation depends very weakly on their precise value and therefore their variation does not change significantly the final result.

First of all we construct the uncertainty band for the approximation of the NNLO gluon coefficient function. The central values of the parameters are those of Eqs.~(\ref{ABNNLO}) and (\ref{CDNNLO}). Observe that in the construction of the approximation at NNLO we do not have approximate ingredients. Therefore, at NNLO the uncertainty band will not have the contribution coming from the uncertainty of unknown ingredients.
Then, we have to observe that the approximation that we constructed depends more strongly on the parameters $C$ and $D$, that govern the parameter $k$ in \eqref{damp:k}, and more weakly on the parameters $A$ and $B$, that govern the parameter $h$ in \eqref{damp:h}. Therefore, $C$ and $D$ will be varied by $30\%$ above and below their central value, while $A$ and $B$ will be varied by a factor of 3. It means that we define
\begin{align}
    A_+&=3A,\quad A_-=A/3, \quad B_+=3B, \quad B_-=B/3, \label{varAB}\\
    C_+&=(1+0.3)C, \quad C_-=(1-0.3)C,\quad D_+=(1+0.3)D \quad D_-=(1-0.3)D,\label{varCD}
\end{align}
and the variations of the approximate coefficient function will be obtained using these values as the parameters of the damping functions.
Then, we plot all the curves in which we vary one parameter at the time, the curve in which all the parameters are varied together above their central value and the curve in which all the parameters are varied together below their central value. In this way we get a set of curves that lie around our approximation. Our uncertainty band will be the envelope of such curves. It means that the two extremes of our band will be the curves that in each point will differ more from the central value of our approximation.
\begin{figure}[!ht]
\centering
\includegraphics[width=\textwidth]{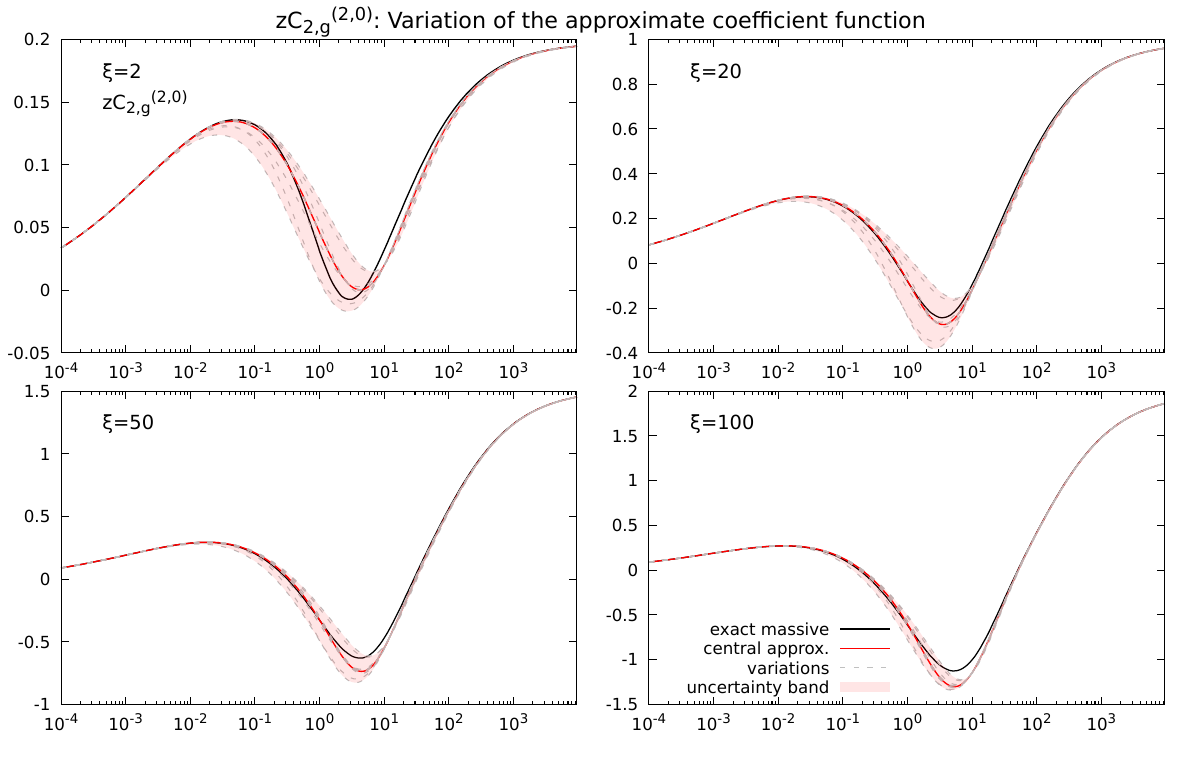}
\caption{Construction of the uncertainty band of the approximation of the NNLO gluon coefficient function via the variation of the parameters of the damping functions. The band is compared with its central value and with the NNLO exact coefficient function.}
\label{variationNNLO}
\hspace{1em}
\includegraphics[width=\textwidth]{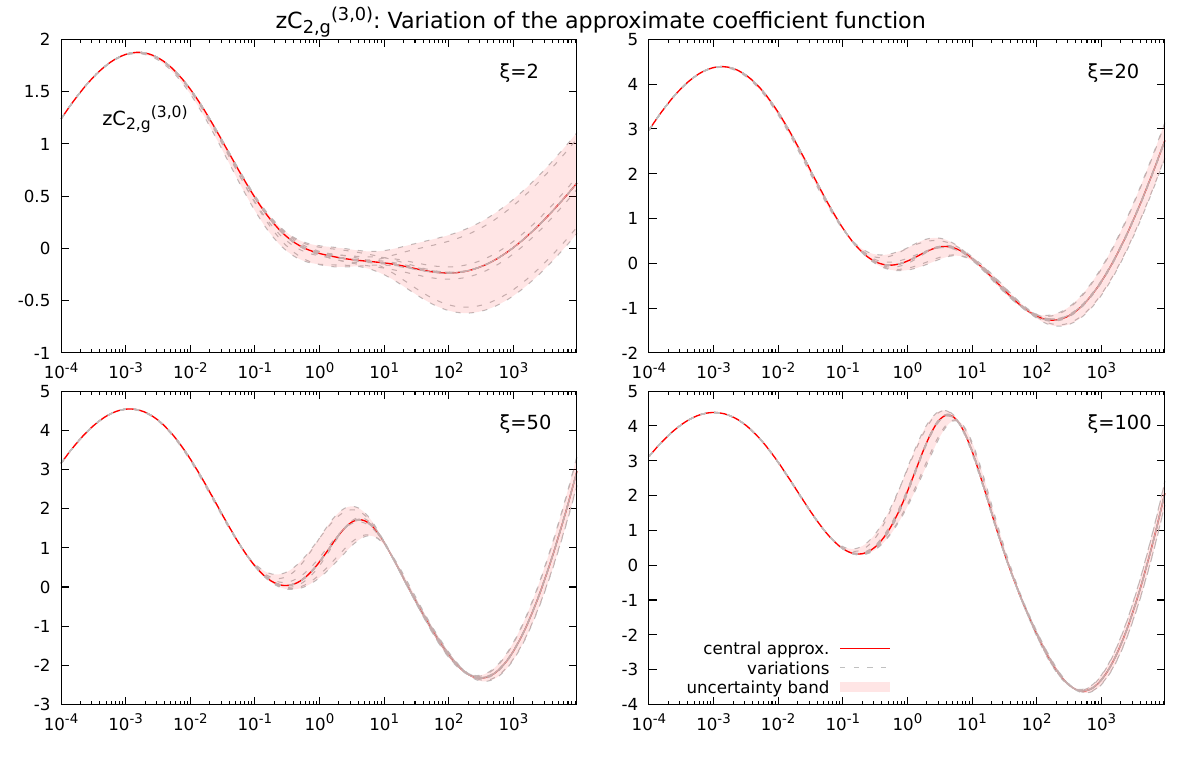}
\caption{Construction of the uncertainty band of the approximation of the N$^3$LO gluon coefficient function via the variation of the parameters of the damping functions. The band is compared with its central value. }
\label{variationN3LO}
\end{figure}
\figref{variationNNLO} shows the results that we obtained.
We can observe that in the threshold region the exact coefficient function lies inside our band for any $\xi$. Instead, in the asymptotic and in the interpolation regions we have that the exact function does not always lie inside the uncertainty band. 
%Moreover, despite our approximation exactly approaches the exact coefficient function at high-$\eta$, we still have a rather big uncertainty band in that region, especially at low-$\xi$.
However, this problem will not affect the approximation at N$^3$LO since in that case the uncertainty in the asymptotic region is increased by the one coming from the unknown term of the high-scale coefficient function, as we will see. 
Anyway, the fact that the exact coefficient function does not lie inside our uncertainty band for some values of $\eta$ does not represent a big problem. In fact, the band we constructed is a rough estimate of the uncertainty of our approximation and therefore in some points it can be a bit inaccurate. However, the agreement between our approximation and the exact NNLO coefficient function is good and thus we think that globally our construction can provide enough accuracy.
In the end, we can observe that, due to the fact that our approximation tends to the exact function at high-$\xi$, the uncertainty band correctly gets smaller at high-$\xi$. Moreover, since at any $\xi$ our approximation approaches the exact function in the asymptotic and in the threshold limits, the band becomes narrower in these two limits.

Now we can construct the uncertainty band of the approximation of the N$^3$LO gluon coefficient function. The central values of the parameters are those of Eqs.~(\ref{ABN3LO}) and (\ref{CDNNLO}). They are varied according to Eqs.~(\ref{varAB}) and (\ref{varCD}). It means that in the construction of the uncertainty band we will plot the curves varied in the same way we did for the NNLO, i.e.\ varying one parameter at the time and varying all the parameters together.
At N$^3$LO, in addition to varying the parameters of the damping functions, we varied the approximate expressions that appear in the construction of the approximation at $\ord{3}$. 
It means that in the construction of the band we added the curves obtained using the two extremes of the uncertainty bands of such approximate contributions, instead of their central values.
For example, the unknown term for the construction of the high-scale coefficient function at $\ord{3}$, i.e.\ $k_{cg}^{(3)0}$, has been varied using in \eqref{Dapprox:mu:indep} the functions $k_{cg,A}^{(3)0}$ and $k_{cg,B}^{(3)0}$ (given in Eqs.~(\ref{kcgA}) and (\ref{kcgB})) instead of their average value.
The approximate terms are the $\mu$-independent part of the heavy quark-gluon N$^3$LO matching condition, the NLL small-$z$ expansion and the $\beta$-independent part of the threshold limit. While for the first two we have an estimate of their uncertainty, for the unknown term of the N$^3$LO threshold limit we do not have such estimate. Therefore, we varied only the two terms for which we have an uncertainty band.
In the curves in which all the parameters are varied together, also the approximate contributions have been varied.
\figref{variationN3LO} shows the uncertainty band that we obtained. As we mentioned before, the band in the high-$\eta$ (small-$z$) region is dominated by the uncertainty of the high-scale limit. We conclude this section observing that also in this case the uncertainty band reduces its size at high $\xi$ since our construction has to approach the exact coefficient function. However, at high $\xi$, due to the bigger uncertainty in the choice of the point that the center of the interpolation region has to tend to, we have a bigger uncertainty in the central region with respect to the $\ord{2}$ approximation. 
%Moreover, we can observe that the uncertainty band extends in the threshold region down to lower values of $\eta$ with respect to the NNLO. This is positive since the threshold limit at N$^3$LO involves an approximate contribution that we have not varied in the construction of the band.

%%%%%%%%%%%%%%%%%%%%%%%%%%%%%%%%%%%%%%%%%%%%%%%%%%%%%%%%%%%%%%%%%%%%%%%%%%%
\section{Comparison with other approximations}\label{comparing}

So far in this chapter we have constructed an approximation for the $\mu$-independent part of the N$^3$LO gluon coefficient function for $F_2$. It has been done combining the the high-scale, high-energy and threshold limits of the exact coefficient function. As we have said previously, in the literature there are other approximations for the N$^3$LO unknown term that are constructed starting from the same three limits.
Our approximation procedure differs from the other ones in the way these three known limits are combined together. In this section we will compare our results with the approximation of the N$^3$LO gluon coefficient function proposed in Ref.~\cite{Kawamura_2012}. Also in that case the accuracy of the approximation is tested applying it to the NNLO coefficient function and comparing the known exact curve with the approximate one. So we will compare our approximation with the one of Ref.~\cite{Kawamura_2012} also at NNLO.
The NNLO approximation that is proposed in Ref.~\cite{Kawamura_2012} is the central value of the two extremes of the uncertainty band given by the functions
\begingroup
\allowdisplaybreaks
\begin{align}
    C_{2,g}^{[3](2,0) \rm approx,A}\Bigl(z,\frac{m^2}{Q^2}\Bigr)={}& C_{2,g}^{[3](2,0) \rm thresh}\Bigl(z,\frac{m^2}{Q^2}\Bigr)-C_{2,g}^{[3](2,0) \rm const}\Bigl(z,\frac{m^2}{Q^2}\Bigr) \notag \\
    &+\bigl(1-f(\xi)\bigr)\beta C_{2,g}^{[3,0](2,0)}\Bigl(z,\frac{m^2}{Q^2}\Bigr)\notag \\
    &+f(\xi)\beta^3\biggl[C_{2,g}^{[3](2,0) \rm LL}\Bigl(z,\frac{m^2}{Q^2}\Bigr)\frac{\eta^\gamma}{C+\eta^\gamma}\biggr],\label{NNLO:vogtA}\\
    C_{2,g}^{[3](2,0) \rm approx,B}\Bigl(z,\frac{m^2}{Q^2}\Bigr)={}& C_{2,g}^{[3](2,0) \rm thresh}\Bigl(z,\frac{m^2}{Q^2}\Bigr) +\bigl(1-f(\xi)\bigr)\beta^3 C_{2,g}^{[3,0](2,0)}\Bigl(z,\frac{m^2}{Q^2}\Bigr)\notag \\
    &+f(\xi)\beta^3\biggl[C_{2,g}^{[3](2,0) \rm LL}\Bigl(z,\frac{m^2}{Q^2}\Bigr)\frac{\eta^\delta}{D+\eta^\delta}\biggr],\label{NNLO:vogtB}\\
\end{align}
\endgroup
where 
\begin{equation}
    f(\xi)=\frac{1}{1+\exp \bigl(2(\xi -4)\bigr)},
\end{equation}
and
\begin{align}
    \gamma&=1.0, \quad C=42.5, \\
    \delta&=0.8, \quad D=19.4.
\end{align}
The function $C_{2,g}^{[3](2,0) \rm const}$ is the $\beta$-independent part of the threshold limit of the NNLO gluon coefficient function, i.e.\ 
\begin{align}
    C_{2,g}^{[3](2,0) \rm const}=\frac{C_{2,g}^{[3](1)}\Bigl(z,\frac{m^2}{Q^2}\Bigr)}{4\pi}\Bigl(c_0(\xi) + 36C_A \log^2 2 - 60 C_A \log 2\Bigr),
\end{align}
where we have omitted the term proportional to $L_\mu$ since we are interested in the $\mu$-independent contribution. 
From Eqs.~(\ref{NNLO:vogtA}) and (\ref{NNLO:vogtB}) we can observe that this approximation treats the three kinematic limits as three different ingredients, each of which with its own damping function (the threshold limit is not multiplied by a damping since it goes to zero at high-$\eta$). Instead, in our approximation we use just two ingredients, since the high-scale and the high-energy limits are combined together in the construction of the asymptotic limit. 
%For this reason our approximation procedure is simpler
\begin{figure}[!ht]
\centering
\includegraphics[width=\textwidth]{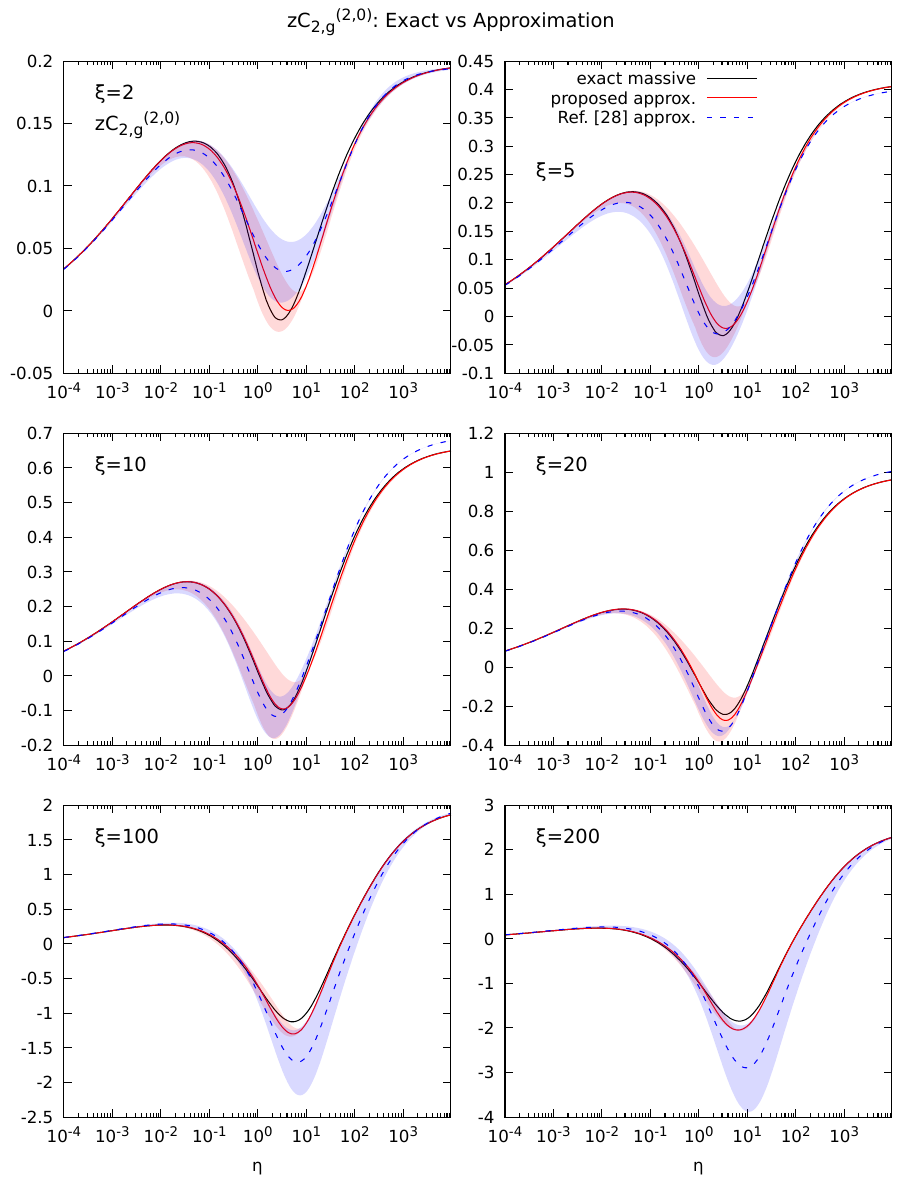}
\caption{Comparison between the $\mu$-independent part of the NNLO exact massive coefficient function of the gluon for $F_2$ multiplied by $z$ (solid orange) computed from the parametrization in Ref.~\cite{Riemersma_1995}, the approximation that we propose (solid red) in \eqref{Dapprox:mu:indep}, and the approximation given in Ref.~\cite{Kawamura_2012} (dashed blue), Eqs.~(\ref{NNLO:vogtA}) and (\ref{NNLO:vogtB}), as a function of $\eta$. Six relevant values of $\xi=Q^2/m^2$ are shown.}
\label{vogtNNLO}
\end{figure}
\figref{vogtNNLO} shows the comparison between the $\mu$-independent part of the NNLO exact massive gluon coefficient function for $F_2$, the approximation that we propose in \eqref{Dapprox:mu:indep} and the approximations given in Ref.~\cite{Kawamura_2012}, that is the average value of the two extremes given in Eqs.~(\ref{NNLO:vogtA}) and (\ref{NNLO:vogtB}). 
First of all, we can observe that for any $\xi$ this approximation does not overlap accurately with the exact coefficient function for intermediate values of $\eta$, while our approximation does. 
Second, while this approximation correctly approach the exact coefficient function in the threshold limit ($\eta \rightarrow 0$) for every value of $\xi$, it does not approach perfectly the exact curve in the high-$\eta$ region. Instead, our approximation tends to the exact result at large $\eta$ for any $\xi$. This is due to the better way of combining the high-scale limit and the high-energy limit into the asymptotic limit.
Last, while the upper extreme of the band gives a good description of the exact function for $\xi \gg 1$, the lower extreme does not, as it can be clearly observed. It follows that the average of the two extremes does not give a good description of the exact function at high-$\xi$.

At N$^3$LO the approximation for the $\mu$-independent part of the gluon coefficient function proposed in Ref.~\cite{Kawamura_2012} is the average of the two extremes of the uncertainty band given by the expressions
\begingroup
\allowdisplaybreaks
\begin{align}
     C_{2,g}^{[3](3,0) \rm approx,A}\Bigl(z,\frac{m^2}{Q^2}\Bigr)={}& C_{2,g}^{[3](3,0) \rm thresh'}\Bigl(z,\frac{m^2}{Q^2}\Bigr) + \bigl(1-f(\xi)\bigr)\beta C_{2,g}^{[3,0](3,0)\rm A}\Bigl(z,\frac{m^2}{Q^2}\Bigr)\notag \\
    &+f(\xi)\beta^3\biggl[-C_{2,g}^{[3](3,0) \rm LL}\Bigl(z,\frac{m^2}{Q^2}\Bigr)\frac{\log \eta}{\log z}\notag \\
    &+C_{2,g}^{[3](3,0) \rm NLL,A}\Bigl(z,\frac{m^2}{Q^2}\Bigr)\frac{\eta^\gamma}{C+\eta^\gamma}\biggr],\label{N3LO:vogtA}\\
    C_{2,g}^{[3](3,0) \rm approx,B}\Bigl(z,\frac{m^2}{Q^2}\Bigr)={}& C_{2,g}^{[3](3,0) \rm thresh'}\Bigl(z,\frac{m^2}{Q^2}\Bigr)-2f(\xi)C_{2,g}^{[3](3,0) \rm const}\Bigl(z,\frac{m^2}{Q^2}\Bigr) \notag \\
    &+\bigl(1-f(\xi)\bigr)\beta^3 C_{2,g}^{[3,0](3,0)\rm B}\Bigl(z,\frac{m^2}{Q^2}\Bigr)\notag \\
    &+f(\xi)\beta^3\biggl[-C_{2,g}^{[3](3,0) \rm LL}\Bigl(z,\frac{m^2}{Q^2}\Bigr)\frac{\log \eta}{\log z}\notag \\
    &+C_{2,g}^{[3](3,0) \rm NLL,B}\Bigl(z,\frac{m^2}{Q^2}\Bigr)\frac{\eta^\delta}{D+\eta^\delta}\biggr],\label{N3LO:vogtB}
\end{align}
\endgroup
where
\begin{align}
    \gamma&=1.0, \quad C=20.0, \\
    \delta&=0.8, \quad D=10.7.
\end{align}
\begin{figure}[!ht]
\centering
\includegraphics[width=\textwidth]{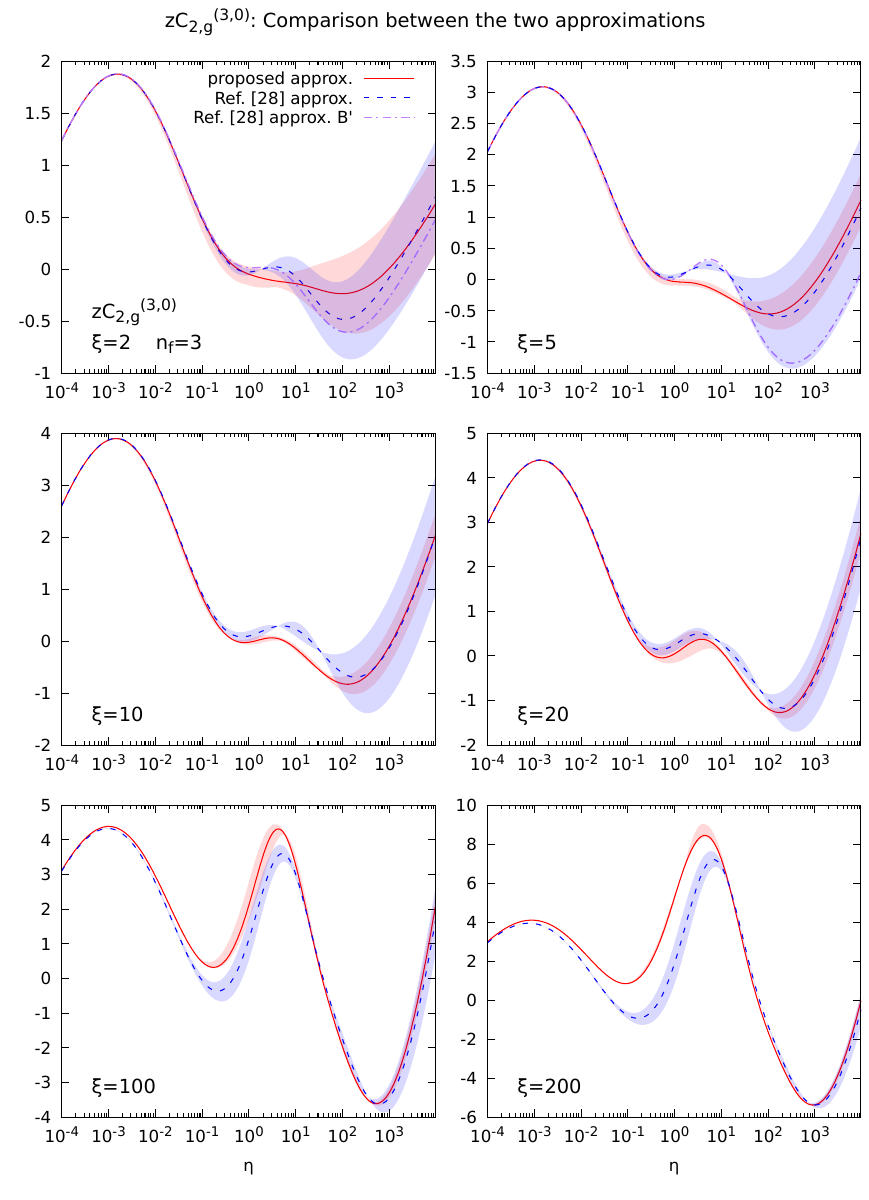}
\caption{Comparison between the $\mu$-independent part of the approximation that we propose (solid red) in \eqref{Dapprox:mu:indep} for the N$^3$LO massive gluon coefficient function for $F_2$, and the approximation given in Ref.~\cite{Kawamura_2012} (dashed blue), Eqs.~(\ref{N3LO:vogtA}) and (\ref{N3LO:vogtB}), as a function of $\eta$. Six relevant values of $\xi=Q^2/m^2$ are shown.}
\label{vogtN3LO}
\end{figure}In this case $C_{2,g}^{[3](3,0) \rm thresh'}$ is defined without the contribution $C_{2,g}^{[3](3,0) \rm const}$, that is the $\beta$-independent part of the threshold limit and is given in \eqref{c-const} (omitting the terms that depend on $\mu$).
$C_{2,g}^{[3,0](3,0)\rm A}$ and $C_{2,g}^{[3,0](3,0)\rm B}$ are the high-scale coefficient functions that use as $k_{cg}^{(3)0}$ the functions $k_{cg,A}^{(3)0}$ and $k_{cg,B}^{(3)0}$ given respectively in Eqs.~(\ref{kcgA}) and (\ref{kcgB}). The functions $C_{2,g}^{[3](2,0) \rm NLL,A}$ and $C_{2,g}^{[3](2,0) \rm NLL,B}$ are the two extremes of an uncertainty band of an approximation for the next-to-leading logarithm high-energy expansion of the gluon coefficient function. They are given by, converting to our normalization, i.e.\ multiplying a factor $4\xi/z$,
\begin{align}
    C_{2,g}^{[3](3,0)\rm NLL,A}={}&\frac{4\xi}{z}\Bigl[0.0007\Bigl(\frac{\log \xi}{\log 5}\Bigr)^4 -0.28\Bigr], \\
    C_{2,g}^{[3](3,0)\rm NLL,B}={}&\frac{4\xi}{z}\Bigl[0.055\Bigl(\frac{\log \xi}{\log 5}\Bigr)^2 -0.423\Bigr].
\end{align}
These expressions are extracted from the high-scale limit and from the high-scale limit of the high-energy limit of the exact coefficient function.
Hence, this approximation differs from the one we constructed also in the way the next-to-leading logarithm of the small-$z$ limit of the gluon coefficient function is estimated.
Fig.~\ref{vogtN3LO} shows the comparison between the approximation that we propose and the approximation proposed in Ref.~\cite{Kawamura_2012}. We used $n_f=3$, as we did previously.
The curve B$'$ is obtained using in \eqref{N3LO:vogtB} the function
\begin{equation}\label{B'}
    C_{2,g}^{[3](3,0)\rm NLL,B'}=\frac{C_A}{C_F}C_{2,q}^{[3](3,0)\rm NLL,B},
\end{equation}
with,
\begin{equation}
    C_{2,q}^{[3](3,0)\rm NLL,B}=\frac{4\xi}{z}\Bigl[0.0245\Bigl(\frac{\log \xi}{\log 5}\Bigr)^2 -0.17\Bigr],
\end{equation}
instead of $C_{2,g}^{[3](3,0)\rm NLL,B}$. \eqref{B'} is a correction of the approximation B of the N$^3$LO next-to-leading logarithm term for low $\xi$, while for high $\xi$ it is indistinguishable from it. Therefore, it is plotted only in the cases $\xi=2$ and $\xi=5$.

Comparing the two approximations, we can observe a perfect agreement in the small-$\eta$ region for the smallest values of $\xi$. However, this agreement slightly deteriorates as $\xi$ increases.
In the high-$\eta$ region (and therefore for small-$z$) our approximation, with its uncertainty band, always lies inside the band of the approximation of Ref.~\cite{Kawamura_2012}, but it slightly differs from its central value. Instead, for the intermediate values of $\eta$ the two approximations are quite different, especially at large $\xi$.

In conclusion, the comparison between our construction at NNLO and the one proposed in Ref.~\cite{Kawamura_2012} shows that at high-$\eta$ the first one is more accurate. Moreover, our approximation shows a better agreement with the NNLO exact function also for the intermediate values of $\eta$ and for $\xi \gg 1$. Therefore, we are confident that at N$^3$LO our approximation for the gluon coefficient function provides more accurate results. Anyway, having both results will surely provide an improvement of the accuracy of the knowledge of the N$^3$LO gluon coefficient function. In fact, combining both results we can construct a bigger uncertainty band, that will likely contain the exact result.

%%%%%%%%%%%%%%%%%%%%%%%%%%%%%%%%%%%%%%%%%%%%%%%%%%%%%%%%
\chapter{Conclusions}\label{concl}
The argument of this thesis is the construction of an approximation for the $\mu$-independent part of the DIS massive gluon coefficient function for $F_2$ in heavy quark pair production, that nowadays is still unknown. This contribution is a crucial ingredient for the construction of a VFNS at N$^3$LO, which in turn is needed for the extraction of the PDFs at N$^3$LO. 

Our approximation was constructed combining the known limits of the exact coefficient function in three kinematic limits, i.e.\ high-scale ($Q^2 \gg m^2$), high-energy ($z\rightarrow 0$, where $z$ is the argument of the coefficient function) and threshold ($z\rightarrow z_{\rm max}=1/(1+4m^2/Q^2)$), similarly to previous works in the literature.
In order to do so we constructed an asymptotic limit for the exact coefficient function: it approximates the exact curve in the small-$z$ region for all the values of $Q^2$. Such limit has been constructed reinserting in the high-scale coefficient function the power terms that had been neglected, in the small-$z$ limit. In this way we assured that for $z\rightarrow 0$ the asymptotic limit approaches the exact curve for all the values of $Q^2$. The power terms in the small-$z$ limit have been estimated from the high-energy limit and its $Q^2\gg m^2$ limit. In fact, subtracting these two functions, we found exactly the power terms in the limit $z\rightarrow 0$.

Once we constructed the asymptotic limit, in order to get our final approximation, we have combined it with the threshold limit using two damping functions. In this way our approximate curve approaches the exact coefficient function both for $z\rightarrow 0$ and for $z\rightarrow z_{\rm max}$. For intermediate values of $z$, the final curve is an interpolation between the two limits. The accuracy of our approximation in this region depends on the choice of the two damping functions. In order to choose the form of such functions, and to tune their parameters, we applied our approximation procedure to the NLO and NNLO massive coefficient function, that are exactly known. Thus, we chose the form that provided the best agreement between the approximate and the exact curves.

However, as we have already observed, even if we tuned precisely the approximation on the lower orders, we cannot be sure that at N$^3$LO the accuracy will be the same, since the N$^3$LO exact coefficient function will be a different function with respect to the NLO and NNLO.
For this reason, at N$^3$LO we expect a bigger uncertainty.
Moreover, while at NLO and NNLO all the ingredients we need are known, at N$^3$LO there are some contributions that are missing. In fact, at N$^3$LO the $\mu$-independent part of the the matching condition between heavy quark and gluon (that is needed for the construction of the high-scale limit), the term independent from $\beta$ in the threshold limit and the next-to leading logarithm small-$z$ expansion are all still unknown. Therefore, for these contributions we had to use approximate forms. This leads to a bigger uncertainty in the final approximate coefficient function. 
As soon as the exact expressions for the unknown terms will be available, it will be possible to update them in our construction, and this will improve the accuracy of our approximation.

Despite the lack of some contributions, we expect a good accuracy in the final approximation. In fact, with our procedure the approximate coefficient function approaches the exact function in the small-$z$ limit for all the values of $Q^2$, while other approximations that can be found in the literature do not. Moreover, our approximation for the NNLO gives a good description of the exact curve for all the values of $\xi$, so we expect the same to be true also at N$^3$LO.
%Furthermore, we have said that, being the N$^3$LO curve a different function with respect to the NLO and NNLO coefficient functions, we cannot be sure that the accuracy of our approximation will be as good as the one that it had on the lower orders in which we tuned the damping functions. Therefore, since our approach consist in a much more simple construction with respect to the other available approximations, we expect good precision also at N$^3$LO.

In conclusion, in this thesis we focused on the gluon coefficient function for $F_2$. In fact, since the gluon PDF is dominant at small-$z$ and the structure function $F_2$ is dominant with respect to $F_L$ in the whole range of $z$, the gluon coefficient function for $F_2$, convoluted with the gluon PDF, is one of the largest contribution in the computation of the hadronic cross sections. However, we can apply our approximation to the quark coefficient function and to the structure function $F_L$ in the same way we applied it to the gluon coefficient function for $F_2$. In fact, also the massive quark coefficient function at N$^3$LO is an important ingredient for the construction of any VFNS at $\ord{3}$. Moreover, in this case we need the N$^3$LO light quark-gluon matching condition, that is exactly known \cite{Alekhin_2017}. 
Once we will have constructed such coefficient functions, the accuracy of our description of DIS will be further increased since this result represents a fundamental ingredient for the determination of a future generation of PDFs at N$^3$LO.

%%%%%%%%%%%%%%%%%%%%%%%%%%%%%%%%%%%%%%%%%%%%%%%%%%%%%%%%%%%%%%%%%%%%%%%%%%%
%%%%%%%%%%%%%%%%%%%%%%%%%%%%%%%%%%%%%%%%%%%%%%%%%%%%%%%%%%%%%%%%%%%%%%%%%%%

\appendix
\chapter{Plus distribution}\label{plus:appendix}
In this appendix we will discuss the plus distribution and its main properties. It is needed to cancel the divergences for $z\rightarrow 1$, i.e.\ the soft divergences. 

\section{Definition}

Given a function $f(z)$ which is divergent and non integrable in $z=1$, then the distribution $\bigl[f(z)\bigr]_+$ is defined as
\begin{equation}\label{plus:dist}
    \int_0^1dz\,\bigl[f(z)\bigr]_+g(z)=\int_0^1dz\,f(z)\bigl(g(z)-g(1)\bigr),
\end{equation}
so that the divergences that arise due to the non integrability of $f(z)$ in $z=1$, exactly cancel for any test function $g(z)$ that is regular in $z=1$ or that diverges at most as
\begin{equation}
    \frac{1}{(1-z)^{\alpha}}, \quad \alpha < 2.
\end{equation}
In particular all the functions of the form
\begin{equation}\label{logk}
    f(z)=\frac{\log^k(1-z)}{1-z},
\end{equation}
with $k \geq 0$, are regularized.
From the definition in \eqref{plus:dist}, it follows that
\begin{equation}
    \bigl[f(z)\bigr]_+=f(z)-\delta(1-z)\int_0^1 dy\,f(y).
\end{equation}
Since $f(z)$ diverges as $z \rightarrow 1$, this definition makes sense only as a limit:
\begin{equation}
     \bigl[f(z)\bigr]_+=\lim_{\eta\rightarrow 0^+} \Bigl[ \theta\bigl(1-\eta -z\bigr)f(z) - \delta(1-z) \int_0^{1-\eta} dy\, f(y)  \Bigr],
\end{equation}
where the limit must be performed after the integration in $y$.

\section{Convolution}

Now we can discuss the convolution between a plus distribution and a test function. This is not trivial since \eqref{plus:dist} is written in terms of an integral from 0 to 1, while the definition of convolution, given in \eqref{conv}, is written in terms of an integral from $x$ to 1.
Such convolution is performed in the following way:
\begingroup
\allowdisplaybreaks
\begin{align}\label{convolution:plus}
    ([f]_+ \otimes g)(x) = &\int_x^1\frac{dz}{z}\,[f(z)]_+g\Bigl(\frac{x}{z}\Bigr)  = \notag \\
    =&\int_0^1dz\,\int_0^1dy\,[f(z)]_+g(y)\delta (x-zy) =  \notag \\
    =& \int_0^1dz\,\int_0^1dy \,f(z) g(y) \Bigl(\delta(x-zy)-\delta(x-y)\Bigr)=\notag \\
    =&\int_0^1dz\,f(z) \biggl(\int_0^1dy\,g(y) \delta(x-zy) - g(x) \biggr)= \notag \\
    =&\int_x^1\frac{dz}{z}\,f(z)g\Bigl(\frac{x}{z}\Bigr) - g(x)\int_0^1dz\,f(z).
\end{align}
\endgroup
Since \eqref{convolution:plus} is composed by two divergent quantities, it must be interpreted as a limit, i.e.\
\begin{equation}
    ([f]_+ \otimes g)(x) = \lim_{\eta \rightarrow 0^+} \biggl[ \int_x^{1-\eta}\frac{dz}{z}\,f(z)g\Bigl(\frac{x}{z}\Bigr) - g(x)\int_0^{1-\eta} dz\,f(z) \biggr].
\end{equation}
\eqref{convolution:plus} is usually written in the form
\begin{equation}\label{conv:plus}
    ([f]_+ \otimes g)(x)=\int_x^1 dz f(z)\biggl(\frac{1}{z}g\Bigl(\frac{x}{z}\Bigr)-g(x)\biggr)-g(x)\int_0^xdzf(z).
\end{equation}
The advantage of writing it in this way is the following: the two integrals of \eqref{convolution:plus} are both divergent because $f(z)$ is not integrable in $z=1$. For this reason we had to define \eqref{convolution:plus} as a limit.
However, the final result is finite so the divergences must exactly cancel. 
Instead, the two integrals of \eqref{conv:plus} are both finite: in fact the second one is trivially finite because we have that $x<1$ and therefore the integral doesn't extends up to the point $z=1$, while the first one is finite because near the point $z=1$ we have that, expanding in $z$, the integrand is
\begin{equation*}
    f(z)\biggl[g(x)+\Bigl(-g(x)-xg'(x)\Bigr)(z-1)-g(x)\biggr]=f(z)\Bigl[-g(x)-xg'(x)\Bigr](z-1).
\end{equation*}
The term $(z-1)$ simplifies with a term $\frac{1}{1-z}$ that is always present inside $f(z)$ (we will consider only functions like the one in \eqref{logk}), so that the first integral in \eqref{conv:plus} is convergent for any function $g(z)$ that is derivable for $0<z<1$.

\section{Convolution between \texorpdfstring{$D_c^{[4](1)}(m=0)$}{Dc1} and \texorpdfstring{$K_{cg}^{(1)}$}{Kcg3}}

In order to show that computing the convolution of a plus distribution with a smooth function, using the form in \eqref{conv:plus}, gives a perfectly finite result, we will compute the convolution between $D_c^{[4](1)}(m=0)$ and $K_{cg}^{(1)}$. 
So, what we want to compute is
\begin{equation}\label{conv:CK}
    ( D_c^{[4](1)}(0)\otimes K_{cg}^{(1)})(x)=\int_x^1 \frac{dz}{z}D_c^{[4](1)}(z,0)K_{cg}^{(1)}\Bigl(\frac{x}{z},\frac{m^2}{Q^2}\Bigr).
\end{equation}
Such computation is present in the high-scale coefficient function of the gluon at $\ord{2}$ for $\mu=Q$, see \eqref{2ord:m0}. $D_c^{[4](1)}(m=0)$ is given in \eqref{Cq1}, while the expression of $K_{cg}^{(1)}$ is
\begin{equation}
     K_{cg}^{(1)}\Bigl(z,\frac{m^2}{Q^2}\Bigr)=\frac{2T_F}{4\pi}(z^2+(1-z)^2)\log\frac{Q^2}{m^2}.
\end{equation}
For simplicity we will consider the case $\mu=Q$ but the generalization to $\mu \neq Q$ is straightforward.
In order to compute \eqref{conv:CK} we can use \eqref{conv:plus} for the terms that involve the plus distribution (i.e. the first two terms of \eqref{Cq1}), and \eqref{conv} to compute the regular terms.

For the computation of the first two terms of \eqref{Cq1} we can proceed as follows:
we have to apply \eqref{conv:plus} with $f(z)=\frac{\log(1-z)}{1-z}$ or $f(z)=\frac{1}{1-z}$ and with $g(z)=z^2+(1-z)^2$. Since we have that
\begin{align}
\frac{1}{1-z}&=(1-z)^{\eta-1} \big|_{\eta=0} ,\\
   \frac{\log(1-z)}{1-z}&=\frac{d}{d\eta}(1-z)^{\eta-1} \big|_{\eta=0},
\end{align}
we will compute \eqref{conv:plus} with $f(z)=(1-z)^{\eta-1}$ and then we will take the zeroth and the first order expansions around $\eta=0$. Therefore we want to compute 
\begin{equation}\label{conv:eta}
    \int_x^1 dz\,(1-z)^{\eta-1} \biggl[\frac{1}{z}\Bigl(2\frac{x^2}{z^2}-2\frac{x}{z}+1\Bigr)-(2x^2-2x+1)\biggr]-(2x^2-2x+1)\int_0^xdz\,(1-z)^{\eta-1}.
\end{equation}
Let's focus first on the last integral of \eqref{conv:eta}: with the change of variable $t=1-z$ we can write that
\begin{align*}
    \int_0^xdz\,(1-z)^{\eta-1}=\int_{1-x}^1dt\, t^{\eta-1}=\frac{t^\eta}{\eta}\bigg|_{1-x}^1=\frac{1-(1-x)^\eta}{\eta}=
\end{align*}
expanding in powers of $\eta$
\begin{align*}
    &=\frac{1}{\eta}\biggl[1-\Bigl(1+\eta \log(1-x)+\frac{\eta^2}{2}\log^2(1-x)+\frac{\eta^3}{3!}\log^3(1-x)+\dots\Bigr) \biggr]= \\
    &=-\log(1-x)-\frac{\eta}{2}\log^2(1-x)-\frac{\eta^2}{3!}\log^3(1-x)-\dots
\end{align*}
In the end what we have found is
\begin{equation*}
    \int_0^xdz\, \frac{\log^k(1-z)}{1-z}=
    \begin{cases}
    -\log(1-x)\quad &k=0 \\
   -\frac{1}{2}\log^2(1-x) \quad &k=1 \\
   -\frac{1}{3}\log^3(1-x) \quad &k=2 \\
    \qquad\quad \vdots &\quad\vdots \\
    \end{cases}
\end{equation*}
so for a general $k$ we have that
\begin{equation}
    \int_0^xdz\, \frac{\log^k(1-z)}{1-z}= -\frac{1}{k+1} \log^{k+1}(1-x).
\end{equation}

Now let's see the first integral of \eqref{conv:eta}: with some easy mathematical manipulations we can write it in the form
\begin{equation*}
    \int_x^1 dz\,(1-z)^{\eta-1}\biggl(2x^2\frac{(1-z)(1+z+z^2)}{z^3}-2x\frac{(1-z)(1+z)}{z^2}+\frac{1-z}{z}\biggr),
\end{equation*}
so that a common term $1-z$ factorizes and cancels with the $(1-z)^{-1}$ in front. Therefore the integral becomes
\begin{align*}
    \int_x^1 dz\,(1-z)^{\eta} \biggl(2\frac{x^2}{z^3}+\frac{2x^2-2x}{z^2}+\frac{2x^2-2x+1}{z}\biggr).
\end{align*}
This last expression shows that we can limit ourselves to the evaluation of integrals of the form 
\begin{equation*}
    \int_x^1 dz\,(1-z)^{\eta}z^k,
\end{equation*}
with $k=-1,-2,-3$.
For $f(z)=\frac{1}{1-z}$ we can take $\eta=0$ before the integration so that we are left with 
\begin{equation*}
    \int_x^1dz\,z^k,
\end{equation*}
that is trivial. So for this term the result is
\begin{align}
    \biggl( \Bigl(\frac{1}{1-z}\Bigr)_+\otimes \bigl(z^2+(1-z)^2\bigr)\biggr)(x)=&-1 + 4 x - 
 3 x^2 + (1 - 2 x + 2 x^2) \log(1 - x) +\notag \\
 &+(-1 + 2 x - 2 x^2) \log(x).
\end{align}
For $f(z)=\frac{\log(1-z)}{1-z}$, using the properties of the Gamma function, we find
\begin{equation}\label{Gamma}
    \int_0^1dz(1-z)^\eta z^k=\frac{\Gamma(\eta+1)\Gamma(1+k)}{\Gamma(2+k+\eta)}-\beta(x,1+k,1+\eta),
\end{equation}
where $\beta(x,1+k,1+\eta)$ is the incomplete beta function, defined as
\begin{equation}
    \beta(x,a,b)=\int_0^xdt\, t^{a-1}(1-t)^{b-1}.
\end{equation}
Computing the first order expansion around $\eta=0$ of \eqref{Gamma}, sending $k$ to $-1,-2,-3$ and then putting everything together, we find that
\begin{align}
     \biggl( \Bigl(\frac{\log(1-z)}{1-z}\Bigr)_+\otimes \bigl(z^2+(1-z)^2\bigr)\biggr)(x)={}& -(1-x)x-\frac{1}{6}\pi^2(2x^2-2x+1)+x(3x-2)\log x \notag \\
     &+(2x^2-2x+1){\rm Li}_2(x) +\Bigl[\frac{1}{2}-(1-x)x\Bigr]\log^2(1-x)\notag \\
     & +\Bigl[-1+(4-3x)x\Bigr]\log(1-x) ,
\end{align}
where ${\rm Li}_2(x)$ is the dilogarithm, defined as
\begin{equation}
    {\rm Li}_2(x)=-\int_0^xdt\,\frac{\log(1-t)}{t}.
\end{equation}

Now that we have computed the singular terms of \eqref{Cq1}, we can apply the standard definition of convolution, i.e \eqref{conv}, in order to compute the regular ones. In conclusion the full result of \eqref{conv:CK} is 

\begin{align}
    ( D_c^{[4](1)}(0)\otimes K_{cg}^{(1)})(x)={}&\frac{2C_F2T_F}{(4\pi)^2}\biggl\{ -\frac{5}{2} + 2 (3 - 4 x) x +\frac{\pi^2}{6} (-1 + 2 x - 4 x^2)  \notag\\
   & +\log^2(1 - x) \Bigl[1 - 2 (1 - x) x\Bigr]  - 
   \frac{1}{2}\log(1 - x) \biggl[7 + 4 x (3 x-4) \notag \\
   &+ \Bigl(4 -8 (1 - x) x\Bigr) \log(x)\biggr] +\frac{1}{2} \log(x) \Bigl[-1 + 4 x (3x-2)\notag \\
   & + (1 - 2 x + 4 x^2) \log(x)\Bigr] + 
   (2x-1) {\rm Li}_2(x) \biggr\}\log \frac{Q^2}{m^2}.
\end{align}

The same procedure can be applied to the computation of the convolution $D_c^{[4](2)}(0)\otimes K_{cg}^{(1)}$ that is present in \eqref{3ord:m0}. In this case we still have to compute \eqref{conv:plus} but we have also the contribution from the terms $f(z)=\frac{\log^2(1-z)}{1-z}$ and $f(z)=\frac{\log^3(1-z)}{1-z}$ \cite{Vermaseren_2005}. This is done observing that
\begin{align}
    \frac{\log^2(1-z)}{1-z}&=\frac{d^2}{d\eta^2}(1-z)^{\eta-1} \big|_{\eta=0}, \\
    \frac{\log^3(1-z)}{1-z}&=\frac{d^3}{d\eta^3}(1-z)^{\eta-1} \big|_{\eta=0},
\end{align}
and proceeding in an analogous way.

\chapter{Approximate NLL small-\textit{z} expansion of the gluon coefficient function}\label{high:energy:NLL}
In this appendix we will construct an approximation for the NLL small-\textit{z} expansion of the gluon coefficient function. Moreover, we will construct an error band associated to it. We will follow the approach presented in Ref.~\cite{Bonvini:2018xvt, Bonvini_2018}.
Considering just the gluon contribution, in Mellin space the hadronic structure function $F_2$ can be written as
\begin{equation}
    F_{2,N}(\xi)=\hat{F}_{2,N}(\xi) f_{g,N}(\mu^2),\label{impact}
    %=\alpha e_q^2C_{2,g}(N,\mu^2,\alpha_s)f_{g,N}(\mu^2)
\end{equation}
with 
\begin{equation}\label{Fpart:mell}
   \hat{F}_{2,N}(\xi)= K_{2,N}(\xi,\gamma)h(\gamma)\Bigl(\frac{m^2}{\mu^2}\Bigr)^\gamma,
\end{equation} 
where $K_{2,N}(\xi)$ and $h(\gamma)$ are given in Ref.~\cite{Kawamura_2012} and read
\begin{align}
    h(\gamma)={}&\pi \alpha_{em} \alpha_s e_q^2 \biggl[\frac{7}{9} + \frac{41}{27}\gamma +\frac{244}{81}\gamma^2 +\mathcal{O}(\gamma^3)\biggr], \\
    K_{2,N}(\xi,\gamma)={}&\Bigl(1+\frac{\xi}{4}\Bigr)^{-N}\frac{3}{(7-5\gamma)(1+2\gamma)}\biggl[\frac{2}{\xi}(1+\gamma) + \Bigl(1+\frac{\xi}{4}\Bigr)^{\gamma-1}\Bigl(2+3\gamma \notag \\
    &- 3\gamma^2 -\frac{2}{\xi}(1+\gamma)\Bigr) {}_2F_1\Bigl(1-\gamma,\frac{\xi}{\xi +4}\Bigr)\biggr]. \label{K_2}
\end{align}
The function ${}_2F_1$ satisfies the expansion 
\begin{align}
    {}_2F_1\bigl(1-\gamma,z^2\bigr)={}&\frac{1}{2z}\biggl\{L(z) + \gamma \biggl[H(-,+,z)+L(z)\log(1-z^2)\biggr] - \frac{1}{2}\gamma^2\biggl[H(-,+,-,z) \notag \\
    &-H(-,+,z)\log(1-z^2)-L(z)\log^2(1-z^2)\biggr]+\mathcal{O}(\gamma^3)\biggr\},
\end{align}
where $z=\sqrt{\xi/(\xi+4)}$ and $L(z)$, $H(-,+,z)$ and $H(-,+,-,z)$ are defined in Eqs.~(\ref{Lz}), (\ref{Hpmz}) and (\ref{Hmpmz}).
Now we can expand \eqref{Fpart:mell} for small $\gamma$ up to second order. What we find is
\begin{equation}\label{Fpart:mell:exp}
    \hat{F}_{2,N}(\xi)= \hat{F}_{2,N}^{(0)}+\gamma \hat{F}_{2,N}^{(1)}+ \gamma^2 \hat{F}_{2,N}^{(2)}+\mathcal{O}(\gamma^3).
\end{equation}
Then we make the substitution \cite{Ball:2007ra, Altarelli:2008aj}
\begin{align}
   \gamma^0 & \rightarrow \bigl[\gamma^0\bigr] =1 ,\\
   \gamma^1 & \rightarrow \bigl[\gamma^1\bigr] =\gamma=\alpha_s \gamma_0 +\alpha_s^2 \gamma_1 +\mathcal{O}(\alpha_s^3) ,\\
   \gamma^2 & \rightarrow \bigl[\gamma^2\bigr]=\gamma(\gamma-\alpha_s\beta_0) =\alpha_s^2(\gamma^2_0- \gamma_0\beta_0) +\mathcal{O}(\alpha_s^3) .
\end{align}
With these definitions \eqref{Fpart:mell:exp} becomes
\begin{equation}\label{Fpart:exp2}
    \hat{F}_{2,N}(\xi)= \hat{F}_{2,N}^{(0)}+\alpha_s \hat{F}_{2,N}^{(1)} \gamma_0+ \alpha_s^2\Bigl(\hat{F}_{2,N}^{(1)} \gamma_1 + \hat{F}_{2,N}^{(2)}\bigl(\gamma_0^2-\gamma_0\beta_0\bigr)\Bigr).
\end{equation}
Observe that since $h(\gamma)$ is $\ord{}$, then all the $\hat{F}_{2,N}^{(k)}$ are $\ord{}$. It means that the term proportional to $\alpha_s^2$ in \eqref{Fpart:exp2} is the $\ord{3}$ expansion of the partonic structure function of the gluon in Mellin space.
Moreover, observe that this procedure must be done in the limit $N \ll \gamma$, and therefore the term $\left(1+\frac{\xi}{4}\right)^{-N}$ in Eq.~\eqref{K_2} can be replaced with 1.
In order to find our approximate result we will use the values \cite{Bonvini:2018xvt, Bonvini_2018}
\begin{align}
    \gamma^{\rm NLL}_0&=\frac{a_{11}}{N}+\frac{a_{10}}{N+1}, \\
    \gamma^{\rm NLL}_1&=\frac{a_{21}}{N}-\frac{2a_{21}}{N+1},
\end{align}
where the coefficients $a_{11}$, $a_{10}$ and $a_{21}$ are given by 
\begin{align}
    a_{11}&=\frac{C_A}{\pi}, \\
    a_{10}&=-\frac{11C_A + 2n_f(1-2C_F/C_A)}{12\pi}, \\
    a_{21}&=n_f\frac{26C_F - 23C_A}{36 \pi^2}.
\end{align}
Now we have to transform back from Mellin space to $z$-space. Using that the Mellin transform is defined as
\begin{equation}
    \mathcal{M}\bigl[f(z)\bigr]\equiv f(N)=\int_0^1dz\,z^{N-1}f(z),
\end{equation}
one can show that
\begin{align}
    \mathcal{M}\bigl[1\bigr]&=\frac{1}{N}, \\
    \mathcal{M}\bigl[z\bigr]&=\frac{1}{N+1}, \\
    \mathcal{M}\bigl[\log(z)\bigr]&=-\frac{1}{N^2}, \\
    \mathcal{M}\bigl[z\log(z)\bigr]&=-\frac{1}{(N+1)^2}. 
\end{align}
We find that, neglecting terms proportional to $z$ and $z\log(z)$, and then dividing a factor $z$ in order to find the coefficient function, our approximation of the next-to-leading logarithm expansion of the $\ord{3}$ gluon coefficient function for $F_2$ is
\begingroup
\allowdisplaybreaks
\begin{align}
		z(4\pi)^3\, &C_{2,g}^{[3](3) \rm NLL}\Bigl(z,\frac{m^2}{Q^2},\frac{m^2}{\mu^2}\Bigr)= \notag \\
		& a_{11}^2\Biggl[-\frac{147}{27} - \frac{8}{3} K(\xi) \left(\frac{1}{\xi}-1\right) + \frac{8}{27} J(\xi) \left(\frac{92}{\xi} - 71\right) + I(\xi) \left(\frac{8}{3} L_\xi \left(\frac{1}{\xi}-1\right) + \frac{8}{9} \left(\frac{10}{\xi} - 13\right)\right) \notag\\
		& \qquad+ \left(-\frac{160}{9} + \frac{16}{3} I(\xi) \left(\frac{1}{\xi}-1\right) + \frac{8}{9} J(\xi) \left(\frac{10}{\xi} - 13\right)\right) L_\mu + \left(-\frac{16}{3} + \frac{8}{3} J(\xi) \left(\frac{1}{\xi}-1\right)\right) L_\mu^2\Biggr] \log(z) \notag\\
		& + a_{21} \left[\frac{16}{9} - \frac{16}{3} I(\xi) \left(\frac{1}{\xi}-1\right) - \frac{8}{9} J(\xi) \left(\frac{10}{\xi} - 13\right)\right] \notag\\
		& + a_{10} a_{11} \Bigl[\frac{2944}{27} + \frac{16}{3} K(\xi) \left(\frac{1}{\xi}-1\right) - \frac{16}{27} J(\xi) \left(\frac{92}{\xi} - 71\right) \notag \\
  &\qquad + I(\xi) \left(-\frac{16}{3} L_\xi \left(\frac{1}{\xi}-1\right) - \frac{16}{9} \left(\frac{10}{\xi} - 13\right)\right)\Bigr] \notag\\
		& + a_{11} \beta_0 \Bigl[-\frac{1472}{27} - \frac{8}{3} K(\xi) \left(\frac{1}{\xi}-1\right) + \frac{8}{27} J(\xi) \left(\frac{92}{\xi} - 71\right) \notag \\
  &\qquad + I(\xi) \left(\frac{8}{3} L_\xi \left(\frac{1}{\xi}-1\right) + \frac{8}{9} \left(\frac{10}{\xi} - 13\right)\right)\Bigr] \notag\\
		& + \Biggl[a_{21} \left(\frac{32}{3} - \frac{16}{3} J(\xi) \left(\frac{1}{\xi}-1\right) \right) + a_{10} a_{11} \left(\frac{320}{9} - \frac{32}{3} I(\xi) \left(\frac{1}{\xi}-1\right) - \frac{16}{9} J(\xi) \left(\frac{10}{\xi} - 13\right)\right) \notag\\
		& \qquad+ a_{11} \beta_0 \left(-\frac{160}{9} + \frac{16}{3} I(\xi) \left(\frac{1}{\xi}-1\right) + \frac{8}{9} J(\xi) \left(\frac{10}{\xi} - 13\right)\right) \Biggr] L_\mu \notag\\
		& + \left[ a_{10} a_{11} \left(\frac{32}{3} - \frac{16}{3} J(\xi) \left(\frac{1}{\xi}-1 \right)\right) + a_{11} \beta_0 \left(-\frac{16}{3} + \frac{8}{3} J(\xi) \left(\frac{1}{\xi}-1\right)\right)\right] L_\mu^2 , \label{CNLL}
\end{align}
\endgroup
where $L_\mu=\log(m^2/\mu^2)$, $L_\xi=\log(1+\xi/4)$ with $\xi=Q^2/m^2$.
The functions $I(\xi)$, $J(\xi)$ and $K(\xi)$ are defined in Eqs.~(\ref{I}-\ref{K}).
Taking the $Q^2 \gg m^2$ limit of \eqref{CNLL} we find its high-scale limit, that is
\begingroup
\allowdisplaybreaks
\begin{align}
   z(4\pi)^3\,& C_{2,g}^{[3,0](3) \rm NLL}\Big(z,\frac{m^2}{Q^2},\frac{m^2}{\mu^2}\Bigr)= \notag \\
   & a_{11}^2 \Biggl[\frac{16}{27} \left(13 \pi^2 -92- 72 \zeta_3\right) -\frac{32}{27} \left(3 \pi^2 - 71\right) L_Q - \frac{208}{9} L_Q^2 + \frac{32}{9} L_Q^3 \notag \\
		& \qquad + L_\mu \left(\frac{32}{9} \left( \pi^2 - 5\right) + \frac{416}{9} L_Q - \frac{32}{3} L_Q^2 \right) + L_\mu^2 \left(-\frac{16}{3} + \frac{32}{3} L_Q\right)\Biggr] \log(x) \notag\\
		& - \frac{32}{27} a_{10} a_{11} \left(13 \pi^2 -92- 72 \zeta_3\right) + \frac{16}{27} a_{11} \beta_0 \left(13 \pi^2 -92- 72 \zeta_3\right) -\frac{32}{9} a_{21} (\pi^2 - 5)\notag\\
		& + \left(-\frac{416}{9} a_{21} + \frac{64}{27} a_{10} a_{11} (3 \pi^2 - 71) - \frac{32}{27} a_{11} \beta_0 \left(3 \pi^2 - 71\right)\right) L_Q \notag\\
		& + \left( \frac{416}{9} a_{10} a_{11} + \frac{32}{3} a_{21} - \frac{208}{9} a_{11} \beta_0 \right) L_Q^2 + \left(-\frac{64}{9} a_{10} a_{11} + \frac{32}{9} a_{11} \beta_0 \right) L_Q^3 \notag\\
		& + L_\mu \Biggl( \frac{32}{3} a_{21} - \frac{64}{9} a_{10} a_{11} \left(\pi^2 - 5\right) + \frac{32}{9} a_{11} \beta_0 \left( \pi^2 - 5\right)  \notag\\
		& \qquad + \left(-\frac{832}{9} a_{10} a_{11} - \frac{64}{3} a_{21} + \frac{416}{9} a_{11} \beta_0 \right)L_Q+ \left( \frac{64}{3} a_{10} a_{11} - \frac{32}{3} a_{11} \beta_0\right) L_Q^2\Biggr) \notag\\
		& + L_\mu^2 \left( \frac{32}{3} a_{10} a_{11} - \frac{16}{3} a_{11} \beta_0 + \left(-\frac{64}{3} a_{10} a_{11} + \frac{32}{3} a_{11} \beta_0 \right) L_Q \right), \label{CNLL:hs}
\end{align}
\endgroup
where $L_Q=\log(m^2/Q^2)$.

Now we have to construct the uncertainty band for this approximation. It will be done using \eqref{Fpart:exp2} with \cite{Bonvini:2018xvt, Bonvini_2018}
\begin{align}
    \gamma_0^{\rm LL'}&=\frac{a_{11}}{N}+\frac{a_{10}}{N+1}, \\
    \gamma^{\rm LL'}_1&=\beta_0 a_{11}\Bigl(\frac{21}{8}\zeta_3-4\log 2\Bigr)\Bigl(\frac{1}{N}-\frac{4N}{(N+1)^2}\Bigr).
\end{align}
In this way we find the curve 
\begin{align}
     z(4\pi)^3\, & C_{2,g}^{[3](3) \rm NLL'}\Big(z,\frac{m^2}{Q^2},\frac{m^2}{\mu^2}\Bigr) = \notag \\
     & a_{11}^2\Biggl[-\frac{147}{27} - \frac{8}{3} K(\xi) \left(\frac{1}{\xi}-1\right) + \frac{8}{27} J(\xi) \left(\frac{92}{\xi} - 71\right) + I(\xi) \left(\frac{8}{3} L_\xi \left(\frac{1}{\xi}-1\right) + \frac{8}{9} \left(\frac{10}{\xi} - 13\right)\right)\notag \\
		& \qquad + \left(-\frac{160}{9} + \frac{16}{3} I(\xi) \left(\frac{1}{\xi}-1\right) + \frac{8}{9} J(\xi) \left(\frac{10}{\xi} - 13\right)\right) L_\mu \notag \\
  &\qquad + \left(-\frac{16}{3} + \frac{8}{3} J(\xi) \left(\frac{1}{\xi}-1\right)\right) L_\mu^2\Biggr] \log(z) \notag\\
		& + a_{10} a_{11} \Biggl[\frac{2944}{27} + \frac{16}{3} K(\xi) \left(\frac{1}{\xi} - 1\right) - \frac{16}{27} J(\xi) \left(\frac{92}{\xi} - 71\right) \notag \\
  &\qquad + I(\xi) \left(-\frac{16}{3} L_\xi \left(\frac{1}{\xi} - 1\right) - \frac{16}{9} \left(\frac{10}{\xi} - 13\right)\right) \Biggr]\notag \\
		& + a_{11} \beta_0 \Biggl[ -\frac{1472}{27} - \frac{8}{3} K(\xi) \left(\frac{1}{\xi} - 1\right) - \frac{640}{9} \log 2 + \frac{140}{3} \zeta_3 \notag\\
		& \qquad + I(\xi) \left(\frac{8}{3} L_\xi \left(\frac{1}{\xi} - 1\right) + \frac{8}{9} \left(\frac{10}{\xi} - 13\right) + \frac{64}{3} \left(\frac{1}{\xi} - 1\right) \log 2 - 14 \left(\frac{1}{\xi} - 1\right) \zeta_3\right) \notag\\
		& \qquad + J(\xi) \left(\frac{8}{27} \left(\frac{92}{\xi} - 71\right) + \frac{32}{9} \left(\frac{10}{\xi} - 13\right) \log 2 - \frac{7}{3} \left(\frac{10}{\xi} - 13\right) \zeta_3 \right)\Biggr] \notag\\
		& + L_\mu \Biggl[a_{10} a_{11} \left(\frac{320}{9} - \frac{32}{3} I(\xi) \left(\frac{1}{\xi} - 1\right) - \frac{16}{9} J(\xi) \left(\frac{10}{\xi} - 13\right)\right)\notag \\
		& \qquad + a_{11} \beta_0 \Biggl(-\frac{160}{9} + \frac{16}{3} I(\xi) \left(\frac{1}{\xi} - 1\right) - \frac{28}{3} \log 2 + 28 \zeta_3 \notag\\
		& \qquad \qquad+ J(\xi) \left(\frac{8}{9} \left(\frac{10}{\xi} - 13\right) + \frac{64}{3} \left(\frac{1}{\xi} - 1\right) \log 2 - 14 \left(\frac{1}{\xi} - 1\right) \zeta_3 \right)\Biggr)\Biggr]\notag\\
		& + \Biggl[a_{10} a_{11} \left(\frac{32}{3} - \frac{16}{3} J(\xi) \left(\frac{1}{\xi} - 1\right)\right) + a_{11} \beta_0 \left(-\frac{16}{3} + \frac{8}{3} J(\xi) \left(\frac{1}{\xi} - 1\right)\right)\Biggr] L_\mu^2,
\end{align}
whose high-scale limit is
\begingroup
\allowdisplaybreaks
\begin{align}
     z(4\pi)^3\, & C_{2,g}^{[3,0](3) \rm NLL'}\Big(z,\frac{m^2}{Q^2},\frac{m^2}{\mu^2}\Bigr) = \notag \\
     & a_{11}^2 \Biggl[\frac{16}{27} \left(13 \pi^2 -92- 72 \zeta_3\right) -\frac{32}{27} \left(3 \pi^2 - 71\right) L_Q - \frac{208}{9} L_Q^2 + \frac{32}{9} L_Q^3 \notag \\
		& \qquad + L_\mu \left(\frac{32}{9} \left( \pi^2 - 5\right) + \frac{416}{9} L_Q - \frac{32}{3} L_Q^2 \right) + L_\mu^2 \left(-\frac{16}{3} + \frac{32}{3} L_Q\right)\Biggr] \log(x) \notag\\
		& + \frac{4}{27} a_{11} \beta_0 \left(-368 + 52 \pi^2 - 480 \log 2 + 96 \pi^2 \log 2 + 27 \zeta_3 - 63 \pi^2 \zeta_3\right) \notag \\
  &- \frac{32}{27} a_{10} a_{11} \left(13 \pi^2 -92- 72 \zeta_3\right) \notag\\
		& + L_Q \left( \frac{64}{27} a_{10} a_{11} (3 \pi^2 - 71) - \frac{4}{27} a_{11} \beta_0 (-568 + 24 \pi^2 - 1248 \log 2 + 819 \zeta_3) \right) \notag\\
		& + L_Q^2 \left(\frac{416}{9} a_{10} a_{11} - \frac{4}{9} a_{11} \beta_0 \left(52 + 96 \log 2 - 63 \zeta_3 \right)\right) + \left(- \frac{64}{9} a_{10} a_{11} + \frac{32}{9} a_{11} \beta_0 \right) L_Q^3 \notag\\
		& L_\mu \Biggl[-\frac{64}{9} a_{10} a_{11} \left( \pi^2 - 5\right) + \frac{4}{9} a_{11} \beta_0 (-40 + 8 \pi^2 - 96 \log 2 + 63 \zeta_3)\notag \\
		& \qquad + L_Q \left(-\frac{832}{9} a_{10} a_{11} + \frac{8}{9} a_{11} \beta_0 (52 + 96 \log 2 - 63 \zeta_3)\right) + \left(\frac{64}{3} a_{10} a_{11} - \frac{32}{3} a_{11} \beta_0 \right) L_Q^2 \Biggr]\notag \\
		& + L_\mu^2 \Biggl[\frac{32}{3} a_{10} a_{11} - \frac{16}{3} a_{11} \beta_0 + \left(-\frac{64}{3} a_{10} a_{11} + \frac{32}{3} a_{11} \beta_0 \right) L_Q \Biggr].
\end{align}
\endgroup
In conclusion, our uncertainty will be given by 
\begin{equation}
\Delta= \bigl|C_{2,g}^{[3](3) \rm NLL}- C_{2,g}^{[3](3) \rm NLL'}\bigr|,
\end{equation}
and the same holds for the high-scale limit.

%%%%%%%%%%%%%%%%%%%%%%%%%%%%%%%%%%%%%%%%%%%%%%%%%%%%%%%%%%%%%%%%%%%%%%%
\chapter{Functions \texorpdfstring{$c_0(\xi)$}{c0} and \texorpdfstring{$\bar{c}_0(\xi)$}{cbar0}}\label{c:cbar}

In this Appendix we give the functions $c_0(\xi)$ and $\bar{c}_0(\xi)$ appearing in Eqs.~(\ref{thresh:ord2}) and (\ref{thresh:ord2}). They are presented in Ref.~\cite{Kawamura_2012} and their expression is
\begin{align}
    c_0(\xi)&= C_A  \bigg\{50 - \pi^2 + \frac{12}{y}\log\bigl(\sqrt{\xi}(y - 1)/2\bigr) + 4\log^2\bigl(\sqrt{\xi}(y - 1)/2\bigr) + \log^2\bigl(1 + \xi/2\bigr) \notag\\
  &+ 6\log\bigl(2 + \xi/2\bigr)- 4\log^2\bigl(2 + \xi/2\bigr) + 2{\rm Li}_2\Bigl( - \frac{2}{2 + \xi}\Bigr) + \frac{48}{2 + \xi} - 4\frac{\log\bigl(2 + \xi/2\bigr)}{2 + \xi} \notag\\
  &+ 64\frac{\log \bigl(2 + \xi/2\bigr)}{(2 + \xi)^2} - 128\frac{\log\bigl(2 + \xi/2\bigr)}{(2 + \xi)^2(4 + \xi)} - \frac{160}{(2 + \xi)(4 + \xi)} - 64\frac{\log\bigl(2 + \xi/2\bigr)}{(2 + \xi)(4 + \xi)} \notag\\
  &+ \frac{128}{(2 + \xi)(4 + \xi)^2} - 12\frac{4 + \zeta_2}{4 + \xi} - 8\frac{\log^2\bigl(\sqrt{\xi}(y - 1)/2\bigr)}{4 + \xi} + \frac{64}{(4 + \xi)^2} \bigg\} \notag \\
 & + C_F  \bigg\{- 18 - \frac{2}{3} \pi^2 - \frac{24}{y} \log \bigl(\sqrt{\xi}(y - 1)/2\bigr) - 8\log^2\bigl(\sqrt{\xi}(y - 1)/2\bigr) + 2\log^2\bigl(1 + \xi/2\bigr)\notag\\
  &- 6\log\bigl(2 + \xi/2\bigr) + 4{\rm Li}_2\Bigl( - \frac{2}{2 + \xi}\Bigr) - \frac{48}{2 + \xi} + 8\frac{\log \bigl(2 + \xi/2\bigr)}{2 + \xi} + \frac{360}{(2 + \xi)(4 + \xi)}\notag\\
 & + 128\frac{\log\bigl(2 + \xi/2\bigr)}{(2 + \xi)(4 + \xi)} - \frac{544}{(2 + \xi)(4 + \xi)^2} + 48\frac{\log^2\bigl(\sqrt{\xi}(y - 1)/2\bigr)}{4 + \xi}  - 8\frac{\log^2\bigl(1 + \xi/2\bigr)}{4 + \xi} \notag\\
  &+ \frac{44 + 40\zeta_2}{4 + \xi} - 120\frac{\log\bigl(2 + \xi/2\bigr)}{(2 + \xi)^2} + 256\frac{\log\bigl(2 + \xi/2\bigr)}{(2 + \xi)^2(4 + \xi)}  - \frac{16}{(4 + \xi)}{\rm Li}_2\Bigl( - \frac{2}{2 + \xi}\Bigr) \notag \\
  &- \frac{272}{(4 + \xi)^2} \bigg\} ,\label{c0}\\
  \bar{c}_0(\xi)&=4C_A \biggl[2+\log\Bigl(1+\frac{\xi}{4}\Bigr)\biggr] -\frac{4}{3}T_F ,\label{c0bar}
\end{align}
where 
\begin{equation}
    y=\sqrt{1+\frac{4}{\xi}}.
\end{equation}

\backmatter
\cleardoublepage
\phantomsection % Give this command only if hyperref is loaded
\addcontentsline{toc}{chapter}{References}
\bibliographystyle{jhep}

\bibliography{biblio}

\chapter*{Ringraziamenti}

Ci tenevo per prima cosa a ringraziare il mio relatore, Dr. Marco Bonvini, per la disponibilità mostrata in questi mesi, per le videochiamate interminabili e per avermi ``iniziato'' a questo bellissimo campo di studi.

\noindent Poi volevo ringraziare Andrea Barontini con cui ho condiviso questi ultimi sei mesi di Laurea Magistale. Senza tutto l'aiuto reciproco che ci siamo dati e le innumerevoli chiamate che abbiamo fatto probabilmente saremmo ancora con la testa china sui libri.

\noindent Un ringraziamento speciale va a mia madre per avermi sempre supportato e per essere stata una fonte di ispirazione. 

\noindent Ringrazio anche i miei fratelli Jacopo e Simone. 

\noindent Ringrazio Matteo che è praticamente da tutta la vita che mi sopporta (e viceversa!). 

\noindent Infine un ringraziamento va a tutti i miei amici con cui abbiamo condiviso momenti bellissimi. Ogni tanto è bello fare gli scemi tutti insieme.

\end{document}